\documentclass[prd,twocolumn,superscriptaddress,nofootinbib]{revtex4-2}
\usepackage{epsfig}
\usepackage{graphicx}
\usepackage{palatino}
\usepackage[english]{babel}
\usepackage{hyphenat}
\usepackage{amsmath}
\usepackage{amssymb}
\usepackage{slashed}
\usepackage{epstopdf}
\usepackage{xcolor}
\usepackage{booktabs}
\usepackage{lipsum, babel}
\usepackage[toc,page]{appendix}
\definecolor{lcolor}{rgb}{0.,0.0,0.}
\definecolor{citcolor}{rgb}{0,0.,0.5}
\usepackage[breaklinks,colorlinks,urlcolor=blue,citecolor=blue,linkcolor=blue]{hyperref}
\usepackage{multirow}
\usepackage{ltablex}
\usepackage{bbm}
\usepackage{soul}
\usepackage{ulem} 
\RequirePackage{xspace}

\begin{document}
\title{Isolating perturbative QCD splittings in heavy-ion collisions}
\author{Leticia Cunqueiro}
\affiliation{Sapienza Universit\`a di Roma \& INFN, Piazzale Aldo Moro 5, Roma 00185, Italy }
\author{Daniel Pablos}
\affiliation{INFN, Sezione di Torino, via Pietro Giuria 1, I-10125 Torino, Italy}
\affiliation{Departamento de F\'isica, Universidad de Oviedo, Avda. Federico Garc\'ia Lorca 18, 33007 Oviedo, Spain}
\affiliation{Instituto Universitario de Ciencias y Tecnolog\'ias Espaciales de Asturias (ICTEA), Calle de la Independencia 13, 33004 Oviedo, Spain}
\author{Alba Soto-Ontoso}
\affiliation{CERN, Theoretical Physics Department, CH-1211 Geneva 23, Switzerland}
\author{Martin Spousta}
\affiliation{Institute of Particle and Nuclear Physics, Faculty of Mathematics and Physics, Charles University, V Hole\v sovi\v ck\' ach 2, 180 00 Prague 8, Czech Republic}
\author{Adam Takacs}
\affiliation{
Department of Physics and Technology, University of Bergen, Allegaten 55, 5007 Bergen, Norway}
\affiliation{
Institute for Theoretical Physics, Heidelberg University, Philosophenweg 16, 69120 Heidelberg, Germany}
\author{Marta Verweij}
\affiliation{Institute for Gravitational and Subatomic Physics (GRASP), Utrecht University, Utrecht, Netherlands}
\affiliation{Nikhef, National institute for subatomic physics, Amsterdam, Netherlands}

\begin{abstract}
    We define a new strategy to scan jet substructure in heavy-ion collisions. The scope is multifold: (i) test the dominance of vacuum jet dynamics at early times, (ii) capture the transition from coherent to incoherent jet energy loss, and (iii) study elastic scatterings in the medium, which are either hard and perturbative or soft and responsible for jet thermalisation.
    To achieve that, we analyse the angular distribution of the hardest splitting, $\theta_{\rm hard}$, above a transverse momentum scale, $k_t^{\rm min}$, in high-$p_t$ jets. 
    Sufficiently high values of $k_t^{\rm min}$ target the regime in which the observable is uniquely determined by vacuum-like splittings and energy loss, leaving the jet substructure unmodified compared to proton-proton collisions. 
    Decreasing $k_t^{\rm min}$ enhances the sensitivity to the relation between energy loss and the intra-jet structure and, in particular, to observe signatures of colour decoherence at small angles. At wider angles it also becomes sensitive to hard elastic scatterings with the medium and, therefore, the perturbative regime of medium response. Choosing $k_t^{\rm min}\approx 0$ leads to order one effects of non-perturbative origin such as hadronisation and, potentially, soft scatterings responsible for jet thermalisation. We perform a comprehensive analysis of this observable with three state-of-the-art jet-quenching Monte Carlo event generators. Our study paves the way for defining jet observables in heavy-ion collisions dominated by perturbative QCD and thus calculable from first principles. 
\end{abstract}

\maketitle

%%%%%%%%%%%%%%%%%%%%%%%%%%%%%
%%%%%%%%%%%%%%%%%%%%%%%%%%%%%
\section{Introduction}
\label{sec:intro}
%%%%%%%%%%%%%%%%%%%%%%%%%%%%%
%%%%%%%%%%%%%%%%%%%%%%%%%%%%%

Collisions of heavy ions at ultra-relativistic energies produce a myriad of particles ($\mathcal{O}(10^3)$). The presence of a quark-gluon plasma (QGP) phase during the system evolution leaves an imprint on the characteristic pattern of final state multi-particle correlations. The ultimate goal of the heavy-ion program is to characterise the QGP in terms of its transport properties and microscopic structure. One of the main challenges is to design and measure observables that are both QGP standard candles and amenable to theoretical computations. A successful approach to this problem is the use of jet observables to measure their modification compared to jets in proton-proton collisions. Experimental overviews on jet modification are provided in Refs.~\cite{Armesto:2015ioy,Connors:2017ptx,Cunqueiro:2021wls,Apolinario:2022vzg,ALICE:2022wpn}, while for theory see Refs.~\cite{Casalderrey-Solana:2007knd,Majumder:2010qh,Mehtar-Tani:2013pia}. 
While early measurements used jets as monolithic objects, most recent measurements focus on the internal structure of jets (see Refs.~\cite{Connors:2017ptx,Cunqueiro:2021wls,Apolinario:2022vzg} for more details).
The main model-agnostic conclusion that can be drawn from these measurements is that in-medium jets lose around $10\%$ of their energy, i.e. are quenched, by means of copious out-of-cone radiation induced by the QGP and that this energy degradation is sensitive to the jet substructure. 

The theoretical description of some of the key processes in jet formation in heavy-ion collisions is in the realm of perturbative QCD and is thus calculable. Effects that pertain to this category are, e.g. nuclear-modified parton distribution functions~\cite{Ethier:2020way,Eskola:2021nhw,AbdulKhalek:2022fyi}, the initial hard-scattering (prior to QGP formation), and early vacuum-like QCD radiation~\cite{Kurkela:2014tla,Caucal:2018dla}. Medium-induced emissions, triggered by the interaction between the jet particles and the QGP colour fields~\cite{Baier:1996kr, Zakharov:1996fv, Gyulassy:1999zd,Wiedemann:2000za}, in most cases, are assumed to be perturbative. 
Their description requires phenomenological input, namely a model of the QGP. A pQCD description of jet evolution breaks down when the partonic cascade reaches energy scales around the QGP temperature, that is, $\mathcal{O}(\Lambda_{\rm QCD})$. Then, other non-perturbative effects such as thermalisation and hadronisation become relevant.
Thermalisation can be studied with QCD kinetic theory~\cite{Ghiglieri:2015zma,Schlichting:2019abc,Berges:2020fwq} (using an artificially big coupling), while the formation of hadrons rely on phenomenological modelling~\cite{Andersson:1983ia, Webber:1983if,Beraudo:2011bh}. Alternatively, strongly coupled descriptions of parton energy loss have also been studied using the gauge/gravity duality~\cite{Chesler:2008uy,Gubser:2008as,Hatta:2008tx,Arnold:2010ir,Chesler:2014jva}. While they offer a natural scenario for jet hydrodynamisation in a strongly coupled plasma, the underlying QFT is not QCD, but $\mathbf{N}=4$ SYM, implying the presence of large theoretical uncertainties in the extrapolation of the results to the system actually produced in heavy-ion collisions.

Due to the intricate interplay among the aforementioned effects and the multi-scale nature of the process, an end-to-end analytic approach to in-medium jet evolution that matches the experimental precision is currently beyond reach. As a consequence, the theoretical interpretation of jet observables relies almost exclusively on phenomenological modelling by means of Monte Carlo event generators. Several implementations of in-medium parton showers have been proposed in the literature~\cite{Lokhtin:2005px,Zapp:2008gi,Armesto:2009fj,Casalderrey-Solana:2011fza,Schenke:2009gb,Majumder:2013re,Wang:2013cia,Casalderrey-Solana:2014bpa,Caucal:2018dla,Bierlich:2018xfw,Ke:2020clc}. They differ not only in the precision in which they describe the individual ingredients of jet evolution but also in the way they assemble them. A paradigmatic example, that will be relevant in this paper, is the interleaving of vacuum and medium-induced emissions. While some approaches implement either a partial or an exact factorisation between vacuum and medium-induced emissions~\cite{Caucal:2018dla,JETSCAPE:2023hqn}, others include them on equal footing in their evolution equations~\cite{Armesto:2009ab, Armesto:2009fj,Zapp:2012ak}. Experimental measurements have not yet been able to pin down which is the correct approach. Consolidating the theoretical description of in-medium jet evolution thus requires experimental guidance by means of more differential and/or precise measurements.

This paper aims to disentangle different stages of jet evolution by combining two jet substructure observables. Our proposal strongly relies on the use of high-$p_t$ jets to ensure a clear separation between perturbative and medium-related scales. A key point in our study is the introduction of an auxiliary, intermediate energy scale that allows us to sweep through different phases of jet evolution in a controlled fashion. By means of state-of-the-art Monte Carlo simulations, we quantitatively address the following fundamental questions:
\begin{itemize}
    \item Is there a regime of pure vacuum evolution in the in-medium development of a parton shower? 
    \item Does energy loss depend on the opening angle of the splitting? If so, at which energy scale does this effect become relevant? 
    \item Are elastic scatterings with the medium visible in jet substructure observables?
\end{itemize}  

The underlying philosophy and technical details of the proposed observable together with its connection to previous measurements are explained in Section~\ref{sec:analysis}. The proton-proton baseline is studied in Sec.~\ref{sec:vacuum} including a comparison between state-of-the-art pQCD and the vacuum prediction of jet quenching event generators. Quantitative results showing the discriminating power of the proposed observable in heavy-ion collisions can be found in Sec.~\ref{sec:medium}. We end up with a brief summary of our results in Sec.~\ref{sec:summary}. The experimental feasibility of this measurement with the upcoming Runs 3 and 4 of the LHC, some analytic considerations, the impact of energy loss prescriptions and medium response are studied in Appendices~\ref{app:exp}, \ref{app:analytics}, \ref{app:lres-dep}, and \ref{app:med-response}, respectively. 

%%%%%%%%%%%%%%%%%%%%%%%%%%%%%
%%%%%%%%%%%%%%%%%%%%%%%%%%%%%
\section{Analysis strategy}
\label{sec:analysis}
%%%%%%%%%%%%%%%%%%%%%%%%%%%%%
%%%%%%%%%%%%%%%%%%%%%%%%%%%%%
\begin{figure}
    \centering
    \includegraphics[width=0.8\columnwidth]{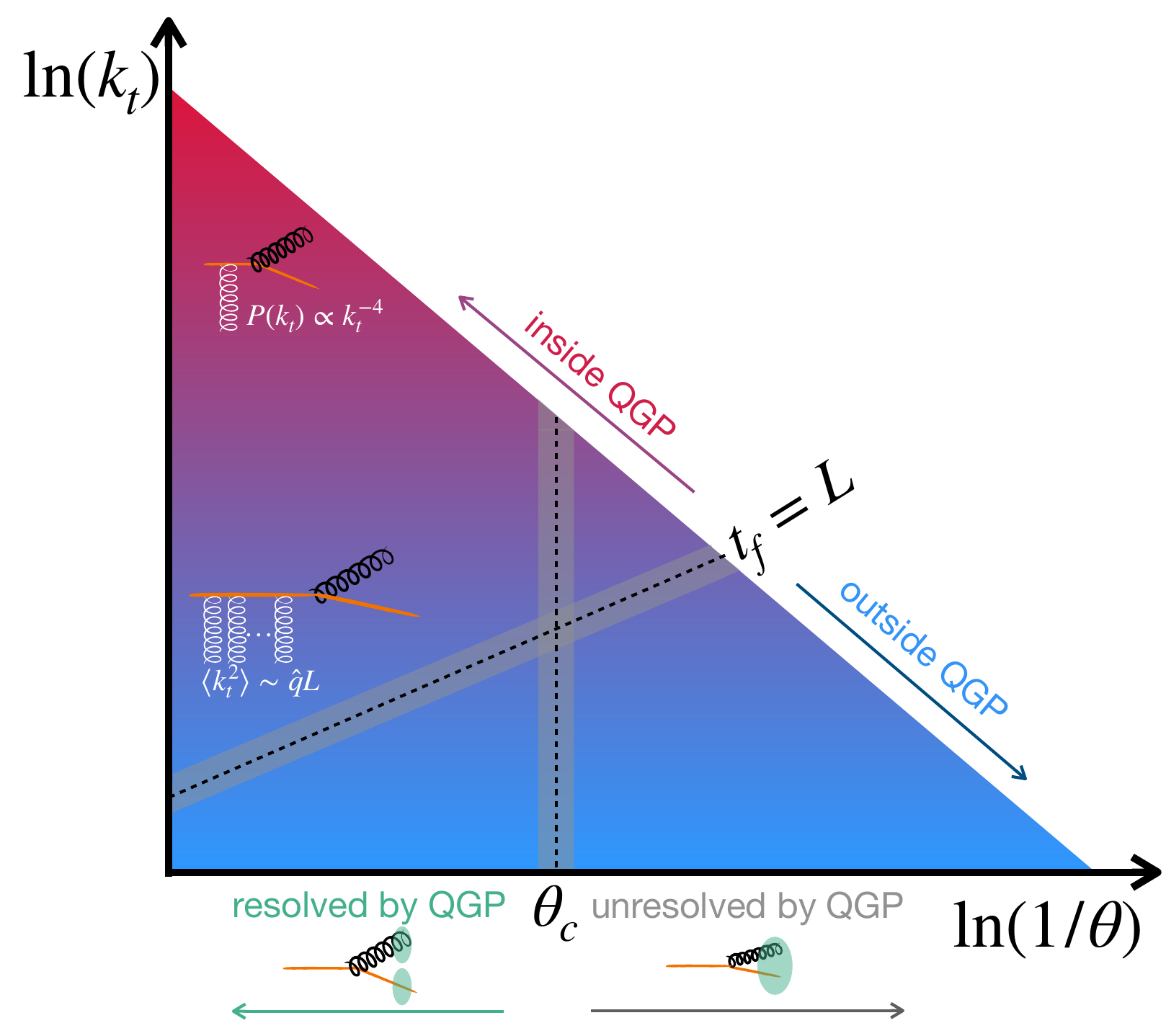}
    \caption{Sketch depicting the different regimes of in-medium 
    jet evolution in a Lund-plane style representation. The gray bands around $\theta=\theta_c$ and $t_f=L$ indicate that these two scales fluctuate on an event-by-event basis.}
    \label{fig:sketch-lund}
\end{figure}
Our goal is to design and study an observable to probe certain corners of the radiation phase-space of an in-medium jet, as sketched in a Lund-plane fashion in Fig.~\ref{fig:sketch-lund}. Before entering into the precise definition of the observable, let us briefly comment on the most relevant features of this radiation phase space.
%%%%%%%%%%%%%%%%%%%%%%%%%%%%%
\subsection{Brief reminder of in-medium jet evolution}
%%%%%%%%%%%%%%%%%%%%%%%%%%%%%
For a given jet with transverse momentum $p_t$ and cone size $R$, we characterise its branchings by their opening angle $\theta$ and their relative transverse momentum $k_t$.\footnote{We follow the standard Lund jet plane definitions, $\theta = \sqrt{(y_1-y_2)^2+(\phi_1-\phi_2)^2}$, $k_t=\min (p_{t1}, p_{t2})\theta$, where the two prongs of the splitting are denoted with subscripts 1 and 2~\cite{Dreyer:2018nbf}.} The maximum relative transverse momentum is $k_t<p_tR$ ($\mathcal{O}(10^2)$ GeV in this study). Another important kinematic variable is the formation time of the splitting $t_f=2/(k_t\theta)$, which describes the quantum mechanical extent of a branching~\cite{Dokshitzer:1991wu}.

Jet constituents interact elastically and inelastically with the QGP colour fields. The former results in transverse momentum broadening, while the latter induces additional radiation. Therefore, splittings in an in-medium parton shower can be either vacuum-like (i.e. described by the Altarelli-Parisi splitting function~\cite{Altarelli:1977zs}) or medium-induced, with a branching probability that depends on the medium properties. For example, modelling the jet-QGP interaction in the multiple soft scattering approximation~\cite{Baier:1996kr,Zakharov:1996fv,Wang:1994fx} results into Gaussian broadening with accumulated transverse momentum $\langle k_t^2\rangle\sim\hat q L$, with $\hat q$ the so-called quenching parameter and $L$ the propagation length. Since the typical value of $\hat q$ is around $1$ to $3$ GeV$^{2}$/fm~\cite{Andres:2016iys,Feal:2019xfl,JETSCAPE:2021ehl}, elastic and inelastic medium processes mainly contribute in the soft regime ($k_t\sim\mathcal O(1)$ GeV) of Fig.~\ref{fig:sketch-lund}. Another important feature is that these medium-induced emissions typically occur at wide angles, drifting energy out of the jet cone. Alternatively, hard collisions with the medium can occur above $\langle k_t^2\rangle>\hat qL$ and induce emissions with a Coulomb-like transverse momentum distribution, i.e. $\propto 1/k_t^4$ tail~\cite{Gyulassy:1999zd,Wiedemann:2000za}. These Molière/higher-twist scatterings~\cite{Wang:2001ifa,DEramo:2018eoy} compete with vacuum-like radiation in the high-$k_t$ regime, although their occurrence is relatively rare for dense media. 

Another medium-related scale highlighted in Fig.~\ref{fig:sketch-lund} is the de-coherence angle, $\theta_c\sim (\sqrt{\hat q L^3})^{-1}$~\cite{Mehtar-Tani:2010ebp, Casalderrey-Solana:2011ule}. If the opening angle of the splitting is small, the medium cannot resolve the two daughters through soft scatterings, and emissions are coherently induced on this composite, $2$-parton object. Above $\theta_c$, the daughters are resolved, and both source medium-induced radiation. 

To illustrate the importance of $\theta_c$, consider the extreme case in which the first splitting of an angular-ordered parton shower has $\theta < \theta_c$. All emissions inside the jet will be unresolved by the medium, its substructure remains unmodified, and the jet loses energy as a single object (just the initiator is quenched). In turn, when a splitting satisfies $\theta>\theta_c$, both prongs act as independent radiators of medium-induced emissions, resulting in a larger energy loss. This angular dependence of energy loss translates into a sizeable modification of jet substructure observables by means of a selection bias, or survivor bias effect. Due to the presence of a steeply falling jet $p_t$ spectrum, inclusive jet ensembles at any given $p_t$ in heavy-ion collisions will be dominated by those jets that on average lost the least amount of energy. Therefore, the number of jets with wide-angle substructure, $\theta>\theta_c$, is suppressed compared to proton-proton collisions in any inclusive jet observable where unavoidably minimum jet $p_t$ thresholds are imposed~\cite{Rajagopal:2016uip,Brewer:2018dfs, Brewer:2020och,Brewer:2021hmh,Takacs:2021bpv,Pablos:2022mrx}.\footnote{Selection biases are also present even if energy loss is independent of the jet substructure. For example, gluon jets are more suppressed since, on average, they lose more energy than quark jets due to their larger colour factor. As gluon jets are typically wider, this bias would also cause some narrowing of the final quenched jet ensemble. This is sometimes referred to as a bias in the $q/g$-fraction~\cite{Spousta:2015fca,Qiu:2019sfj,Ringer:2019rfk}. Additionally, the selection bias mixed with colour resolution (i.e. small angle splittings do not source energy loss) gives even stronger narrowing and it is necessary to explain jet measurements~\cite{Caucal:2021cfb,Pablos:2022mrx,JETSCAPE:2023hqn,ATLAS:2022vii}.\label{foot:qg_fraction}} 

An important observation is that varying the jet reconstruction parameters, ($p_t, R$), enables the separation of medium dynamics from vacuum physics. For example, raising the jet $p_t$ increases the phase space for high-energy emissions (i.e. the upper diagonal of the emission phase space in Fig.~\ref{fig:sketch-lund} moves to higher values). Since $\hat q$ depends only mildly on energy~\cite{Liou:2013qya,Blaizot:2014bha}, the scale at which medium-induced emissions dominate ($k^2_t\sim \hat q L$) will remain the same and the new phase-space region should be mainly populated by vacuum emissions. Similarly, increasing the jet radius $R$ further separates the $\theta=R$ boundary from the $\theta_c$ scale (again almost independent of the jet energy). We will exploit these facts in the definition of our observable.

%%%%%%%%%%%%%%%%%%%%%%%%%%%%%
\subsection{Observable definition}
%%%%%%%%%%%%%%%%%%%%%%%%%%%%%

We cluster events with the anti-$k_t$ algorithm~\cite{Cacciari:2008gp,Cacciari:2011ma} using a jet radius of $R=0.2$ and select all jets that satisfy $p_t>400$ GeV, and $|\eta|<2.8$. This subset of jets is then reclustered with the Cambridge/Aachen algorithm~\cite{Dokshitzer:1997in,Wobisch:1998wt}, effectively reordering the branching history in decreasing angles. For each of these jets, we find the hardest splitting, i.e. the splitting with maximum $k_t=\min(p_{t1},p_{t2})\theta$. If the splitting satisfies $k_t>k_t^{\rm min}$, we record the opening angle of this hardest branching, that we denote $\theta_{\rm hard}$.

Let us now justify each of the choices that we have made in the previous paragraph and comment on their experimental feasibility.
\begin{itemize}
    \item $R=0.2$: there are certain advantages of using jets with small cone sizes. The physics associated with vacuum-like emissions and coherence that we are interested in the present work typically appear at small angles ($<0.1$). The reduced jet area decreases the contamination from the uncorrelated fluctuating background and notably improves experimental systematic uncertainties. Moreover, smaller jet cones also reduce the contribution from medium response, which typically populates the larger angle region, thus further improving the signal.
    
    \item $p_t>400$ GeV: the larger the jet $p_t$ is, the more pQCD dominated the observable is. This guarantees not only calculability but also a large separation of scales, breaking up the problem of in-medium jet evolution into simpler pieces. Going to higher momenta, however, significantly reduces the statistics. The expected number of jets in PbPb collisions during Runs 3 and 4 at the LHC is discussed in Appendix~\ref{app:exp}.
    
    \item $k^{\rm min}_t$: this auxiliary variable is central to our approach. Its role is to slice the radiation phase space, separating different regimes of jet evolution.  
    We explore three values: $20$, $5$, and $1$ GeV. The highest $k^{\rm min}_t$ selects splittings at the top of the emission phase space (Fig.~\ref{fig:sketch-lund}), with formation times less than $0.4$ fm, and addresses the question of whether vacuum splittings indeed dominate the early evolution of jets. In the multiple soft scatterings approximation, medium-induced emissions will typically appear at $k_t^2\approx\hat qL\approx 2^2$ GeV$^2$ and $\theta>\theta_c$. Therefore, if the value of the cut is large enough $k_t^{\min}\gg\sqrt{\hat qL}$, one expects a vacuum-like substructure even in the medium. We refer to this as the factorisation (or separation) of vacuum-like and medium physics. Lowering $k^{\rm min}_t$ (but still keeping it far from non-perturbative scales $\mathcal{O}(1)$ GeV) opens up the phase space for perturbative collinear splittings below $\theta_c$, and the observable becomes sensitive to splittings being resolved or unresolved by the medium. In addition, it could potentially be affected by perturbative elastic scatterings and medium-induced emissions, whose modelling changes in different jet quenching models. Finally, the lowest value of $k^{\rm min}_t$ maps the region dominated by non-perturbative physics. 

    \item $\theta_{\rm hard}$: the angular distribution of the hardest splitting is known to be sensitive to the substructure-dependence of energy loss~\cite{Caucal:2021cfb}, e.g. as caused by $\theta_c$~\cite{Mehtar-Tani:2010ebp,Casalderrey-Solana:2012evi}. It is also motivated by the simplified theoretical picture in which jet energy loss depends on an angular resolution scale, namely $\theta_c$~\cite{Mehtar-Tani:2010ebp,Casalderrey-Solana:2012evi}. From the point of view of distinguishing vacuum from medium-induced emissions, this choice is also reasonable since the latter typically occur at large angles as a result of transverse momentum broadening.  

\end{itemize}

In the landscape of other jet substructure studies, the proposed algorithm is unique in the sense that it first selects the hardest splitting above the $k_t^{\rm min}$ cut and then studies its angular distribution. 
The main difference with respect to \textit{dynamical grooming} observables with $a=1$~\cite{Mehtar-Tani:2019rrk} is precisely $k_t^{\rm min}$. Predictions for dynamically groomed observables in heavy-ion collisions have been put forward in Refs.~\cite{Caucal:2021cfb,Wang:2022yrp,ALICE:2022hyz}, demonstrating its sensitivity to colour coherence. Given our interest in studying the earliest stages of the shower, selecting the splitting with the shortest formation time might a priori be appealing~\cite{Mehtar-Tani:2019rrk,Apolinario:2020uvt,Apolinario:2022guz}. Nevertheless, these splittings can have an arbitrarily low $k_t$, and therefore the observable is polluted with non-perturbative corrections. Including a transverse momentum cut was also explored in the \textit{Late-$k_t$} approach~\cite{Cunqueiro:2022svx}, where instead of the hardest, the most collinear splitting above a certain $k_t^{\rm min}$ is selected. This option is not well suited for our purposes since we want to explore the full angular distribution. 
Furthermore, \textit{SoftDrop}~\cite{Larkoski:2014wba} with $\beta=1$ and $z_{\rm cut}=k_t^{\min}/(p_tR)$ would probe the same phase space as we do, but it also biases the angular information by selecting the widest instead of the hardest splitting. So far, the SoftDrop setup mostly explored in heavy-ion collisions considers $\beta=0$~\cite{Mehtar-Tani:2016aco,Chien:2016led,Milhano:2017nzm, Casalderrey-Solana:2019ubu,CMS:2017qlm, STAR:2020ejj,ATLAS:2022vii,ALICE:2022hyz,CMS:2018fof}. Setting $\beta=0$ does not restrict the $k_t$ of branchings and, therefore, non-perturbative effects might still be sizeable. Instead of selecting the hardest splitting per jet, one could choose all primary splittings with $k_t>k_t^{\rm min}$, corresponding to a slice of the \textit{primary jet Lund plane}~\cite{Dreyer:2018nbf,Andrews:2018jcm}. This option would explore all the physics mechanisms we are interested in but, for the sake of simplicity, we stick to the one-splitting-per-jet case. Finally, resolution effects on \textit{energy-correlators} are also currently under study~\cite{Andres:2022ovj,Andres:2023xwr}. 

%%%%%%%%%%%%%%%%%%%%%%%%%%%%%
%%%%%%%%%%%%%%%%%%%%%%%%%%%%%
\section{Proton-proton baseline}
\label{sec:vacuum}
%%%%%%%%%%%%%%%%%%%%%%%%%%%%%
%%%%%%%%%%%%%%%%%%%%%%%%%%%%%
\begin{figure}
    \centering
    \includegraphics[width=0.99\columnwidth]{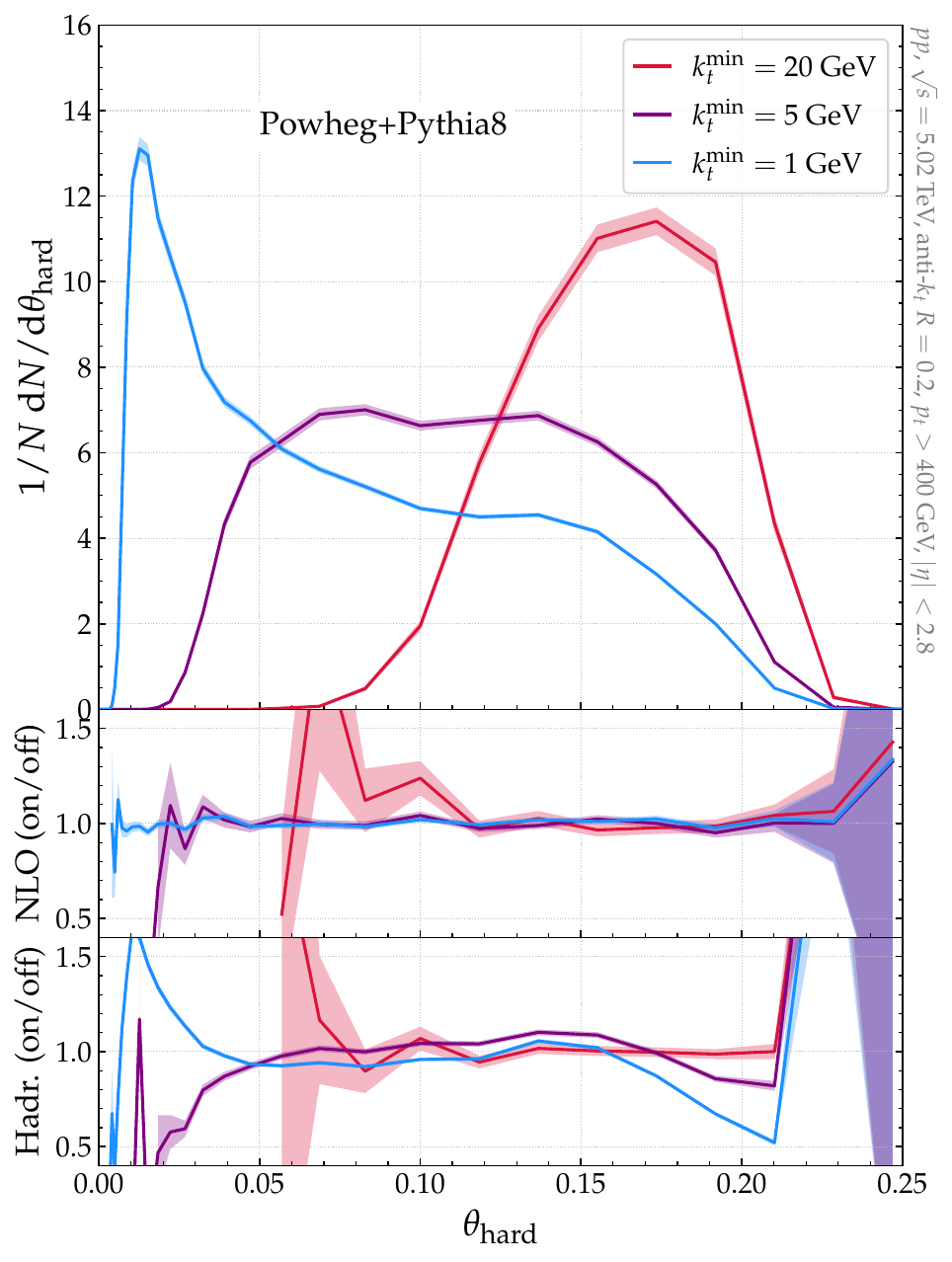}
    \caption{The angular distribution of the hardest-$k_t$ splitting inside inclusive, pp jets for different $k^{\rm min}_t$ values in NLO Powheg+Pythia8. The bottom panels show: (i) the impact of NLO matching, and (ii) the impact of hadronisation on the pure shower samples, i.e. not matched to NLO. In all cases, the band represents the statistical uncertainty. 
    }
    \label{fig:pythia8}
\end{figure}

\begin{figure*}
    \centering
    \includegraphics[width=0.33\textwidth]{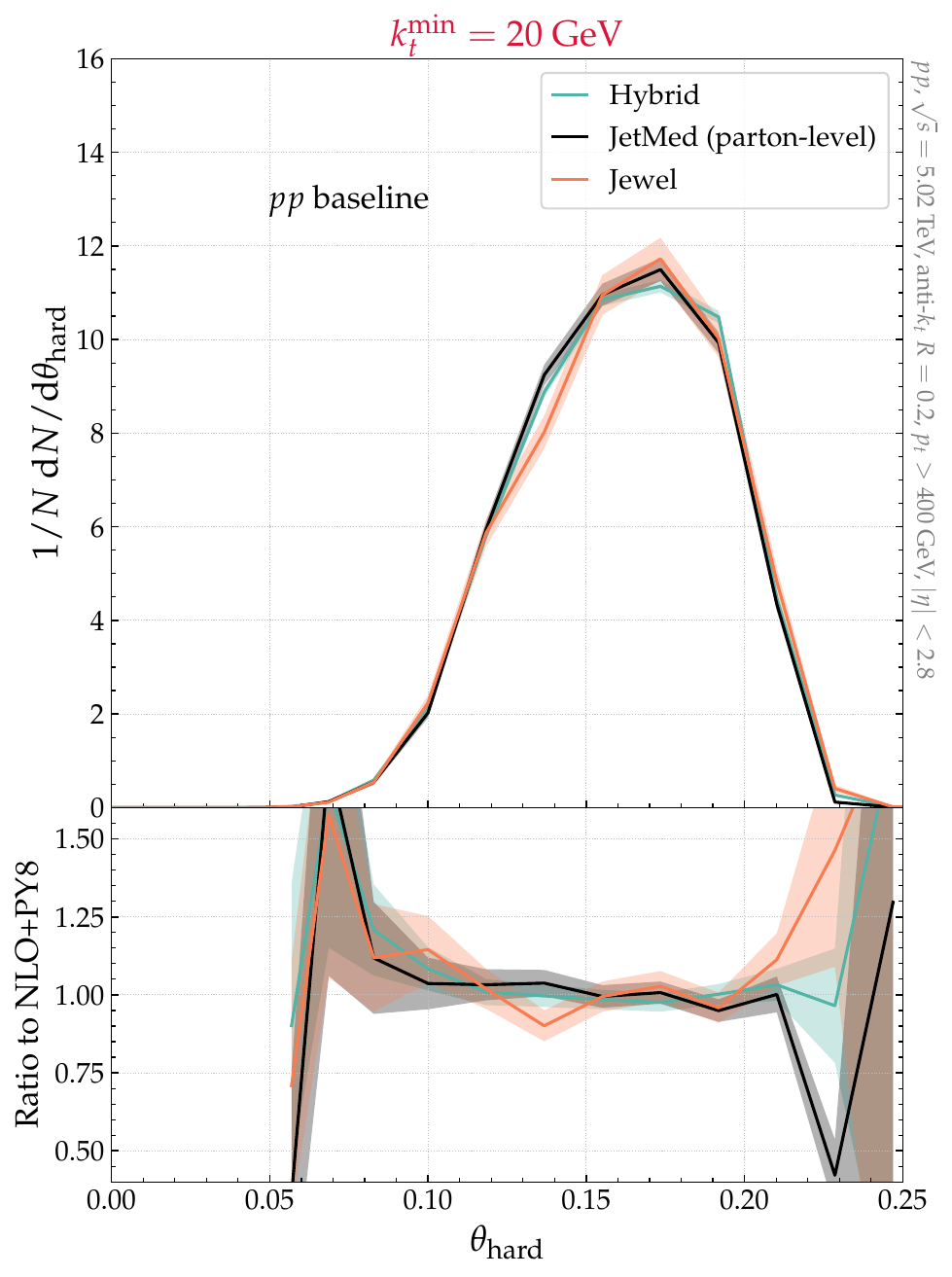}
    \includegraphics[width=0.33\textwidth]{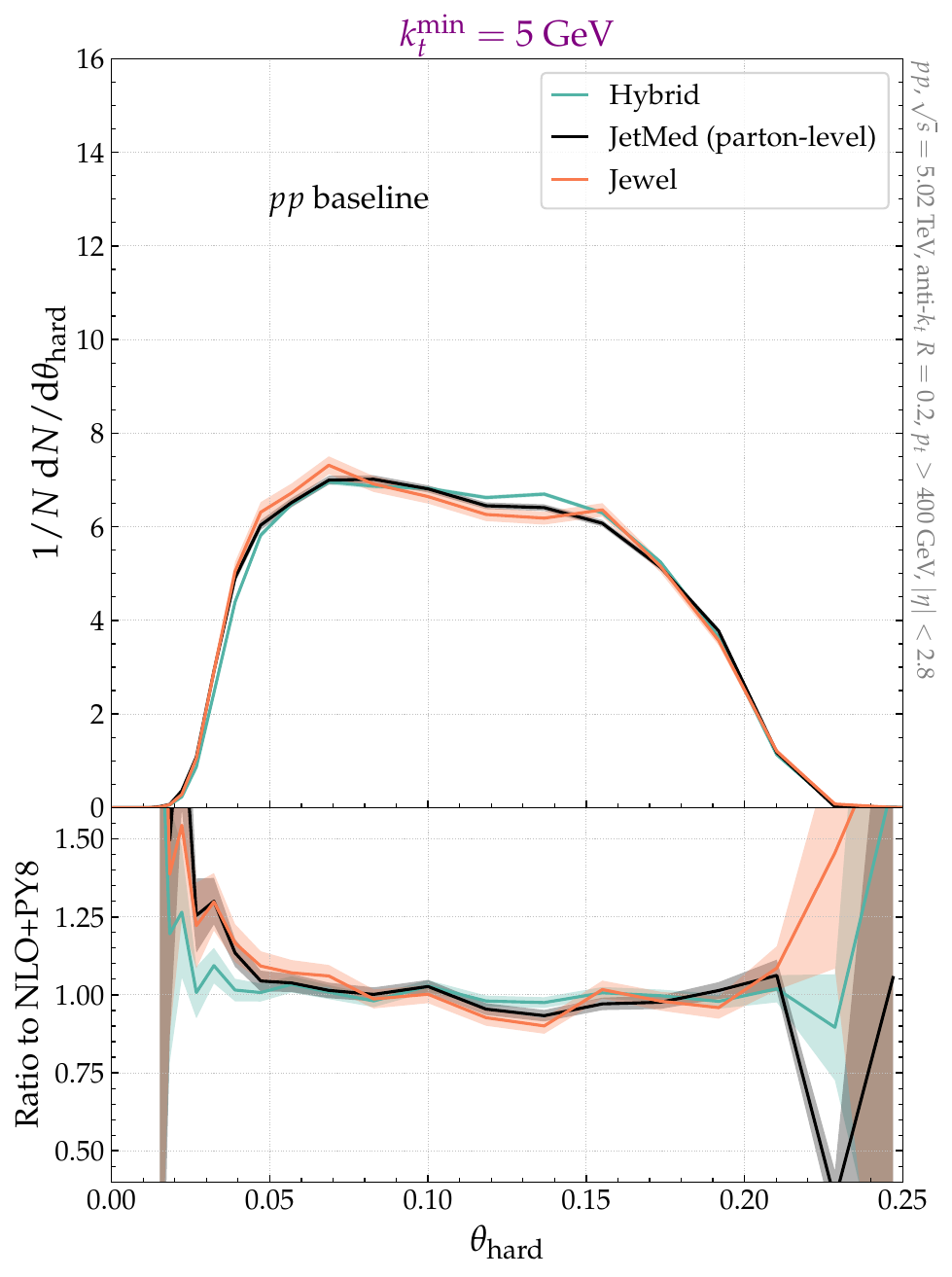}
    \includegraphics[width=0.33\textwidth]{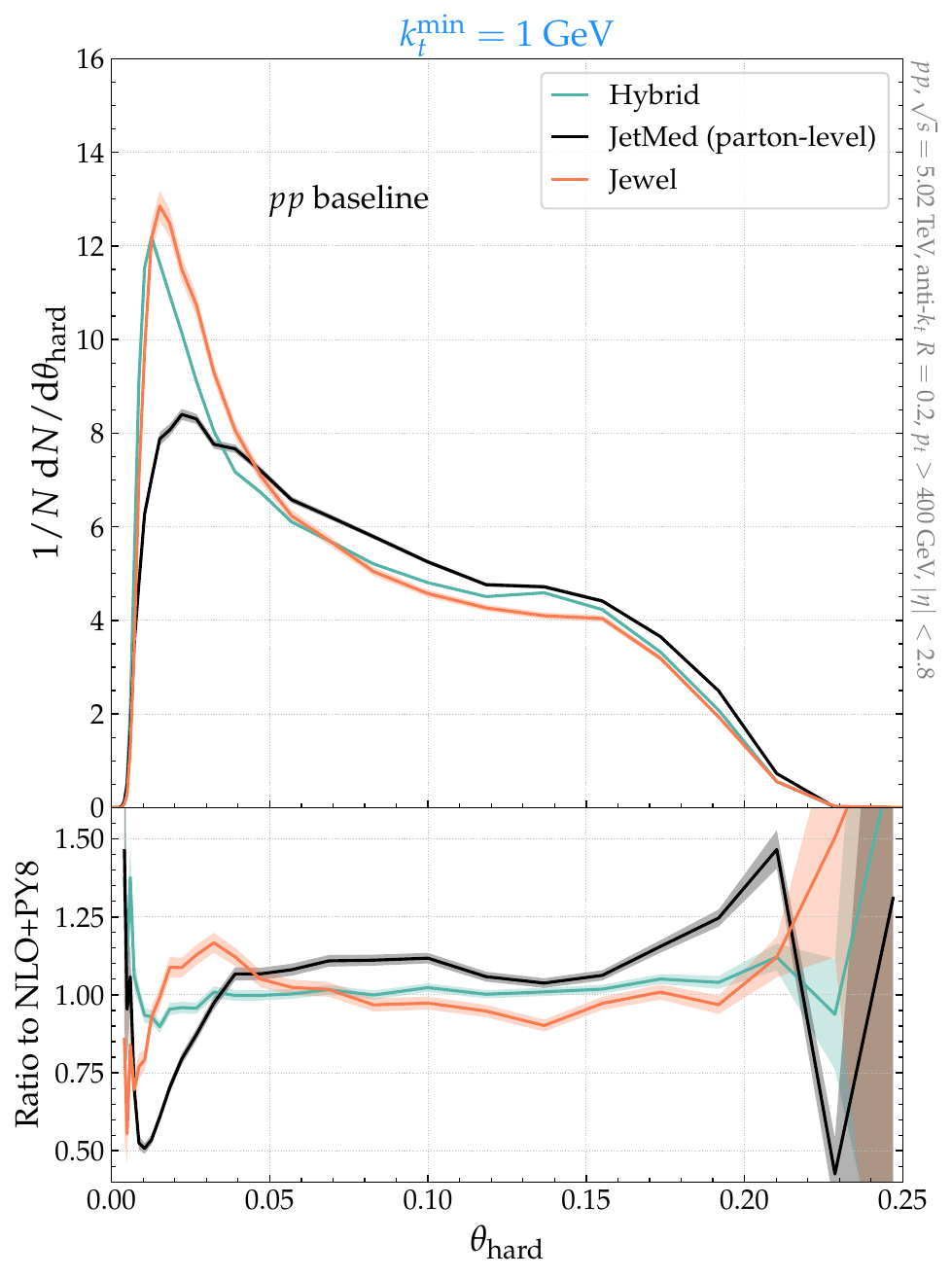}
    \caption{The pp baseline prediction for $\theta_{\rm hard}$ by different jet quenching Monte Carlos. From right to left the value of $k^{\rm min}_t$ decreases. The bottom panel displays the ratio to the state-of-the-art prediction, namely Powheg+Pythia8.}
    \label{fig:pp-MCs}
\end{figure*}

We begin our analysis by exploring the behaviour of the observable in proton-proton collisions at $\sqrt s=5.02$ TeV. In Fig.~\ref{fig:pythia8}, we show the self-normalised $\theta_{\rm hard}$-distribution for different values of $k^{\rm min}_t$ as obtained by matching, in a Powheg approach~\cite{Alioli:2010xa}, the exact $pp\to jj$ next-to-leading order (NLO) matrix element to the Pythia8 parton shower~\cite{Bierlich:2022pfr}. We observe a clear shift towards smaller angles when decreasing $k^{\rm min}_t$. This feature can be easily understood from Fig.~\ref{fig:sketch-lund}, since phase-space in the collinear regime opens up when lowering the $k^{\rm min}_t$ constraint. The lowest accessible angle is given by $\theta^{\rm min}_{\rm hard}\propto k^{\rm min}_{t}/p_t$.\footnote{Notice that our jet selection is $p_t>400$ GeV, and therefore there is no hard cutoff on the smallest angle.} We anticipate that the net angular reach for a given $k^{\rm min}_t$ will be an important aspect when including medium effects. On the wide-angle side, we note that angles larger than $R$ are possible due to the C/A reclustering, but they are highly unlikely. The error bands correspond to statistical uncertainties only (combined in the ratio plots).

In the lower panels of Fig.~\ref{fig:pythia8} we quantify the impact of NLO corrections and hadronisation.\footnote{We also performed NLO studies with MadGraph5\_aMC@NLO+Herwig~\cite{Alwall:2014hca,Bellm:2015jjp}, resulting in similar curves, although with a slower statistical convergence.}\textsuperscript{,}\footnote{Multi-parton interaction effects were also studied and they are negligible.} The former is expected to be more relevant for the largest $k^{\rm min}_t$ since, in that regime, the true matrix element notably differs from the soft-and-collinear approximation used by the parton shower. Indeed, we observe that for $k^{\rm min}_t=1,5$ GeV, NLO corrections are practically irrelevant. In turn, we find a moderate effect of about $20\%$ for $k^{\rm min}_t=20$ GeV. 

Hadronisation effects have the opposite $k^{\rm min}_t$-dependence compared to NLO corrections. The deeper we go into the infrared regime by lowering $k^{\rm min}_t$, the bigger the corrections are. For $k^{\rm min}_t = 20$ GeV we find that hadron and parton curves agree within statistical uncertainty (except in the $\theta_{\rm hard}>R$ region coming from hadrons on the boundary of the jet cone), thus demonstrating the pQCD purity of this region of phase-space. The intermediate value of $k^{\rm min}_t$ receives less than $20\%$ hadronisation corrections in a wide angular region, while they reach up to $50\%$ for $k^{\rm min}_{t}=1$ GeV. 

Next, we compare these state-of-the-art pQCD results with the pp (i.e. vacuum) baseline of the jet quenching Monte Carlo codes that we use in this paper: Hybrid~\cite{Casalderrey-Solana:2014bpa}, JetMed~\cite{Caucal:2018dla,Caucal:2019uvr}, and Jewel~\cite{Zapp:2008gi,Zapp:2012ak}. This comparison is relevant since none of the current jet-quenching MCs implement NLO corrections. It could be argued that deficiencies in vacuum modelling are irrelevant since they would potentially cancel in a medium-to-vacuum ratio. However, this statement is exact only when vacuum and medium physics are entirely decoupled. This is actually not the case in most jet-quenching Monte Carlo event generators. A deficient description of the pp baseline can lead to a misleading interpretation of the medium-modified distributions.  
\begin{figure*}
    \centering
    \includegraphics[width=0.33\textwidth]{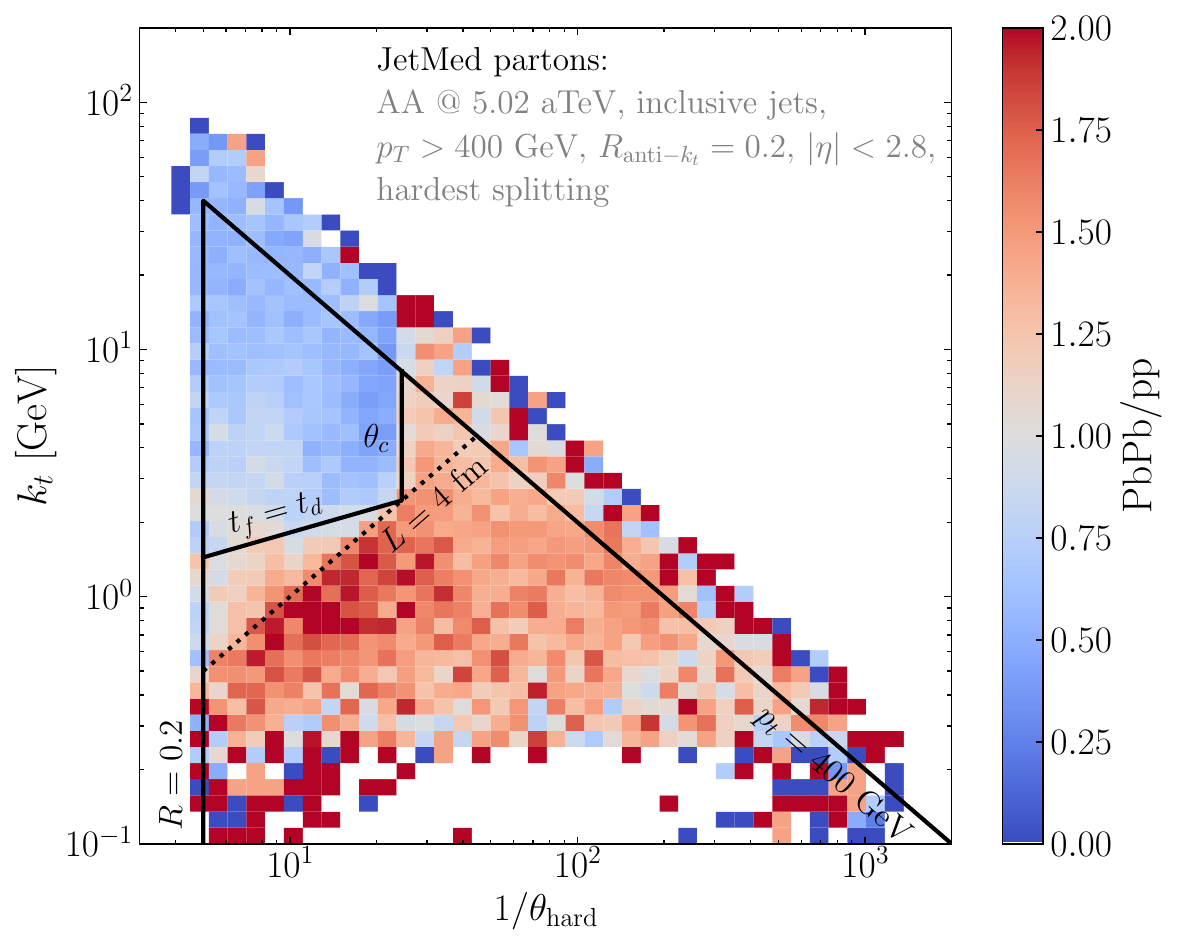}
    \includegraphics[width=0.33\textwidth]{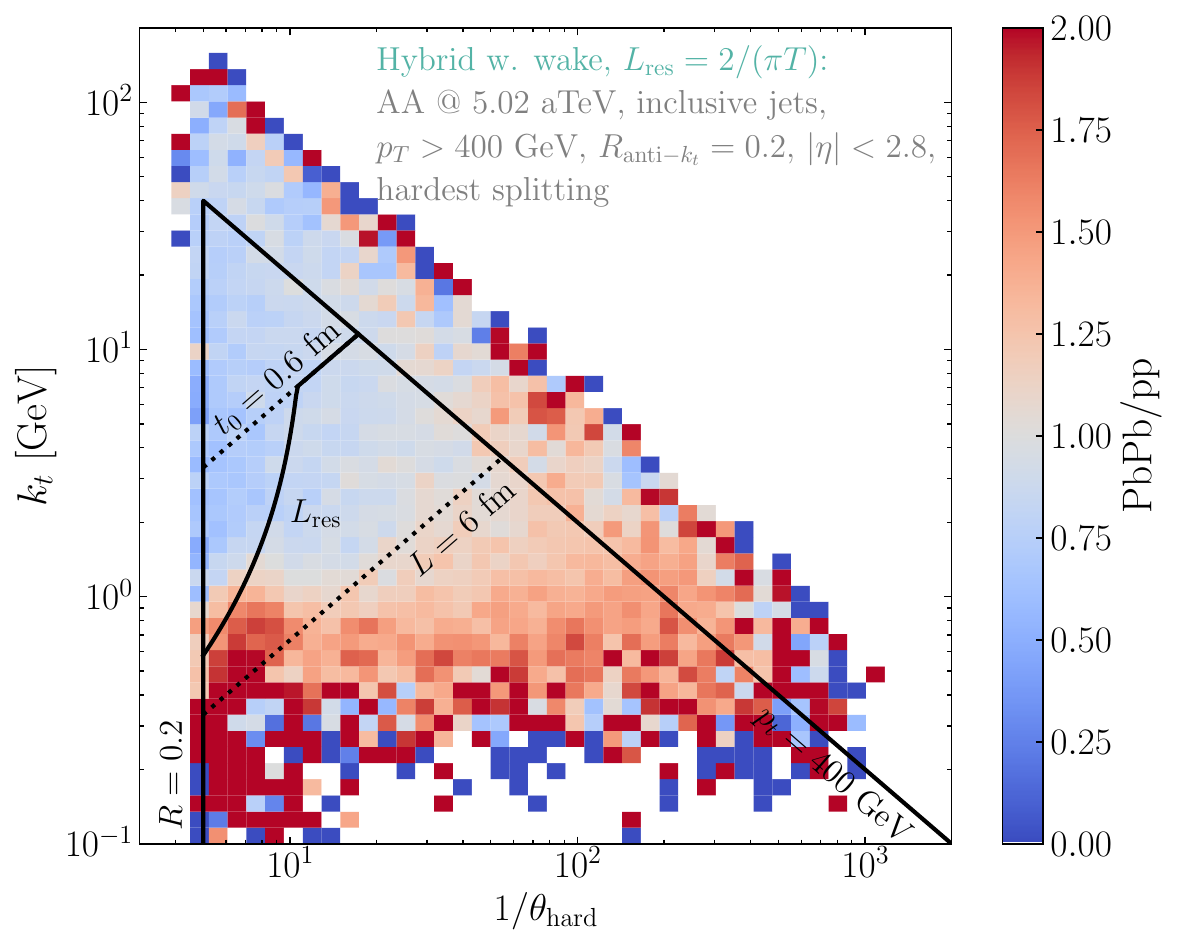}
    \includegraphics[width=0.33\textwidth]{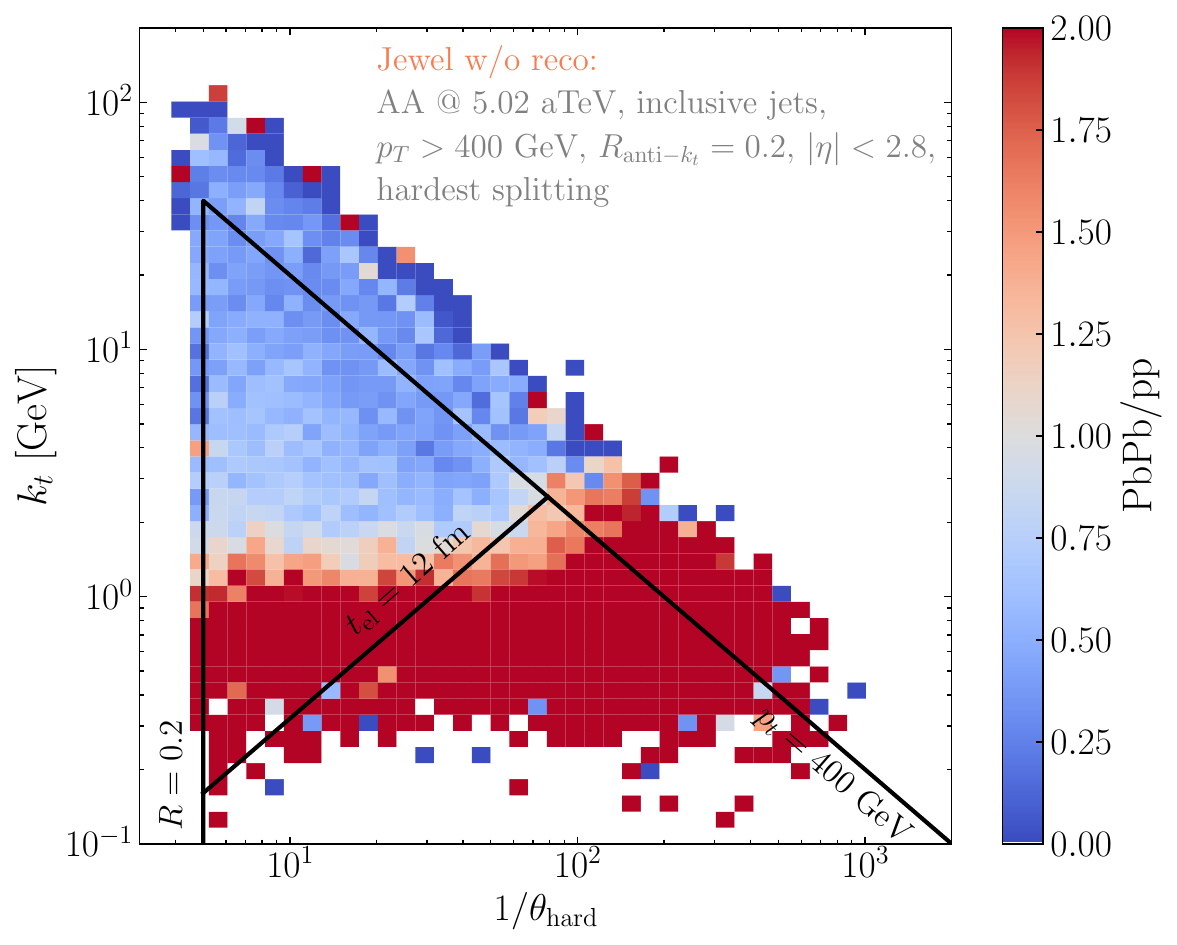}
    \caption{2D distribution of the hardest splitting ($\theta_{\rm hard},k_{t,{\rm hard}}$), using different jet-quenching models (similar to the sketch in Fig.~\ref{fig:sketch-lund}). The ratio to pp illustrates the modification of jets inside the quark-gluon plasma. The approximated resolution phase spaces are denoted with black lines. Note that the ratio extends in the upper diagonal because of our $p_t>400$ GeV selection.}
    \label{fig:PbPb-MCs-LP}
\end{figure*}
The models under comparison have different vacuum showers: Hybrid is built on the default tune of Pythia8.3, while Jewel implements a virtuality-ordered Pythia6-like shower~\cite{Sjostrand:2006za}. They both include initial-state radiation and hadronisation, in contrast to JetMed, which relies on a pure final-state shower at parton level and thus misses many ingredients that might be particularly relevant at low $p_t$ and large jet radii. The three models are compared in Fig.~\ref{fig:pp-MCs} for all values of $k^{\rm min}_{t}$. For the highest $k_t$-cut we observe quite a good agreement ($<10\%$ differences) among the three models. Remarkably, we also find the ratios to the state-of-the-art vacuum baseline to be compatible with unity in the whole angular range, although statistical uncertainties blow up in the edge bins. The fact that the parton-level curve obtained with JetMed agrees with the rest in the bulk of the distribution indicates that the observable is indeed dominated by pQCD physics. This well-defined vacuum baseline is an essential requirement to unambiguously identify medium modifications. For the $k^{\rm min}_{t}=5$ GeV setup, JetMed and Jewel deviate up to $25\%$ in the $\theta_{\rm hard}<0.05$ region from the Powheg+Pythia8 baseline. It is important to keep in mind that this regime of very small angles will play a role in the medium studies. This consensus among models breaks down when $k_{t,\rm min}=1$ GeV. The $50\%$ disagreement in the JetMed case is mainly caused by the lack of hadronisation, as we show in Fig.~\ref{fig:pythia8}. Naturally, the Hybrid model ratio to NLO+Pythia8 is compatible with $1$ since they share the same shower and, in this regime, NLO corrections are negligible. The Pythia6 vs. Pythia8.3 differences result in an up to $20\%$ deviation in the Jewel case. 

After this detailed analysis of the pp baseline in which we have found a remarkably good agreement between Monte Carlo codes for the highest $k^{\rm min}_{t}$ values, we proceed to study medium modifications.

%%%%%%%%%%%%%%%%%%%%%%%%%%%%%
%%%%%%%%%%%%%%%%%%%%%%%%%%%%%
\section{Heavy-ion results}
\label{sec:medium}
%%%%%%%%%%%%%%%%%%%%%%%%%%%%%
%%%%%%%%%%%%%%%%%%%%%%%%%%%%%
In this section we apply our analysis technique for jets in heavy-ion collisions, using some of the most popular jet quenching models. We demonstrate the advantage of slicing the jet substructure by isolating different effects in the medium one by one. These MC studies employ a high number of jets to better illustrate the physics under study. A discussion on the projected experimental statistics and its uncertainties can be found in App.~\ref{app:exp}.

%%%%%%%%%%%%%%%%%%%%%%%%%%%%%
\subsection{Radiation phase-space for MCs under study}
\label{sec:mc-description}
%%%%%%%%%%%%%%%%%%%%%%%%%%%%%
To gain intuition on the medium modifications to the $\theta_{\rm hard}$-observable shown below, we first simulate the analog of Fig.~\ref{fig:sketch-lund} with the three jet-quenching Monte Carlo codes introduced in the previous section: JetMed, Hybrid, and Jewel. That is, we plot in Fig.~\ref{fig:PbPb-MCs-LP} the PbPb-to-pp ratio of the ($k_t$, $\theta_{\rm hard}$) self-normalised density of the hardest splitting within our jet selection. Let us discuss each of the three plots individually and, in particular, the relevant scales therein: 
\begin{itemize}
    \item \textbf{JetMed}~\cite{Caucal:2018dla,Caucal:2019uvr}. The description of the QGP in this model is rather simplified, i.e. it can either run with a Björken expanding media~\cite{Caucal:2020uic} or a brick. In this paper, we use the former with $L=4$ fm and $\hat q = 1.5$ GeV$^2$/fm. The parton-medium interaction is calculated in the multiple soft scatterings approximation. An explicit realisation of colour coherence dynamics is implemented by means of $\theta_c$ (with a constant value event-by-event). The typical relative transverse momentum of induced emissions is $\langle k_t\rangle=\sqrt{\hat q L} \approx 2.4$ GeV, and they appear above the critical angle $\theta>\theta_c=2/\sqrt{\hat qL^3}=0.04$. These induced emissions are responsible for the small enhancement below the $t_f=t_d=[4/(\hat q\theta^2)]^{1/3}$ line.~\footnote{The decoherence time $t_d$ is the time at which a colour-connected dipole gets resolved by the medium, by destroying its colour correlation via multiple colour rotations.} The most prominent feature in Fig.~\ref{fig:PbPb-MCs-LP} is the depletion of emissions in the polygon delimited by the $t_f<t_d$ and $\theta>\theta_c$ boundaries. Vacuum-like emissions in this corner of phase space are resolved by the medium and source more energy loss. Due to the selection bias (described in Sec.~\ref{sec:analysis}) these wide-angle substructure jets are thus suppressed. Another relevant feature of this model is that vacuum-like emissions are forbidden (vetoed) in the $t_d<t_f<L$ region since this is the regime dominated by broadening dynamics and not enhanced by the jet energy. In this context, vacuum and medium-induced cascades do factorise exactly, with the former acting as sources for the latter. Finally, a third stage of vacuum-like showering is also implemented with an extended angular region, i.e., the first emission outside the medium can be emitted at any angle due to the color randomisation suffered by the emitting dipoles during their evolution throughout the medium~\cite{Casalderrey-Solana:2012evi}. This third stage of the shower fills the lowermost part of the phase space, down to $1$ GeV. The hard-coded value of $\theta_c$ helps to quantify the sensitivity of the $\theta_{\rm hard}$ observable to coherent vs. incoherent energy loss. Finally, no medium response and hadronisation are implemented in JetMed. 
    
    \item \textbf{Hybrid}~\cite{Casalderrey-Solana:2014bpa,Casalderrey-Solana:2015vaa}. This model runs a perturbative vacuum shower down to the hadronisation scale and then rewinds through the branching history so as to incorporate an energy loss rate computed at strong coupling~\cite{Chesler:2014jva,Chesler:2015nqz}.
    No medium-induced emissions are considered. The total amount of energy lost by splittings depends on their propagation distance within the QGP, as determined by embedding the Pythia8 event in a realistic heavy-ion simulation using 2+1D viscous hydrodynamics~\cite{Shen:2014vra}. The jet production points are sampled from two overlapped nuclear density profiles using the Glauber model~\cite{Miller:2007ri}.
    Emissions whose formation time is smaller than the hydrodynamic initialization time $t_0=0.6$ fm are not quenched. Another feature of the Hybrid model is that only those partons that are ``separated enough'', i.e. resolved by the medium, lose energy independently.
    The medium-resolution power is controlled by the $L_{\rm res}$ parameter, which depends on medium properties such as the local temperature~\cite{Hulcher:2017cpt}. For two colour-connected legs to lose energy independently, their transverse distance, $r_{\perp,\rm dip}\approx\theta(L-t_f)$, has to be larger than the resolution length $L_{\rm res}$. In order to illustrate the resolved phase-space, we pick a representative value of the medium length in central PbPb collisions, $L\sim6$ fm, and $L_{\rm res}=2/(\pi\cdot0.25{\rm [GeV]})\approx0.5$ fm. All emissions above the $r_{\perp,\rm dip}=(L-t_f)\theta>L_{\rm res}$ (black curve in Fig.~\ref{fig:PbPb-MCs-LP}) are resolved by the medium, lose more energy, and are therefore suppressed by the selection bias. Here, the phase space boundary is not as sharp as for JetMed because both $L$ and $L_{\rm res}(T)$ fluctuate. In what follows, we choose $L_{\rm res}=2/(\pi T)$ and explore other values of $L_{\rm res}$ in App.~\ref{app:lres-dep}.
    The lost energy is then hydrodynamised, producing jet-induced wakes that decay into hadrons at the freeze-out hypersurface of the flowing medium~\cite{Casalderrey-Solana:2016jvj}. These wake particles get clustered with the jet resulting in an excess of soft, wide-angle splittings (lower left corner in Fig.~\ref{fig:PbPb-MCs-LP}). All results shown below include medium response and a detailed study of its impact is presented in App.~\ref{app:med-response}.
    
    \item \textbf{Jewel}~\cite{Zapp:2008gi,Zapp:2012ak}. 
    At each evolution step in the parton shower, the propagating parton can either emit vacuum radiation or undergo elastic ($2$-$2$) scatterings with medium partons. The process with the shorter ``time" (formation time, or virtuality, compared to light-cone time) is realized. Elastic scatterings result in transverse momentum broadening, drifting energy out of the jet cone, and causing energy loss. Scatterings also reset the shower scale and induce collinear emissions without changing the jet energy. To estimate at which time ($t_{\rm el}$) elastic scatterings become dominant, we compare the scattering probability $p_{\rm el}\approx (t_{\rm el}-t_0)/l_{\rm mfp}$ with the radiation one $p_{\rm rad}\approx\int^{t_{\rm el}}_{t_0}dt\int_0^1dzP_{1\to2}(t,z)$. Here, $l_{\rm mfp}$ is the average distance between scatterings, $t_0$ is the initial formation time, and the splitting probability (expressed with formation time) is $P_{1\to2}\approx\alpha_sP(z)/(\pi t)$, where we used the Altarelli--Parisi splitting function $P(z)\approx 2C_i/z$. When $t\gg t_{\rm el}$, elastic scatterings interrupt vacuum radiation and induce extra collinear emissions (small angle, hard enhancement, $z\sim1$ in Fig.~\ref{fig:PbPb-MCs-LP}). Additionally, elastic scatterings broaden soft splittings to wider angles (wide angle enhancement below $1$ GeV in Fig.~\ref{fig:PbPb-MCs-LP}).\footnote{In Fig.~\ref{fig:PbPb-MCs-LP}, induced collinear emissions and broadening overlap. In our numerical tests, we however identified the separation of these two regions.}
    In contrast to the previous two models, an additional notion of colour coherence is not implemented. The medium in Jewel is modelled as a thermal, longitudinally expanding parton gas. The medium response is accounted for by keeping track of the recoiled medium partons after an elastic scattering takes place. These recoilers however free stream without further scatterings or radiation. In what follows we will disregard recoil particles and explore their impact in App.~\ref{app:med-response}. We do this as our current setup of Jewel would overestimate all medium response effects since, after the scattering, their evolution is frozen, i.e. they are non-dynamical. 
     
\end{itemize}
 
%%%%%%%%%%%%%%%%%%%%%%%%%%%%%
\subsection{Early times/vacuum-dominated regime}
\label{sec:early}
\begin{figure}
    \centering
    \includegraphics[width=0.99\columnwidth]{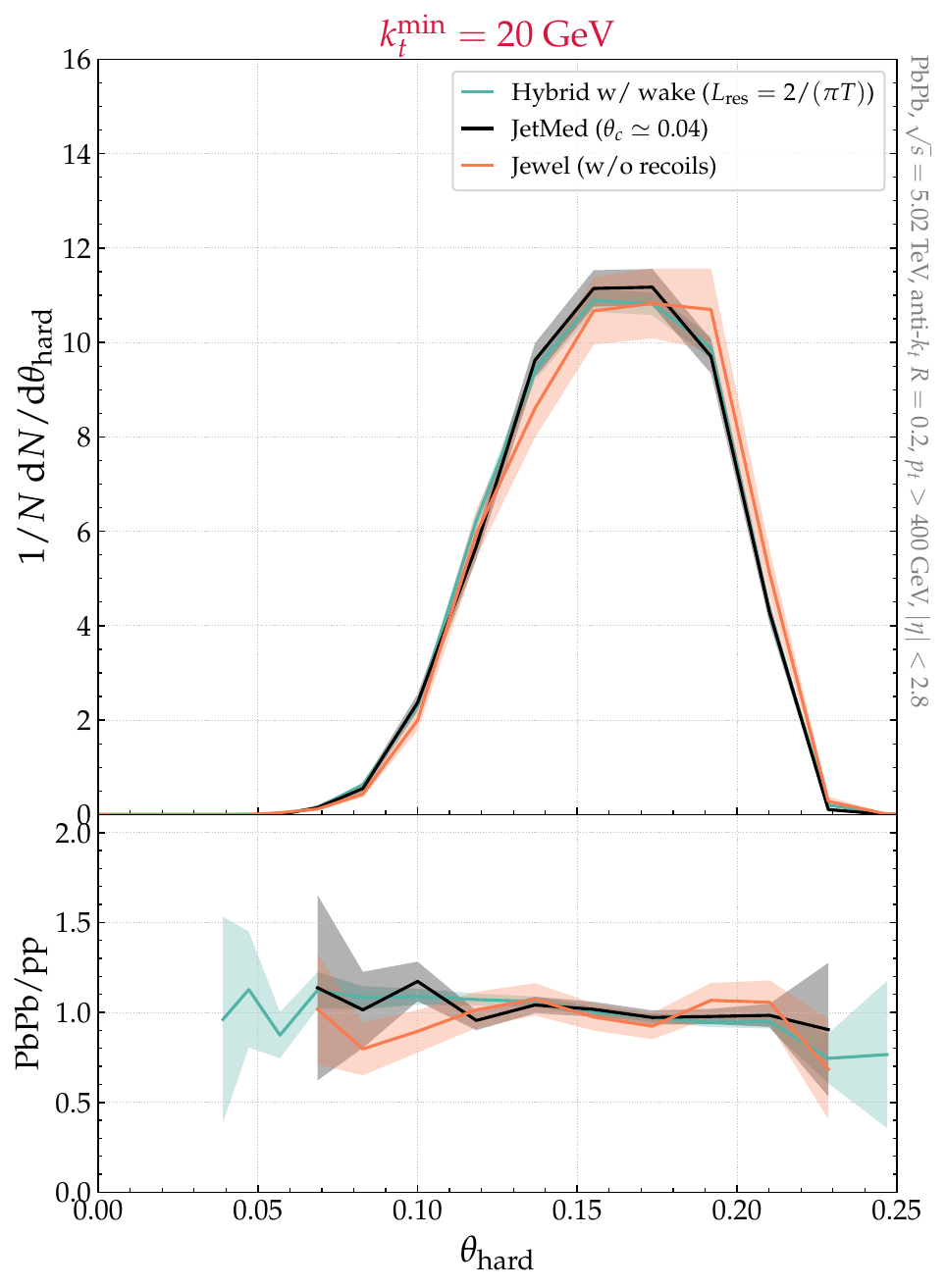}
    \caption{The angular distribution of the hardest-$k_t$ splitting inside inclusive jets in heavy-ion collisions with $k_t^{\rm min}=20$ GeV for the jet quenching Monte Carlo event generators described in Fig.~\ref{fig:PbPb-MCs-LP}. The bottom panel displays the medium-to-vacuum ratio.}
    \label{fig:PbPb-hard}
\end{figure}

%%%%%%%%%%%%%%%%%%%%%%%%%%%%%
We first study jets in PbPb collisions whose hardest splitting satisfies $k_t>20$ GeV. This region corresponds to very short formation times, i.e. parametrically $t_f<2p_t/(k^{\rm min}_t)^2\approx 0.4$ fm, smaller than the mean free path $l_{\rm{mfp}}=(\rho\sigma_{\rm{el}})^{-1}$ (where $\rho$ is the density of scattering centers and $\sigma_{\rm el}$ the total elastic cross section). The $\theta_{\rm hard}$-distribution is displayed in Fig.~\ref{fig:PbPb-hard}. We observe that all models agree within statistical uncertainties and that the medium-to-vacuum ratio is remarkably close to unity.\footnote{Note that these distributions are self-normalised and thus the overall suppression of the jet spectrum is factored out. This choice is justified since our main interest is a potential shape modification.} This result suggests that, in this upper corner of phase space, vacuum-like splittings are likely to be tagged and that its prongs lose the same amount of energy, independently of their opening angle. We show this independence analytically in App.~\ref{app:analytics}. We refer to this as the factorisation (or separation) of vacuum and medium-induced physics. This separation is very general in the sense that it is independent of the medium modelling when one goes to asymptotically high jet-$p_t$ and $k_t^{\rm min}$. In the JetMed context, these results are easily interpreted since the cut excludes the possibility of tagging medium-induced emissions ($k^{\rm min}_t\gg\sqrt{\hat qL}$) and $\theta_{\rm hard}$ is always bigger than $\theta_c$. We would like to note that Hybrid wake effects at $k^{\rm min}=20$ GeV are completely negligible, and recoil particles in Jewel are strongly suppressed (see in App.~\ref{app:med-response}).

We would like to highlight that not all jet-quenching models predict a flat ratio. For example, a potential source of substructure modifications at such high energy scales is the single-hit (or higher-twist) corrections induced by rare, hard interactions with the medium. These effects are included in the SCET$_{g}$ formalism \cite{Ovanesyan:2011xy} and in MATTER+LBT Monte Carlo generator~\cite{Majumder:2013re,Wang:2013cia}. However, the latest implementation of MATTER makes use of a virtuality-dependent $\hat{q}$~\cite{JETSCAPE:2022jer,JETSCAPE:2023hqn} that effectively reduces the impact of such corrections in this high-$k_t$ regime. 

The experimental measurement of the proposed observable has the potential to unambiguously establish whether vacuum physics dominates the early stages of jets in heavy-ion collisions. Knowing that there is a region of phase-space dominated by vacuum splittings implies that all jet quenching models should agree not only among them, but also with the pp baseline, within that given region. This would represent a major step forward in our understanding of jet quenching and is one of the main results of this paper. 

%%%%%%%%%%%%%%%%%%%%%%%%%%%%%
\subsection{Substructure-dependent energy loss}
\label{sec:eloss}
%%%%%%%%%%%%%%%%%%%%%%%%%%%%%
\begin{figure}
    \centering
    \includegraphics[width=0.99\columnwidth]{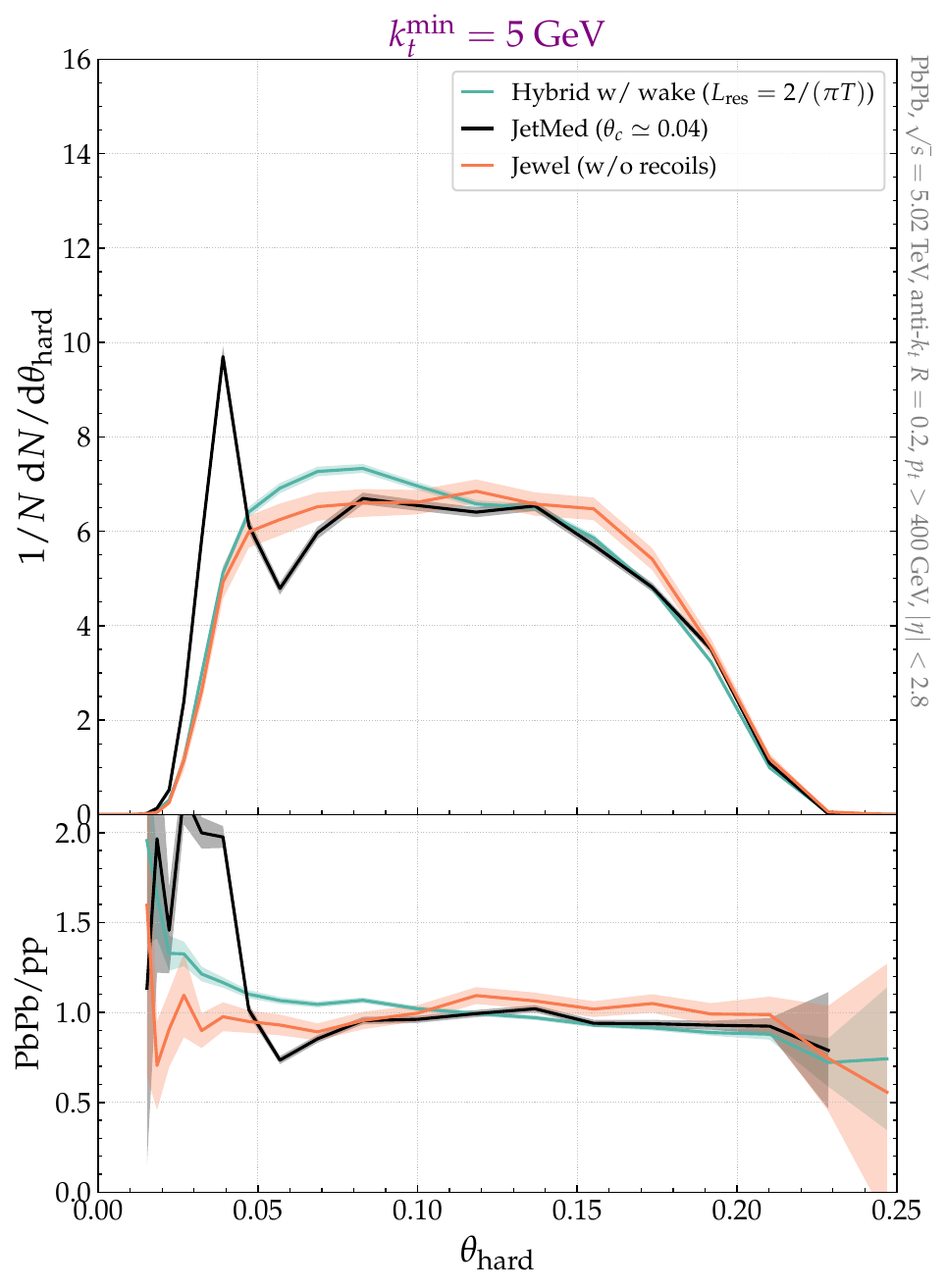}
    \caption{Same as Fig.~\ref{fig:PbPb-hard} but for $k^{\rm min}_{t}=5$ GeV.
    }
    \label{fig:PbPb-semi}
\end{figure}

When lowering the cut to $k_t^{\rm min}=5$ GeV, we enter a domain where one can test the substructure dependence of energy loss. Here, emissions with much longer formation times $\lambda_{\rm{mfp}} \ll t_f<6.4$ fm appear, while soft medium-induced splittings are still suppressed $k_t^{\rm min}>\sqrt{\hat qL}\approx2.4$ GeV. In the colour coherence picture sketched in Fig.~\ref{fig:sketch-lund}, we expect splittings with $\theta<\theta_c$ to appear, which are unresolved by the medium. As we have discussed, jets with wider splittings $\theta>\theta_c$ loose more energy and are suppressed due to selection bias. Consequently, we expect the suppression of $\theta_{\rm hard}>\theta_c$ (or equivalently an enhancement of $\theta_{\rm hard}<\theta_c$). Most modern jet quenching models implement this coherence (or resolution) effect, but they differ in the shape of the phase space boundary as discussed in Sec.~\ref{sec:mc-description} (see in Fig.~\ref{fig:PbPb-MCs-LP}). Theoretically, this boundary is not known beyond the soft-and-collinear limit, and models implement it differently. An experimental measurement in this regime, therefore, could put tighter constraints on the modelling of the resolution phase space.

In Fig.~\ref{fig:PbPb-semi}, we show the angular distribution of the hardest splitting $\theta_{\rm hard}$ for $k_t^{\rm min}=5$ GeV. The PbPb-to-pp ratio in JetMed displays a sharp enhancement at $\theta_c=0.04$, a clear consequence of colour resolution and selection bias. Our expectation is that, on both sides of $\theta_c$, the jet substructure does not change much, it is still vacuum-like (the ratio is flat). For the Hybrid model, the resolution boundary is not as sharp as it gets washed out due to fluctuations in the propagation length, temperature, and parton energy. The overall picture is however similar to JetMed: wide-angle splittings are suppressed. Results for the Hybrid model with different $L_{\rm res}$ values are shown in App.~\ref{app:lres-dep}. We would like to highlight that the angular distribution in the $L_{\rm res}=\infty$ fully coherent limit of the Hybrid model (only the initiator loses energy independent of the substructure) is barely modified as compared to vacuum, as expected. Even though the wake is included in Fig.~\ref{fig:PbPb-semi}, the $5$ GeV cut is still large enough to completely remove its effects (see it in more detail in App.~\ref{app:med-response}). 

The Jewel model is somewhat distinct from the other two models since it lacks an explicit implementation of colour coherence. The competition between elastic scattering and vacuum radiation still outlines a boundary that we referred to as $t_{\rm el}$. However, this boundary is beyond the phase space limited by $k_t^{\min}=5$ GeV (see Fig.~\ref{fig:PbPb-MCs-LP}) and therefore the medium-to-vacuum ratio is compatible with unity. Jets in Jewel of course lose energy, but their internal structure is vacuum-like even at $k_t\sim 5$ GeV. This ratio mildly changes when keeping track of the recoiling partons as we present in App.~\ref{app:med-response}. 

Let us re-emphasise that the proposed measurement, after identifying the vacuum-like dominated region of phase-space, has the potential to pin down the angular dependence of energy loss. In particular, it can quantify the angular resolution power of the QGP, tightly connected with its microscopic, transport properties, while putting tighter constraints on the actual size of the resolved phase space. 

Furthermore, at wider angles ($\theta_{\rm hard} \gg\theta_c$) our observable with $k_t^{\rm min}=5$ GeV is sensitive to the presence of hard recoils produced in perturbative elastic scatterings (also referred to as Moli\`ere scatterings), while remaining insensitive to the presence of the soft hadrons from the jet-induced wake (see the detailed study in App.~\ref{app:med-response}). This exemplifies the discriminating power of future experimental measurements of this observable to distinguish between different physical assumptions on the thermalisation dynamics of jets in the QGP.

%%%%%%%%%%%%%%%%%%%%%%%%%%%%%
\subsection{Non-perturbative regime}
\label{sec:npregime}
%%%%%%%%%%%%%%%%%%%%%%%%%%%%%

\begin{figure}
    \centering
    \includegraphics[width=0.99\columnwidth]{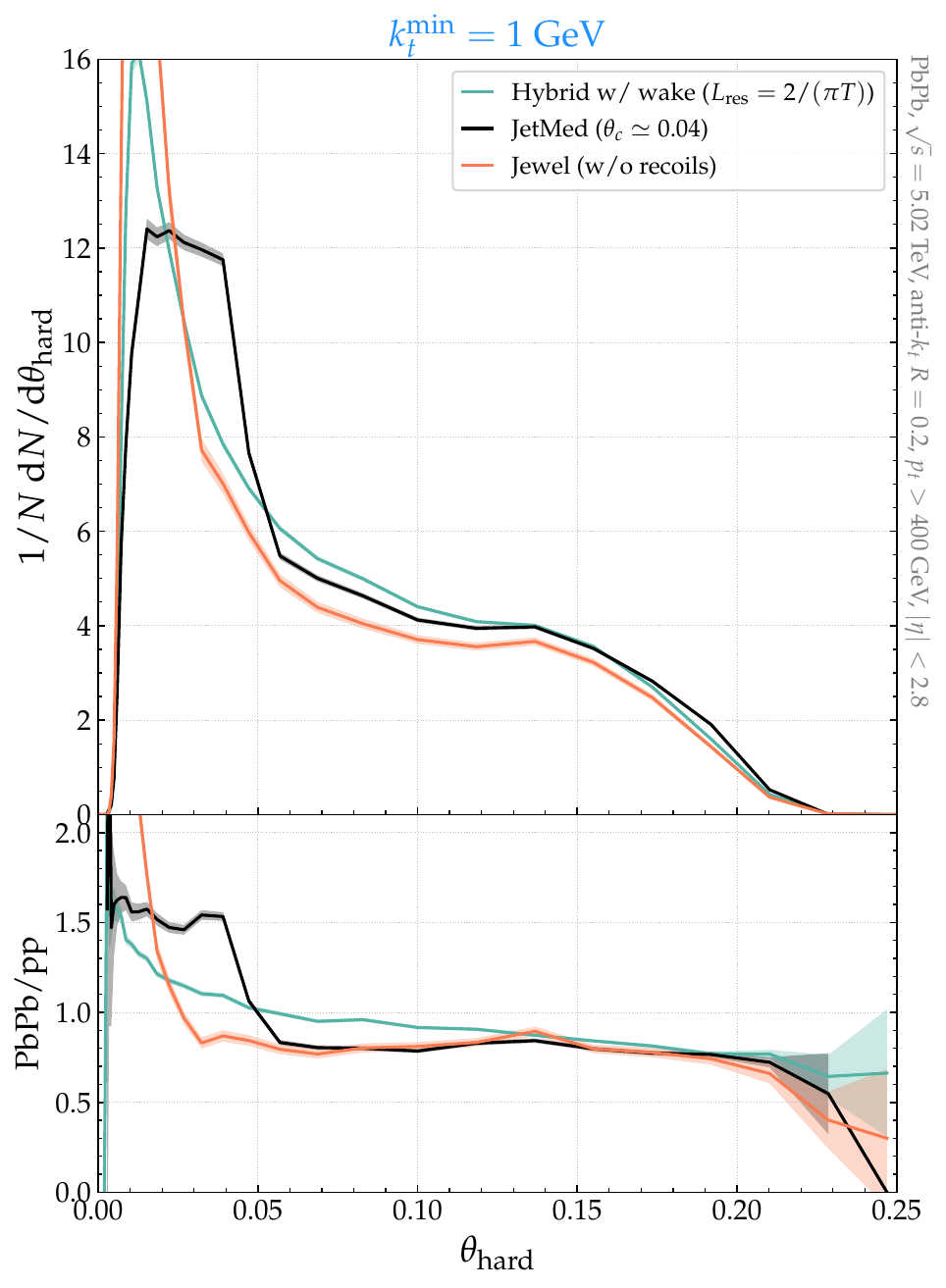}
    \caption{Same as Fig.~\ref{fig:PbPb-hard} but for $k^{\rm min}_{t}=1$ GeV.
    }
    \label{fig:PbPb-soft}
\end{figure}

As a last step, we set a lower selection cutoff $k_t^{\rm min}=1$ GeV. By doing this, we extend the low-angle reach of the distributions and we also enter a region where contributions from non-perturbative effects are expected. In addition, as discussed in App.~\ref{app:exp}, fake splittings might hamper the unfolding of experimental results. Even if it might not be possible to produce particle-level measurements in experiments, we think it is important to show when and how the perturbative picture studied above breaks down. The results are shown in Fig.~\ref{fig:PbPb-soft}. With the extended low-angle reach, the threshold behaviour of the JetMed distribution at $\theta_{\rm hard}=\theta_{c}$ becomes more apparent. The ratio on both sides of $\theta_c$ is flat, showing the dominance of vacuum-like substructure. The contribution of medium-induced emissions seems to be subleading (recall that $k_t^{\rm min}<\sqrt{\hat qL}=2.4$ GeV). An important caveat is that hadronisation corrections can reach up to $50$\% in this regime (see Fig.~\ref{fig:pp-MCs}) and thus JetMed might not be accurate in this regime.\footnote{It is often argued that hadronisation effects cancel in the PbPb-to-pp ratio. This is, however, a model-dependent statement that we have explicitly checked by computing the double ratio between the PbPb and pp distributions at hadron and parton levels. In the Hybrid model, this ratio is consistent with unity for this value of $k^{\rm min}_{t}$, while in Jewel $\mathcal{O}(1)$ hadronisation effects are observed for $\theta<0.05$. This difference originates from the fact that only Jewel changes the number of partons that undergo string fragmentation, while Hybrid either reduces their energy or removes them completely.} 

For the Hybrid model, the narrowing of the distribution persists, which indicates that wake particles, typically manifesting as a bump at large angles (see App.~\ref{app:med-response}), are suppressed by this observable with this set of fiducial cuts, mainly due to the small $R=0.2$ choice. We note that the fact that JetMed and Hybrid ratio curves cross unity at the same angle is coincidental. In fact, while the JetMed curves switch from depletion to enhancement at $\theta_{\rm hard}\sim \theta_c$ for $k^{\rm min}_t=1\to5$ GeV, the crossing point in the Hybrid case changes from $\theta_{\rm hard}\sim 0.1$ to $0.05$ when lowering $k^{\rm min}_t$ from $5\to1$ GeV. Thus, with a high enough angular resolution, the experimental measurement of the crossing point as a function of $k^{\rm min}_t$ could potentially disentangle between selection bias effects. Differentiating in centrality and jet $p_t$ would further prove the existence of a critical resolution angle. 

In the case of Jewel, we also observe an enhancement of narrow structures in PbPb versus pp. As we previously discussed, we attribute this to elastic scatterings, which reset the vacuum shower scale inducing collinear emissions. Further elastic scatterings did not broaden these emissions enough towards wider angles. All three models indicate that, even in this low domain of $k^{\rm min}_{t}$, the prevailing medium modification of the $\theta_{\rm hard}$ distribution is the intra-jet structure dependence of energy loss.

It is worth noting that this distribution is significantly altered when introducing recoil effects in Jewel (see Fig.~\ref{fig:recoils}). At the same time, it turns out to be remarkably resilient to wake particles. One can still study the wake particles and therefore the thermalisation of jets by opening the jet cone to $R=0.4$ (and keeping $k_t^{\rm min}=1$ GeV), see in App.~\ref{app:med-response}.

%%%%%%%%%%%%%%%%%%%%%%%%%%%%%
\section{Summary}
\label{sec:summary}
%%%%%%%%%%%%%%%%%%%%%%%%%%%%%
%%%%%%%%%%%%%%%%%%%%%%%%%%%%%

This study proposes a novel analysis strategy aiming at experimentally accessing different stages of jet evolution in heavy-ion collisions. It explores the angular distribution of the hardest splitting, $\theta_{\rm hard}$, above a transverse momentum threshold $k_t^{\rm min}$. This momentum scale can be tuned so as to enhance or suppress certain features of jet dynamics. For example, selecting hard enough splittings ($k_t^{\rm min}\sim20$ GeV) the observable is dominated by perturbative physics and, in particular, by the vacuum evolution of an in-medium jet. This perturbative region extends to even lower cuts ($\sim 5$ GeV) with the additional sensitivity to the colour resolution power of the medium and hard elastic scatterings. In contrast, when the hardest splitting becomes commensurate to the QGP and confinement scales ($k_t^{\rm min}\sim\Lambda_{\rm QCD},T$), it becomes sensitive to thermalisation and hadronisation effects in the medium. A key observation is that this strategy is optimal when the jet radiation phase-space is so large ($p^{\rm jet}_t=\mathcal O(100)$ GeV) that it admits well-separated regions between vacuum- and medium-dominated dynamics. Although the outlined strategy is completely general/data-driven, we illustrate the potential of this observable by using three of the most popular jet quenching Monte Carlo event generators: Hybrid, JetMed, and Jewel. 

First, we performed a systematic study of the proton-proton baseline using state-of-the-art techniques (Powheg+Pythia8). We compared this baseline to the vacuum prediction of jet-quenching event generators. At sufficiently high $k_t^{\rm min}$ cut, we find remarkably good agreement among all vacuum models, and therefore any potential modifications due to medium effects arise from a well-controlled baseline.
Differences in the vacuum mode of the jet-quenching models gradually appear at low $k_t$, where hadronisation corrections become sizeable (up to 50\% for $k_t^{\rm min}=1$ GeV). 

On the heavy-ion side, we first address the question of whether the early stages of in-medium evolution are dominated by vacuum dynamics. To that end, we selected $k_t^{\rm min}=20$ GeV, which is high enough to ensure perturbativity. The three models under study predict a PbPb-to-pp ratio compatible with unity across all $\theta_{\rm hard}$ values. We interpret this as the result of a vacuum-like splitting being tagged, followed by energy loss that is largely independent of the substructure. We show analytically why the amount of lost energy is independent of the opening angle of the splitting in this hard regime. \textbf{An experimental confirmation of this high $k_t^{\rm min}$ result will show, for the first time, a vacuum-dominated observable in a heavy-ion environment.}

Next, we extend the angular coverage of the observable by lowering to $k_t^{\rm min}=5$ GeV. This intermediate value opens up the possibility of understanding the small-angle dependence of energy loss while staying in the perturbative regime. At small angles, a reduction of energy loss is expected due to colour resolution effects. This angular dependence is then enhanced by selection bias effects, where the jet ensemble contains fewer jets with wider substructure, as they typically lose more energy than narrower ones.\footnote{See footnote~\ref{foot:qg_fraction} for an extended discussion on selection bias effects.} We indeed observe a clear narrowing in the $\theta_{\rm hard}$-distribution in those models where some sort of resolution scale is implemented. In regards to wide-angle enhancement, we observe the imprints of hard elastic scatterings, especially for wider $R=0.4$ jets, in Jewel. For the Hybrid model, there is no signal from the medium response as the implemented thermal wake is removed by the $k_t^{\min}=5$ GeV cut. \textbf{The experimental test of this intermediate $k_t^{\rm min}$ result could reveal the colour resolution scale in the medium, a fundamental property of the QGP. Furthermore, this region also gives experimental access to explore hard (Moli\`ere-like) elastic scatterings in the medium.}

Decoupling selection bias and colour coherence effects might be possible with high enough experimental precision or with complementary measurements in $\gamma$/Z+jet events~\cite{Du:2020pmp,Brewer:2021hmh}, at forward rapidities~\cite{Pablos:2022mrx} or by means of a centrality scan~\cite{Mehtar-Tani:2021fud}. In fact, recent experimental results from CMS in $\gamma$+jet events have shown that the selection bias can be suppressed by reducing $x_{J}=p_{t,{\rm jet}}/p_{t,\gamma}$~\cite{CMS:2023cka}. We leave the study of the $\theta_{\rm hard}$ observable in boson-tagged events for future work.

Finally, we explore the scenario in which $k_t^{\rm min}=1$ GeV, which lies in the non-perturbative regime. We find that in JetMed the $\theta_{\rm hard}$-distribution is narrowed in PbPb due to the colour resolution with selection bias effects and only a few medium-induced emissions get tagged for narrow jets ($R=0.2$).
This spoils the possibility of measuring the medium-induced emission kernel with this observable and endorses the idea of doing so by exploring a regime in which vacuum-like emissions are strongly suppressed, such as the dead-cone~\cite{Armesto:2003jh,Cunqueiro:2022svx,Andres:2023ymw}.
Another effect that this value of $k_t^{\rm min}$ targets is the soft medium response, i.e. jet thermalisation. Wake particles in the Hybrid model leave a weak imprint on this observable mainly due to the use of relatively narrow $R=0.2$ jets. In the Jewel case, recoil medium particles create a broadening of the $\theta_{\rm hard}$-distribution around the jet boundary. This regime, although experimentally challenging to unfold, has the strongest potential to help us understand jet thermalisation and constrain the modeling of medium response in jet-quenching Monte Carlo generators. 

As future steps, we will work on the analytic calculation of this observable in proton-proton collisions at high-logarithmic accuracy by extending the results presented in Ref.~\cite{Caucal:2021bae} to the case of an additional $k_t^{\rm min}$ selection. In regards to the heavy-ion scenario, the present Monte Carlo study has informed us about the feasibility to perturbatively describe the high-$k_t$ regime, which motivates carrying out analytic calculations, including medium-induced effects, that can be sensibly confronted with experimental data. 

We are confident that an experimental realisation of the proposed strategy will provide new key inputs to improve our understanding of the rich interplay between the vacuum and medium scales that govern jet evolution in the quark-gluon plasma.

\section*{Acknowledgements}
The authors would like to thank the organisers of the ``QCD challenges from pp to AA workshop" which took place in Padova, Italy, where this collaboration started. We also thank Jack Holguin for participating in the early stages of this project and his feedback on the manuscript. We are grateful to Paul Caucal for providing the JetMed samples and comments on the manuscript, Luca Rottoli for providing the Powheg samples, and Silvia Ferrario-Ravasio, Aidin Masouminia, and Simon Pl\"atzer for helping with the Herwig results. The work of LCM was supported by the European Research Council project ERC-2020-COG-101002207 QCDHighDensityCMS. The work of DP has received funding from the European Union’s Horizon 2020 research and innovation program under the Marie Sklodowska-Curie grant agreement No. 754496. The work of ASO has been funded by the European Research Council (ERC) under the European Union’s Horizon 2020 research and innovation program (grant agreement No 788223). The work of MS has been supported by the Ministry of Education, Youth and Sports of the Czech Republic under grant ERC-CZ LL2327. The work of AT is supported by the Starting Grant from Trond Mohn
Foundation (BFS2018REK01), the University of Bergen, and by DFG – Project number 496831614. The work of MV was supported by the Dutch Research Council (NWO) - Project number STU.019.019.

%%%%%%%%%%%%%%%%%%%%%%%%%%%%%%%%%%%%%%%%%%
%%%%%%%%%% APPENDIX         %%%%%%%%%%%%%%
%%%%%%%%%%%%%%%%%%%%%%%%%%%%%%%%%%%%%%%%%%
\appendix
%%%%%%%%%%%%%%%%%%%%%%%%%%%%%%%%%%%%%%%%%%
%%%%%%%%%%%%%%%%%%%%%%%%%%%%%%%%%%%%%%%%%%

\begin{figure*}
    \centering
     \includegraphics[width=0.55\textwidth]{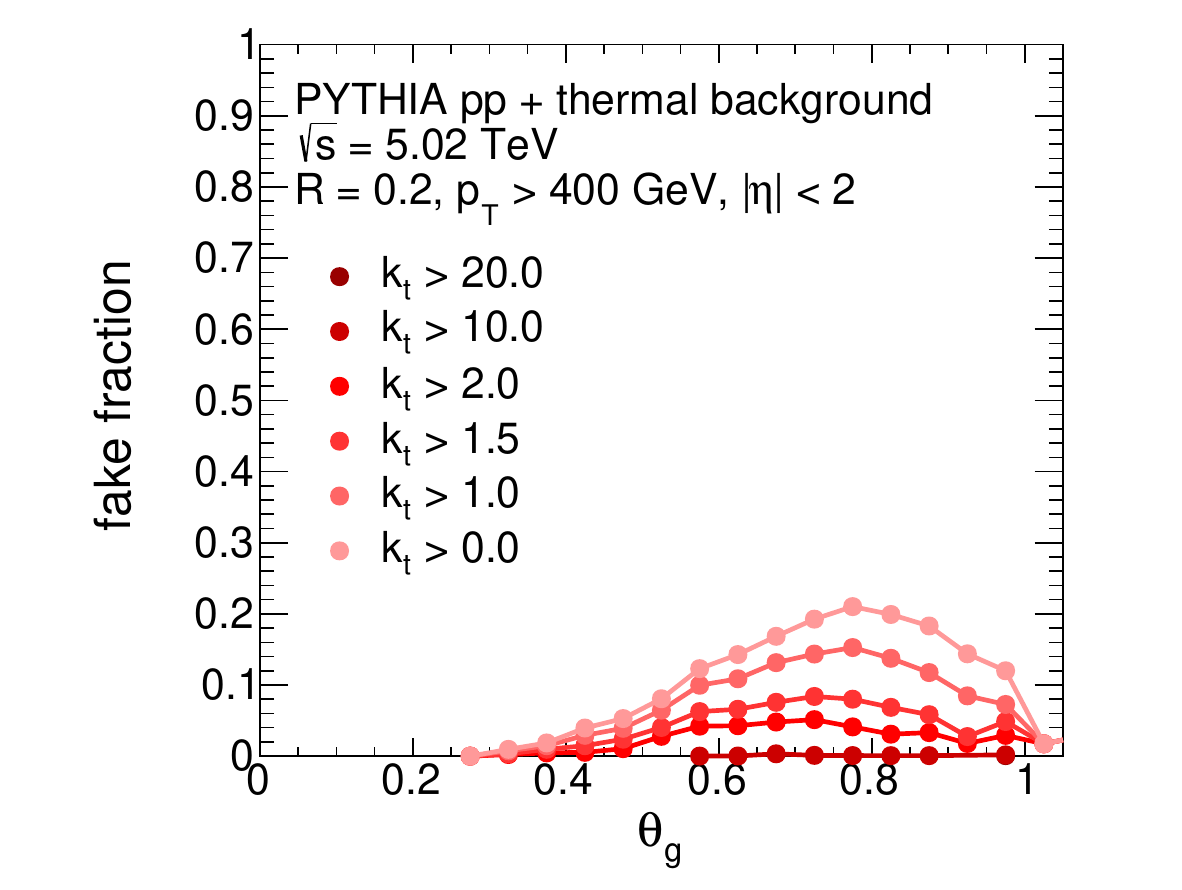}
    \includegraphics[width=0.44\textwidth]{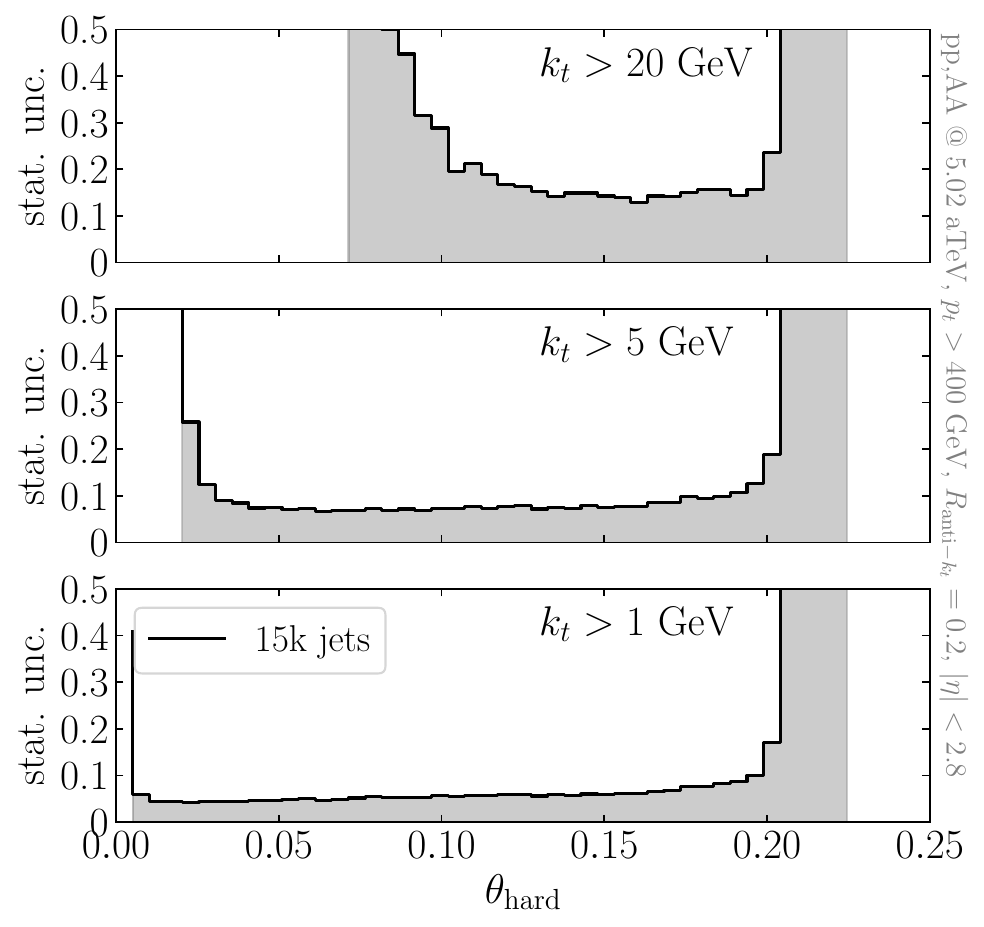}
    \caption{\textit{Left}: Fraction of fake splittings obtained by embedding Pythia8 pp events in a thermal background~\cite{Andrews:2018jcm, Mulligan:2020tim} for various minimum selections of $k_{t}$. \textit{Right}: the relative statistical uncertainty of the $\theta_{\rm hard}$ distribution by considering the experimentally accessible number of jets in runs 3 and 4 of the LHC. 
    }
    \label{fig:exp}
\end{figure*}

%%%%%%%%%%%%%%%%%%%%%%%%%%%%%%%%%%%%%%%%%%
%%%%%%%%%%%%%%%%%%%%%%%%%%%%%%%%%%%%%%%%%%
\section{Experimental considerations}
\label{app:exp}
%%%%%%%%%%%%%%%%%%%%%%%%%%%%%%%%%%%%%%%%%%
%%%%%%%%%%%%%%%%%%%%%%%%%%%%%%%%%%%%%%%%%%

Here we address the experimental feasibility of the proposed observable. To start with, this would be the first triple-differential measurement in heavy-ion collisions, i.e. the measurement will require the simultaneous unfolding of background and detector effects in three quantities, $p_t^\mathrm{jet}$, $k_t^{\rm min}$ and $\theta_{\rm hard}$. However, since only thresholds on jet $p_{t}$ and $k_t$ are imposed, the unfolding problem is simpler than when performing the triple differential measurement over many bins. 

An important aspect when correcting the measurement to particle level is the fraction of fake splittings (splittings originating entirely from upward fluctuations of the underlying event). We have estimated the fraction of fake-splittings in our observable by embedding Pythia8 events into a thermal background mimicking the underlying event of a heavy-ion collision. The results are shown in the left panel of Fig.~\ref{fig:exp} after applying full event constituent subtraction~\cite{Berta:2014eza} to the embedded jets. Jet substructure measurements have been reported for fake fractions smaller than $20\%$. This threshold is met by our observable as long as $k_t^{\rm min} \gtrsim 2$ GeV. 
ALICE studies of the dynamically groomed $k_{t}$ have shown that for $R=0.2$ jets in central collisions, a detector-level cutoff of $k_{t}>1.5$ GeV renders the observable unfoldable. Due to the strong migrations in $k_{T}$ between detector and true levels, the fully corrected results can be reported at $p_{t}=60$ GeV and $k_{t}>3$ GeV in central collisions~\cite{Ehlers:2023jbf}. The higher jet $p_{t}$ required in our study implies higher purities and thus we conclude that unfolding this triple-differential observable should not be accompanied by large systematic uncertainties. 

Other challenges to realize experimentally the scan of the emission phase-space described in the previous Sections concern the ability to measure small-angle splittings. Track-based measurements of the Lund plane by ATLAS~\cite{ATLAS:2020bbn}, CMS~\cite{CMS:2023ovl}, and ALICE~\cite{ALICE:2021yet} have demonstrated the ability to reconstruct splittings down to angles $\theta \approx 0.005$ and should therefore provide sufficient sensibility to colour coherence effects.

In this study, jets were required to have $p_t>400$ GeV. For LHC Run 3 and Run 4, assuming total collected luminosity of 13~nb$^{-1}$~\cite{Chamonix2023}, we estimated the total number of $R=0.2$ jets with $p_t>400$ GeV within $|\eta|<2.8$ to exceed $1.5 \cdot 10^4$ in $0-10$\% central PbPb collisions. The estimate is based on the data published in Ref.~\cite{ATLAS:2023hso}. The right panel of Fig.~\ref{fig:exp} shows the expected statistical uncertainty of the $\theta_{\rm hard}$-distribution with this projected statistics obtained with Monte Carlo simulations. We would like to remark that reducing the jet $p_t$ selection down to $200$ GeV can increase the statistics by a factor of $50$. The exact jet $p_{t}$ selection is thus to be decided by experiments based on the available luminosity and effects seen in the data. 

%%%%%%%%%%%%%%%%%%%%%%%%%%%%%%%%%%%%%%%%%%
%%%%%%%%%%%%%%%%%%%%%%%%%%%%%%%%%%%%%%%%%%
\section{The reduction of bias effects at high $k_t^{\rm min}$}
\label{app:analytics}
%%%%%%%%%%%%%%%%%%%%%%%%%%%%%%%%%%%%%%%%%%
%%%%%%%%%%%%%%%%%%%%%%%%%%%%%%%%%%%%%%%%%%

\begin{figure}
    \centering
    \includegraphics[width=0.99\columnwidth]{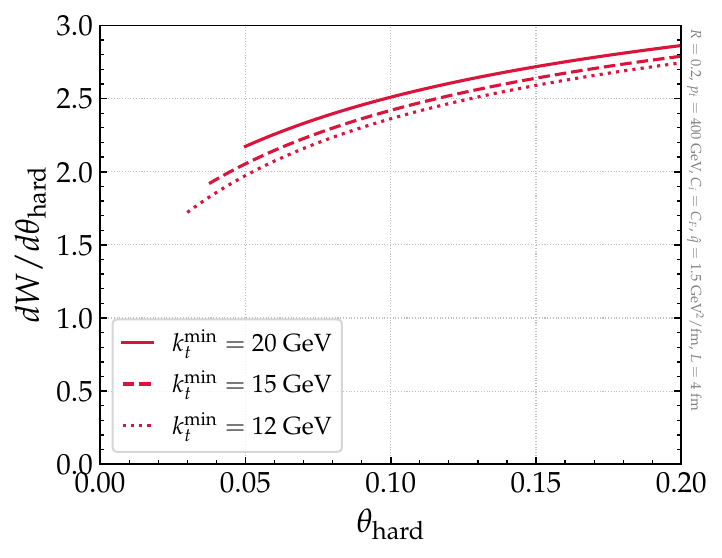}
    \caption{Size of resolved phase-space for jets with $R=0.2$, $p_t>400$ GeV as a function of $\theta_{\rm hard}$ for different $k_t^{\rm min}$ values (corresponding to splittings with $\theta>\theta_c)$ in the soft-and-collinear limit, see Eq.~\eqref{eq:W-hard}.}
    \label{fig:plot-phase-space}
\end{figure}

\begin{figure*}
    \centering
    \includegraphics[width=0.33\textwidth]{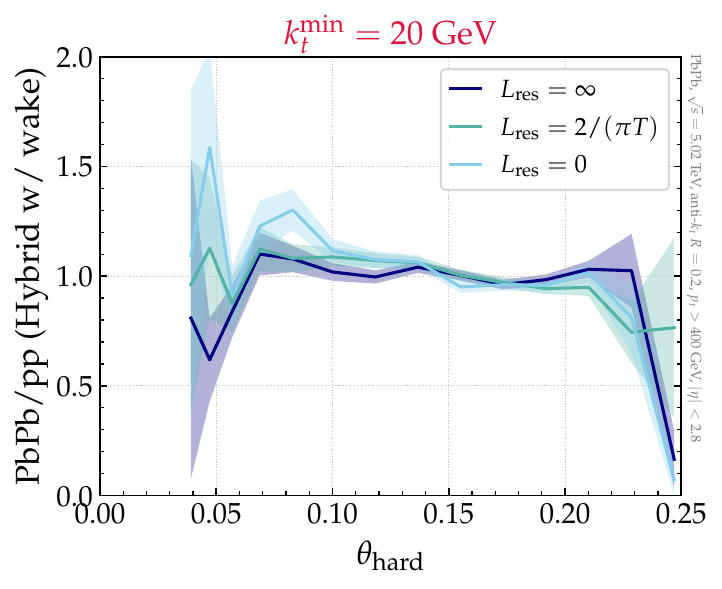}
    \includegraphics[width=0.33\textwidth]{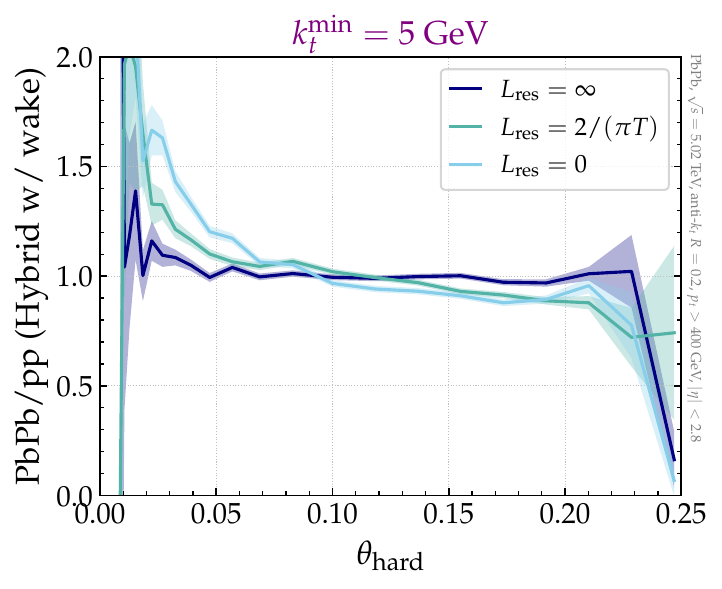}
    \includegraphics[width=0.33\textwidth]{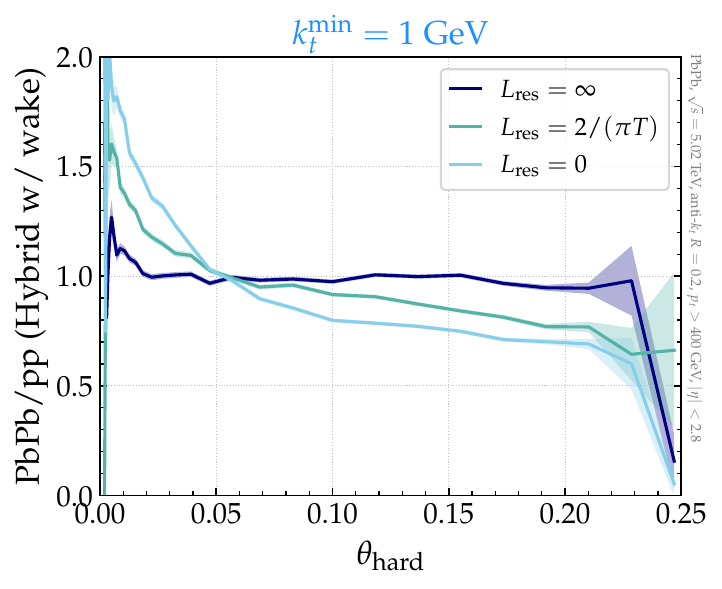}
    \caption{Sensitivity of Hybrid results to the value of the resolution length $L_{\rm res}$. From left to right the value of $k_t^{\rm min}$ decreases.}
    \label{fig:hybrid-lres}
\end{figure*}

The absence of selection bias at high-$k_t^{\rm min}$ values can be understood based on simple considerations. The object of interest is the energy lost by a jet featuring a splitting with $k_{t,\rm hard}>k^{\rm min}_t$ as a function of its opening angle $\theta_{\rm hard}$, that we denote $W(\theta_{\rm hard}, k^{\rm min}_t)$. The amount of energy lost is directly proportional to the size of the resolved phase space~\cite{Mehtar-Tani:2017web}, that at $\mathcal{O}(\alpha_s)$ reads
\begin{align}
\label{eq:omega-kt}
\Omega(k_{t,\rm{hard}})&= \int_0^R d\theta \int_0^{k_{t,\rm hard}} dk_t \frac{2C_i \alpha_s(k_t)}{\pi} \frac{d\sigma}{d k_{t} d\theta} \nonumber \\
& \times \Theta(t_f<L) \Theta (t_f<t_d),
\end{align}
where $d^2\sigma/d k_t d\theta$ is the double-differential cross section for producing a vacuum-like splitting (stripped off colour factor and coupling) and the second line of the equation contains all phase-space constraints required for an emission to be part of the resolved phase-space. In order to estimate the amount of energy loss differential in angle, this quenched phase-space has to be weighted by the cross-section for producing the tagged splitting, i.e.
\begin{align}
\label{eq:W-hard}
    \frac{dW}{d\theta_{\rm hard}}\Big|_{k_t^{\rm min}}=\frac{\displaystyle\int_{k_t^{\rm min}}^{p_t \theta_{\rm hard}}dk_{t,\rm hard} \frac{1}{\sigma}\frac{d\sigma}{dk_{t,\rm hard} d\theta_{\rm hard}}\Omega(k_{t,\rm hard})}{\displaystyle \int_{k_t^{\rm min}}^{p_t \theta_{\rm hard}}dk_{t,\rm hard} \frac{1}{\sigma}\frac{d\sigma}{d k_{t,\rm hard} d\theta_{\rm hard}}}\, .
\end{align}
In the soft-and-collinear approximation and at fixed-coupling, Eq.~\ref{eq:W-hard} can be solved analytically and the result is plotted in Fig.~\ref{fig:plot-phase-space} for a quark-jet. We observe that for $k_t^{\rm min}=20$ GeV, the size of the quenched phase space is largely independent of $\theta_{\rm hard}$ thus confirming the picture drawn by the Monte Carlo generators in Fig.~\ref{fig:PbPb-hard}. We note that only $k_t^{\rm min}$ values for which $\theta^{\rm min}>\theta_c$ are shown in the figure. This is so due to the toy model not being applicable when both resolved and unresolved splittings contribute. 

%%%%%%%%%%%%%%%%%%%%%%%%%%%%%%%%%%%%%%%%%%
%%%%%%%%%%%%%%%%%%%%%%%%%%%%%%%%%%%%%%%%%%
\section{$L_{\rm res}$ dependence of Hybrid results}
\label{app:lres-dep}
%%%%%%%%%%%%%%%%%%%%%%%%%%%%%%%%%%%%%%%%%%
%%%%%%%%%%%%%%%%%%%%%%%%%%%%%%%%%%%%%%%%%%

The Hybrid model allows for smoothly changing the size of the resolved phase-space by varying the $L_{\rm res}$ parameter. In Fig.~\ref{fig:hybrid-lres} we show the $\theta_{\rm hard}$-distribution for all $k^{\rm min}_{t}$ values in three different scenarios: only the jet initiator is quenched, $L_{\rm res}=\infty$ (fully coherent energy loss), all particles are quenched inside the medium, $L_{\rm res}=0$ (fully incoherent energy loss), and the intermediate scenario present in the main text in which splittings whose transverse size satisfy $r_{\perp, \rm dip} > L_{\rm res}=2/(\pi T)$ are resolved and lose energy. For the highest value of $k^{\rm min}_{t}$ we observe that the resulting $\theta_{\rm hard}$-distribution is barely dependent on $L_{\rm res}$ thus confirming that, in this corner of phase-space, energy loss details are not relevant and that the ingredient at test is the vacuum splitting evolution.\footnote{There is still some ordering among different $L_{\rm res}$, which disappears for $R=0.4$ jets, and thus has its origin in statistical fluctuations occurring for $R=0.2$.}
By lowering $k^{\rm min}_{t}$ the observable becomes sensitive to the correlation between energy loss and jet substructure. In fact, $L_{\rm res}=\infty$ yields no in-medium modification as it is independent of the jet substructure. A finite value of $L_{\rm res}$ results in a narrowing of the distribution since more collimated jets tend to lose less energy.
%and thus pass the $p_t$ selection cut more easily because of the selection bias effect. 
This result endorses the potential of the proposed observable to test the presence of an angular scale that controls the degree of quenching, i.e. the medium resolution power. Setting $L_{\rm res}=0$ (corresponding to all splittings with $t_f<L$ losing energy) leads to an even stronger narrowing of the $\theta_{\rm hard}$ distribution. In this case, the resolution boundary does not fluctuate with temperature and/or parton energy, resulting in a sharper enhancement.

%%%%%%%%%%%%%%%%%%%%%%%%%%%%%%%%%%%%%%%%%%
%%%%%%%%%%%%%%%%%%%%%%%%%%%%%%%%%%%%%%%%%%
\section{Medium response in Jewel and Hybrid}
\label{app:med-response}
%%%%%%%%%%%%%%%%%%%%%%%%%%%%%%%%%%%%%%%%%%
%%%%%%%%%%%%%%%%%%%%%%%%%%%%%%%%%%%%%%%%%%

\begin{figure*}
    \centering
    \includegraphics[width=0.49\textwidth]{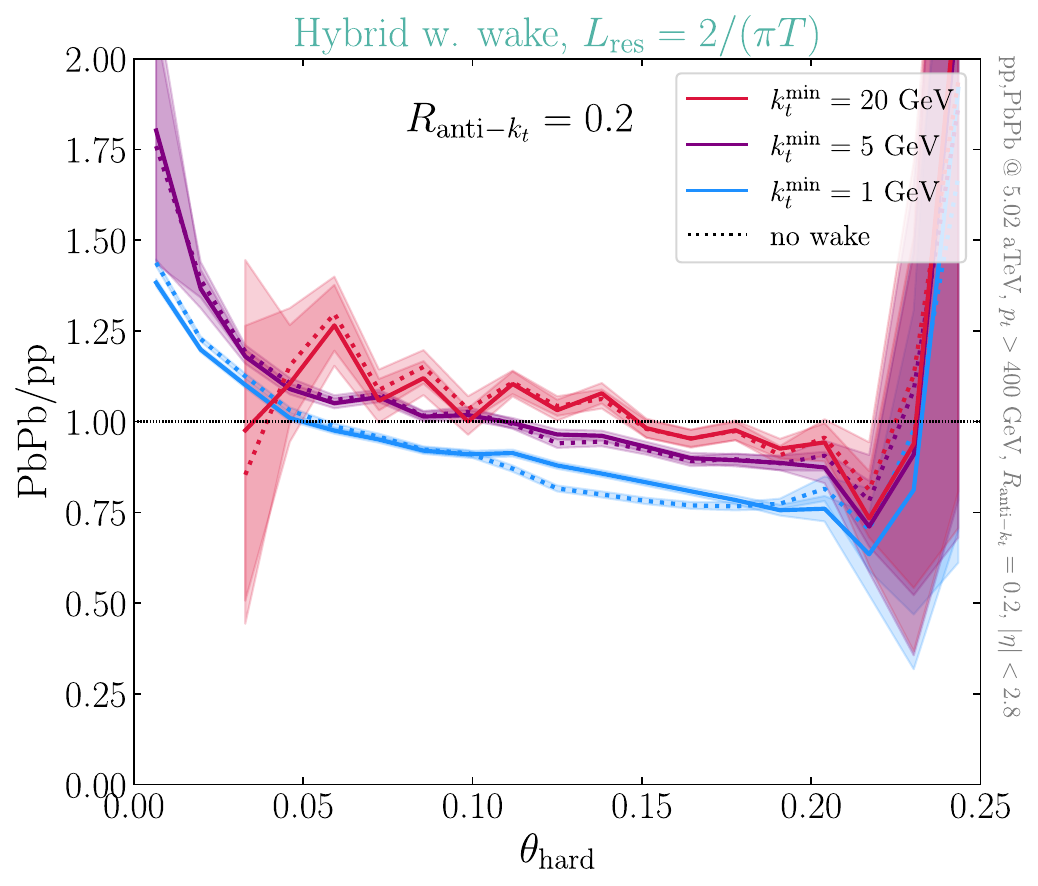}
    \includegraphics[width=0.49\textwidth]{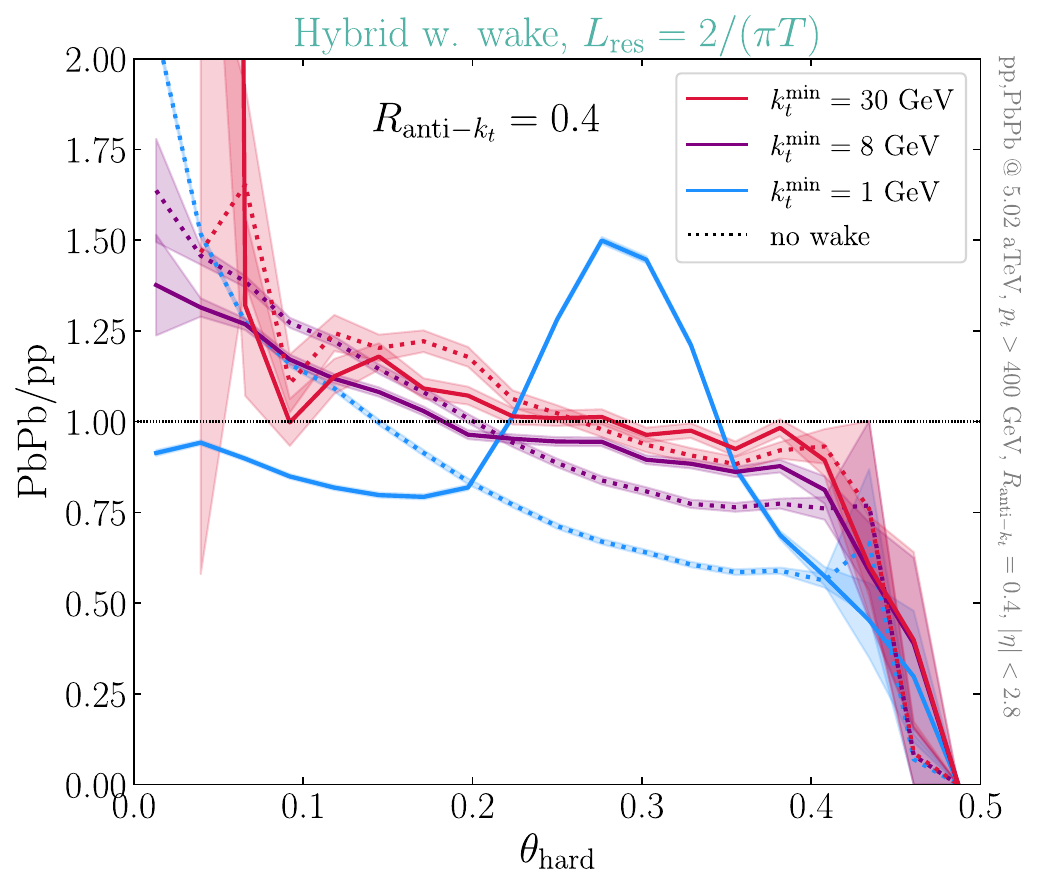}
    \includegraphics[width=0.49\textwidth]{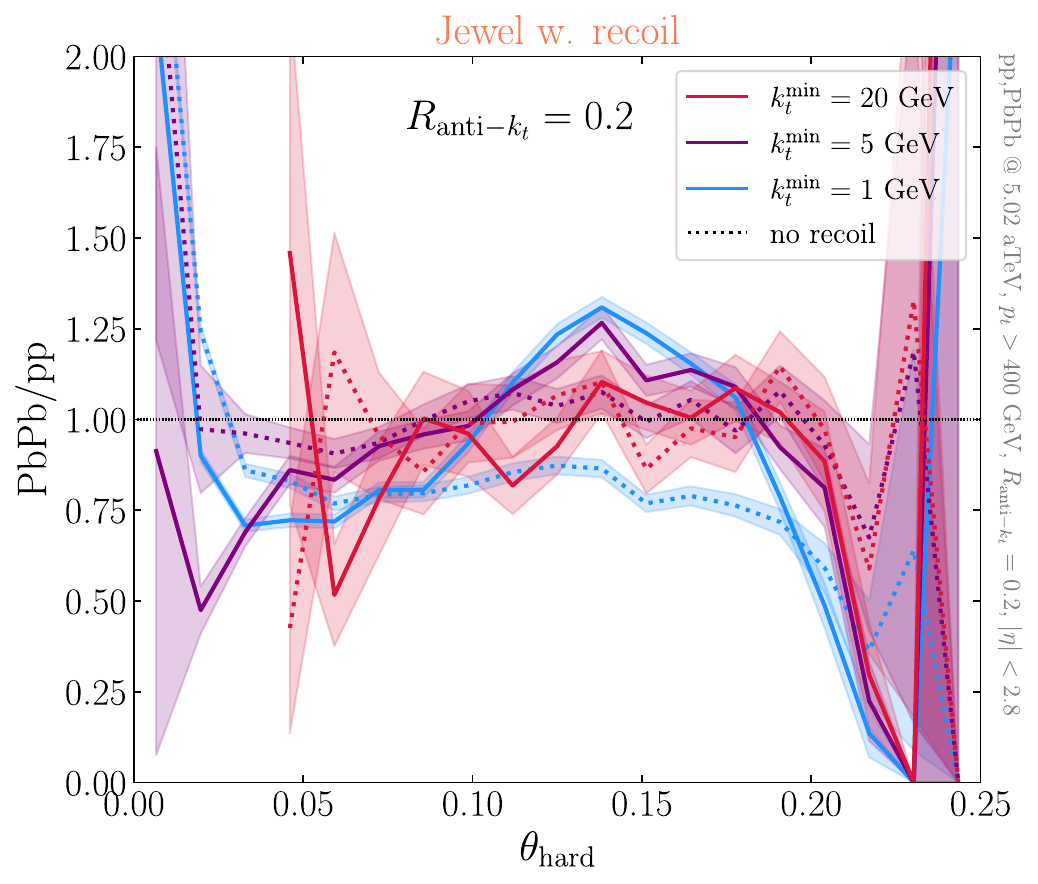}
    \includegraphics[width=0.49\textwidth]{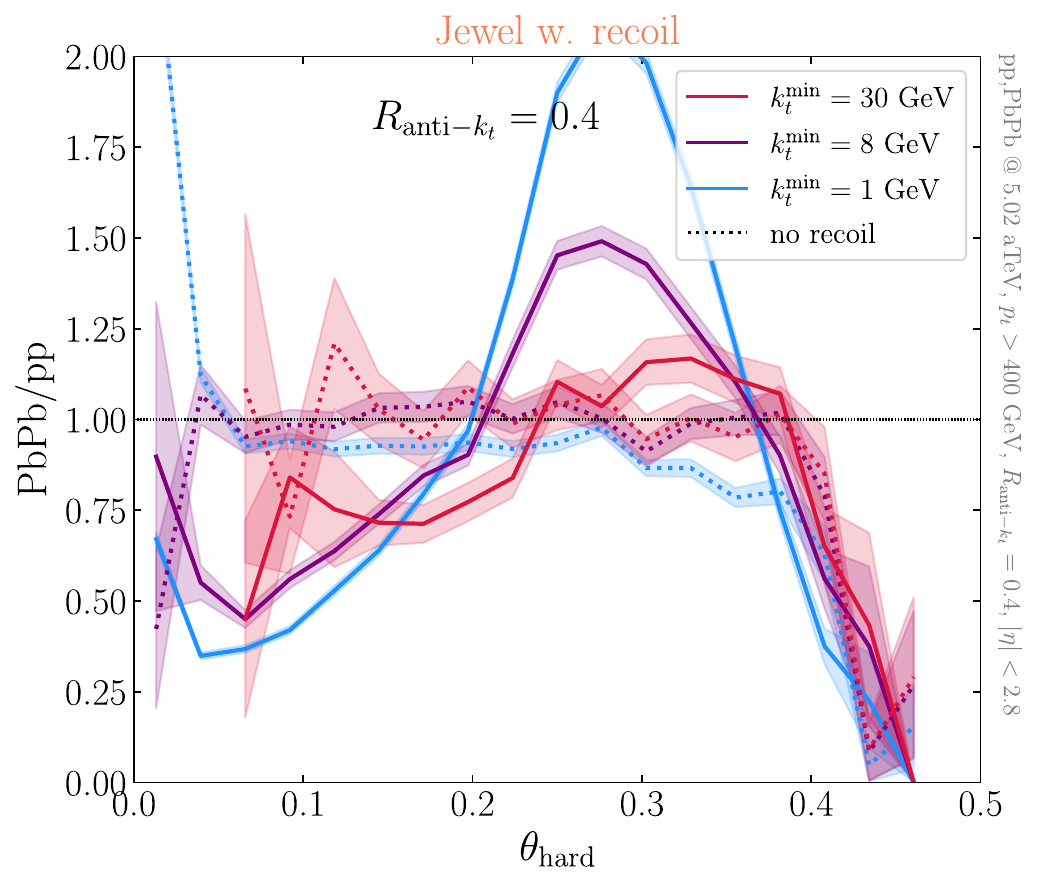}
    \caption{Same as the ratio plots in Figs.~\ref{fig:PbPb-hard},\ref{fig:PbPb-semi}, and~\ref{fig:PbPb-soft}. The upper panels are Hybrid events with and without the wake for different $k_t^{\rm min}$. The lower panels are Jewel events with and without keeping track of recoilers for different $k_t^{\rm min}$. The left and right side are jets with different cone sizes $R=0.2$ and $R=0.4$.}
    \label{fig:recoils}
\end{figure*}

In this section, we study the sensitivity of the proposed observable to medium response effects in the Hybrid and Jewel models. We remind the reader that JetMed does not account for this effect. In the main text figures we included the Hybrid wake, while discarded Jewel recoils.

In Fig.~\ref{fig:recoils} we show the PbPb/pp ratio of the $\theta_{\rm hard}$ distributions with and without medium response for both Hybrid and Jewel models (similar to the ratio plots in Figs.~\ref{fig:PbPb-hard},\ref{fig:PbPb-semi}, and~\ref{fig:PbPb-soft}). The results of the Hybrid model (upper left panel in Fig.~\ref{fig:recoils}) suggest that the medium response effects remain below $10$\% across all $k^{\rm min}_t$ choices for $R=0.2$ jets. The reason is that in the Hybrid model all lost energy is assumed to thermalize instantly and contributes as a source term in the hydrodynamic equations of motion. Perturbative, hard elastic scatterings are neglected in the present version of Hybrid, and the resulting medium response is a soft, wide-angle hadron distribution from the wake. To further illustrate wake effects in Hybrid, we repeat the previous plot for wider, $R=0.4$ jets (upper right panel in Fig.~\ref{fig:recoils}). Here, the wake becomes substantially more important at lower $k_t^{\rm min}$ cuts, implying that the wake appears at wide angles ($\theta>0.2$). We would like to remind the reader that the overall decreasing trend of the PbPb/pp ratio with increasing angle is caused by the substructure-dependent energy loss and selection bias effects as we discussed in the main text. The effect from medium response sits on top of this effect, typically at larger angles.

Jewel implements medium response in a radically different manner. The medium response dynamics can affect the $\theta_{\rm hard}$ distribution even for moderate $k^{\rm min}_t$ value (lower left panel in Fig.~\ref{fig:recoils}). The size of this effect can reach up to $50$\% for $k^{\rm min}_t = 1$ GeV for $R=0.2$. These recoils are thermal particles that become part of correlated background after undergoing elastic $2\to2$ scatterings with the jet particles. These scatterings can be perturbative or not depending on the exchanged momentum.\footnote{$2\to2$ pQCD scatterings are modified in the QGP due to thermal screening effects typically at small momenta~\cite{Blaizot:2001nr,kapusta_gale_2006}. We refer to these soft scatterings as non-perturbative as the plasma temperature is relatively low $T_{\rm QGP}\sim\Lambda_{\rm QCD}$.}Therefore, the medium response in Jewel produces harder particles at smaller angles than the wake in Hybrid. Two technical details on the treatment of these recoils are relevant: we have used the method presented in Ref.~\cite{Milhano:2022kzx} to ensure energy-momentum conservation (subtracting ``holes''). Also, in our runs (default in Jewel), further rescatterings and splittings of recoil particles are neglected, and they free stream. In the most recent version of Jewel, recoil rescatterings can be activated. Due to the associated computational cost, we did not consider it here. Recoil in Jewel therefore implements the other extreme scenario, a completely weakly coupled medium response. To further illustrate the impact of recoil in Jewel, we repeated the analysis for wider $R=0.4$ jets. In this case, even the highest $k_t^{\rm min}=30$ GeV cut is sensitive to recoil particles,
evidencing the stark consequences of the free streaming setup adopted in the present Jewel study. 
This is the reason why we did not include the Jewel recoils in the main text. 
Indeed, it would be very instructive to test more realistic modelling of medium response, where re-scatterings can occur, for instance using kinetic theory~\cite{Wang:2013cia,Mehtar-Tani:2022zwf}. Differences in the PbPb/pp ratio between Hybrid and Jewel in the absence of medium response are due to the different energy loss mechanisms adopted by each model, where in particular one observes a flatter ratio in Jewel than in Hybrid.

To sum up, our observable not only separates perturbative from non-perturbative effects in terms of radiation and energy loss, but it is also sensitive to the physics of perturbative scatterings and the thermalisation process of jets. The choice of using jets with high-enough $p_t$ is in fact crucial, as it allows enough phase space for large-momentum transfer, and thus perturbative, scatterings with the thermal QGP constituents to occur.
Measuring this observable in experiments will in this way serve to guide the modelling of medium response in Monte Carlo event generators.

%%%%%%%%%%%%%%%%%%%%%%%%%%%%%%%%%%%%%%%%%%
%%%%%%%%%%%%%%%%%%%%%%%%%%%%%%%%%%%%%%%%%%

\bibliography{references}{}

\begin{thebibliography}{121}%
\makeatletter
\providecommand \@ifxundefined [1]{%
 \@ifx{#1\undefined}
}%
\providecommand \@ifnum [1]{%
 \ifnum #1\expandafter \@firstoftwo
 \else \expandafter \@secondoftwo
 \fi
}%
\providecommand \@ifx [1]{%
 \ifx #1\expandafter \@firstoftwo
 \else \expandafter \@secondoftwo
 \fi
}%
\providecommand \natexlab [1]{#1}%
\providecommand \enquote  [1]{``#1''}%
\providecommand \bibnamefont  [1]{#1}%
\providecommand \bibfnamefont [1]{#1}%
\providecommand \citenamefont [1]{#1}%
\providecommand \href@noop [0]{\@secondoftwo}%
\providecommand \href [0]{\begingroup \@sanitize@url \@href}%
\providecommand \@href[1]{\@@startlink{#1}\@@href}%
\providecommand \@@href[1]{\endgroup#1\@@endlink}%
\providecommand \@sanitize@url [0]{\catcode `\\12\catcode `\$12\catcode
  `\&12\catcode `\#12\catcode `\^12\catcode `\_12\catcode `\%12\relax}%
\providecommand \@@startlink[1]{}%
\providecommand \@@endlink[0]{}%
\providecommand \url  [0]{\begingroup\@sanitize@url \@url }%
\providecommand \@url [1]{\endgroup\@href {#1}{\urlprefix }}%
\providecommand \urlprefix  [0]{URL }%
\providecommand \Eprint [0]{\href }%
\providecommand \doibase [0]{https://doi.org/}%
\providecommand \selectlanguage [0]{\@gobble}%
\providecommand \bibinfo  [0]{\@secondoftwo}%
\providecommand \bibfield  [0]{\@secondoftwo}%
\providecommand \translation [1]{[#1]}%
\providecommand \BibitemOpen [0]{}%
\providecommand \bibitemStop [0]{}%
\providecommand \bibitemNoStop [0]{.\EOS\space}%
\providecommand \EOS [0]{\spacefactor3000\relax}%
\providecommand \BibitemShut  [1]{\csname bibitem#1\endcsname}%
\let\auto@bib@innerbib\@empty
%</preamble>
\bibitem [{\citenamefont {Armesto}\ and\ \citenamefont
  {Scomparin}(2016)}]{Armesto:2015ioy}%
  \BibitemOpen
  \bibfield  {author} {\bibinfo {author} {\bibfnamefont {N.}~\bibnamefont
  {Armesto}}\ and\ \bibinfo {author} {\bibfnamefont {E.}~\bibnamefont
  {Scomparin}},\ }\bibfield  {title} {\bibinfo {title} {{Heavy-ion collisions
  at the Large Hadron Collider: a review of the results from Run 1}},\ }\href
  {https://doi.org/10.1140/epjp/i2016-16052-4} {\bibfield  {journal} {\bibinfo
  {journal} {Eur. Phys. J. Plus}\ }\textbf {\bibinfo {volume} {131}},\ \bibinfo
  {pages} {52} (\bibinfo {year} {2016})},\ \Eprint
  {https://arxiv.org/abs/1511.02151} {arXiv:1511.02151 [nucl-ex]} \BibitemShut
  {NoStop}%
\bibitem [{\citenamefont {Connors}\ \emph {et~al.}(2018)\citenamefont
  {Connors}, \citenamefont {Nattrass}, \citenamefont {Reed},\ and\
  \citenamefont {Salur}}]{Connors:2017ptx}%
  \BibitemOpen
  \bibfield  {author} {\bibinfo {author} {\bibfnamefont {M.}~\bibnamefont
  {Connors}}, \bibinfo {author} {\bibfnamefont {C.}~\bibnamefont {Nattrass}},
  \bibinfo {author} {\bibfnamefont {R.}~\bibnamefont {Reed}},\ and\ \bibinfo
  {author} {\bibfnamefont {S.}~\bibnamefont {Salur}},\ }\bibfield  {title}
  {\bibinfo {title} {{Jet measurements in heavy ion physics}},\ }\href
  {https://doi.org/10.1103/RevModPhys.90.025005} {\bibfield  {journal}
  {\bibinfo  {journal} {Rev. Mod. Phys.}\ }\textbf {\bibinfo {volume} {90}},\
  \bibinfo {pages} {025005} (\bibinfo {year} {2018})},\ \Eprint
  {https://arxiv.org/abs/1705.01974} {arXiv:1705.01974 [nucl-ex]} \BibitemShut
  {NoStop}%
\bibitem [{\citenamefont {Cunqueiro}\ and\ \citenamefont
  {Sickles}(2022)}]{Cunqueiro:2021wls}%
  \BibitemOpen
  \bibfield  {author} {\bibinfo {author} {\bibfnamefont {L.}~\bibnamefont
  {Cunqueiro}}\ and\ \bibinfo {author} {\bibfnamefont {A.~M.}\ \bibnamefont
  {Sickles}},\ }\bibfield  {title} {\bibinfo {title} {{Studying the QGP with
  Jets at the LHC and RHIC}},\ }\href
  {https://doi.org/10.1016/j.ppnp.2022.103940} {\bibfield  {journal} {\bibinfo
  {journal} {Prog. Part. Nucl. Phys.}\ }\textbf {\bibinfo {volume} {124}},\
  \bibinfo {pages} {103940} (\bibinfo {year} {2022})},\ \Eprint
  {https://arxiv.org/abs/2110.14490} {arXiv:2110.14490 [nucl-ex]} \BibitemShut
  {NoStop}%
\bibitem [{\citenamefont {Apolin\'ario}\ \emph
  {et~al.}(2022{\natexlab{a}})\citenamefont {Apolin\'ario}, \citenamefont
  {Lee},\ and\ \citenamefont {Winn}}]{Apolinario:2022vzg}%
  \BibitemOpen
  \bibfield  {author} {\bibinfo {author} {\bibfnamefont {L.}~\bibnamefont
  {Apolin\'ario}}, \bibinfo {author} {\bibfnamefont {Y.-J.}\ \bibnamefont
  {Lee}},\ and\ \bibinfo {author} {\bibfnamefont {M.}~\bibnamefont {Winn}},\
  }\bibfield  {title} {\bibinfo {title} {{Heavy quarks and jets as probes of
  the QGP}},\ }\href {https://doi.org/10.1016/j.ppnp.2022.103990} {\bibfield
  {journal} {\bibinfo  {journal} {Prog. Part. Nucl. Phys.}\ }\textbf {\bibinfo
  {volume} {127}},\ \bibinfo {pages} {103990} (\bibinfo {year}
  {2022}{\natexlab{a}})},\ \Eprint {https://arxiv.org/abs/2203.16352}
  {arXiv:2203.16352 [hep-ph]} \BibitemShut {NoStop}%
\bibitem [{ALI(2022)}]{ALICE:2022wpn}%
  \BibitemOpen
  \bibfield  {title} {\bibinfo {title} {{The ALICE experiment -- A journey
  through QCD}},\ }\href@noop {} {\  (\bibinfo {year} {2022})},\ \Eprint
  {https://arxiv.org/abs/2211.04384} {arXiv:2211.04384 [nucl-ex]} \BibitemShut
  {NoStop}%
\bibitem [{\citenamefont {Casalderrey-Solana}\ and\ \citenamefont
  {Salgado}(2007)}]{Casalderrey-Solana:2007knd}%
  \BibitemOpen
  \bibfield  {author} {\bibinfo {author} {\bibfnamefont {J.}~\bibnamefont
  {Casalderrey-Solana}}\ and\ \bibinfo {author} {\bibfnamefont {C.~A.}\
  \bibnamefont {Salgado}},\ }\bibfield  {title} {\bibinfo {title}
  {{Introductory lectures on jet quenching in heavy ion collisions}},\
  }\href@noop {} {\bibfield  {journal} {\bibinfo  {journal} {Acta Phys. Polon.
  B}\ }\textbf {\bibinfo {volume} {38}},\ \bibinfo {pages} {3731} (\bibinfo
  {year} {2007})},\ \Eprint {https://arxiv.org/abs/0712.3443} {arXiv:0712.3443
  [hep-ph]} \BibitemShut {NoStop}%
\bibitem [{\citenamefont {Majumder}\ and\ \citenamefont
  {Van~Leeuwen}(2011)}]{Majumder:2010qh}%
  \BibitemOpen
  \bibfield  {author} {\bibinfo {author} {\bibfnamefont {A.}~\bibnamefont
  {Majumder}}\ and\ \bibinfo {author} {\bibfnamefont {M.}~\bibnamefont
  {Van~Leeuwen}},\ }\bibfield  {title} {\bibinfo {title} {{The Theory and
  Phenomenology of Perturbative QCD Based Jet Quenching}},\ }\href
  {https://doi.org/10.1016/j.ppnp.2010.09.001} {\bibfield  {journal} {\bibinfo
  {journal} {Prog. Part. Nucl. Phys.}\ }\textbf {\bibinfo {volume} {66}},\
  \bibinfo {pages} {41} (\bibinfo {year} {2011})},\ \Eprint
  {https://arxiv.org/abs/1002.2206} {arXiv:1002.2206 [hep-ph]} \BibitemShut
  {NoStop}%
\bibitem [{\citenamefont {Mehtar-Tani}\ \emph {et~al.}(2013)\citenamefont
  {Mehtar-Tani}, \citenamefont {Milhano},\ and\ \citenamefont
  {Tywoniuk}}]{Mehtar-Tani:2013pia}%
  \BibitemOpen
  \bibfield  {author} {\bibinfo {author} {\bibfnamefont {Y.}~\bibnamefont
  {Mehtar-Tani}}, \bibinfo {author} {\bibfnamefont {J.~G.}\ \bibnamefont
  {Milhano}},\ and\ \bibinfo {author} {\bibfnamefont {K.}~\bibnamefont
  {Tywoniuk}},\ }\bibfield  {title} {\bibinfo {title} {{Jet physics in
  heavy-ion collisions}},\ }\href {https://doi.org/10.1142/S0217751X13400137}
  {\bibfield  {journal} {\bibinfo  {journal} {Int. J. Mod. Phys. A}\ }\textbf
  {\bibinfo {volume} {28}},\ \bibinfo {pages} {1340013} (\bibinfo {year}
  {2013})},\ \Eprint {https://arxiv.org/abs/1302.2579} {arXiv:1302.2579
  [hep-ph]} \BibitemShut {NoStop}%
\bibitem [{\citenamefont {Ethier}\ and\ \citenamefont
  {Nocera}(2020)}]{Ethier:2020way}%
  \BibitemOpen
  \bibfield  {author} {\bibinfo {author} {\bibfnamefont {J.~J.}\ \bibnamefont
  {Ethier}}\ and\ \bibinfo {author} {\bibfnamefont {E.~R.}\ \bibnamefont
  {Nocera}},\ }\bibfield  {title} {\bibinfo {title} {{Parton Distributions in
  Nucleons and Nuclei}},\ }\href
  {https://doi.org/10.1146/annurev-nucl-011720-042725} {\bibfield  {journal}
  {\bibinfo  {journal} {Ann. Rev. Nucl. Part. Sci.}\ }\textbf {\bibinfo
  {volume} {70}},\ \bibinfo {pages} {43} (\bibinfo {year} {2020})},\ \Eprint
  {https://arxiv.org/abs/2001.07722} {arXiv:2001.07722 [hep-ph]} \BibitemShut
  {NoStop}%
\bibitem [{\citenamefont {Eskola}\ \emph {et~al.}(2022)\citenamefont {Eskola},
  \citenamefont {Paakkinen}, \citenamefont {Paukkunen},\ and\ \citenamefont
  {Salgado}}]{Eskola:2021nhw}%
  \BibitemOpen
  \bibfield  {author} {\bibinfo {author} {\bibfnamefont {K.~J.}\ \bibnamefont
  {Eskola}}, \bibinfo {author} {\bibfnamefont {P.}~\bibnamefont {Paakkinen}},
  \bibinfo {author} {\bibfnamefont {H.}~\bibnamefont {Paukkunen}},\ and\
  \bibinfo {author} {\bibfnamefont {C.~A.}\ \bibnamefont {Salgado}},\
  }\bibfield  {title} {\bibinfo {title} {{EPPS21: a global QCD analysis of
  nuclear PDFs}},\ }\href {https://doi.org/10.1140/epjc/s10052-022-10359-0}
  {\bibfield  {journal} {\bibinfo  {journal} {Eur. Phys. J. C}\ }\textbf
  {\bibinfo {volume} {82}},\ \bibinfo {pages} {413} (\bibinfo {year} {2022})},\
  \Eprint {https://arxiv.org/abs/2112.12462} {arXiv:2112.12462 [hep-ph]}
  \BibitemShut {NoStop}%
\bibitem [{\citenamefont {Abdul~Khalek}\ \emph {et~al.}(2022)\citenamefont
  {Abdul~Khalek}, \citenamefont {Gauld}, \citenamefont {Giani}, \citenamefont
  {Nocera}, \citenamefont {Rabemananjara},\ and\ \citenamefont
  {Rojo}}]{AbdulKhalek:2022fyi}%
  \BibitemOpen
  \bibfield  {author} {\bibinfo {author} {\bibfnamefont {R.}~\bibnamefont
  {Abdul~Khalek}}, \bibinfo {author} {\bibfnamefont {R.}~\bibnamefont {Gauld}},
  \bibinfo {author} {\bibfnamefont {T.}~\bibnamefont {Giani}}, \bibinfo
  {author} {\bibfnamefont {E.~R.}\ \bibnamefont {Nocera}}, \bibinfo {author}
  {\bibfnamefont {T.~R.}\ \bibnamefont {Rabemananjara}},\ and\ \bibinfo
  {author} {\bibfnamefont {J.}~\bibnamefont {Rojo}},\ }\bibfield  {title}
  {\bibinfo {title} {{nNNPDF3.0: evidence for a modified partonic structure in
  heavy nuclei}},\ }\href {https://doi.org/10.1140/epjc/s10052-022-10417-7}
  {\bibfield  {journal} {\bibinfo  {journal} {Eur. Phys. J. C}\ }\textbf
  {\bibinfo {volume} {82}},\ \bibinfo {pages} {507} (\bibinfo {year} {2022})},\
  \Eprint {https://arxiv.org/abs/2201.12363} {arXiv:2201.12363 [hep-ph]}
  \BibitemShut {NoStop}%
\bibitem [{\citenamefont {Kurkela}\ and\ \citenamefont
  {Wiedemann}(2015)}]{Kurkela:2014tla}%
  \BibitemOpen
  \bibfield  {author} {\bibinfo {author} {\bibfnamefont {A.}~\bibnamefont
  {Kurkela}}\ and\ \bibinfo {author} {\bibfnamefont {U.~A.}\ \bibnamefont
  {Wiedemann}},\ }\bibfield  {title} {\bibinfo {title} {{Picturing perturbative
  parton cascades in QCD matter}},\ }\href
  {https://doi.org/10.1016/j.physletb.2014.11.054} {\bibfield  {journal}
  {\bibinfo  {journal} {Phys. Lett. B}\ }\textbf {\bibinfo {volume} {740}},\
  \bibinfo {pages} {172} (\bibinfo {year} {2015})},\ \Eprint
  {https://arxiv.org/abs/1407.0293} {arXiv:1407.0293 [hep-ph]} \BibitemShut
  {NoStop}%
\bibitem [{\citenamefont {Caucal}\ \emph {et~al.}(2018)\citenamefont {Caucal},
  \citenamefont {Iancu}, \citenamefont {Mueller},\ and\ \citenamefont
  {Soyez}}]{Caucal:2018dla}%
  \BibitemOpen
  \bibfield  {author} {\bibinfo {author} {\bibfnamefont {P.}~\bibnamefont
  {Caucal}}, \bibinfo {author} {\bibfnamefont {E.}~\bibnamefont {Iancu}},
  \bibinfo {author} {\bibfnamefont {A.~H.}\ \bibnamefont {Mueller}},\ and\
  \bibinfo {author} {\bibfnamefont {G.}~\bibnamefont {Soyez}},\ }\bibfield
  {title} {\bibinfo {title} {{Vacuum-like jet fragmentation in a dense QCD
  medium}},\ }\href {https://doi.org/10.1103/PhysRevLett.120.232001} {\bibfield
   {journal} {\bibinfo  {journal} {Phys. Rev. Lett.}\ }\textbf {\bibinfo
  {volume} {120}},\ \bibinfo {pages} {232001} (\bibinfo {year} {2018})},\
  \Eprint {https://arxiv.org/abs/1801.09703} {arXiv:1801.09703 [hep-ph]}
  \BibitemShut {NoStop}%
\bibitem [{\citenamefont {Baier}\ \emph {et~al.}(1997)\citenamefont {Baier},
  \citenamefont {Dokshitzer}, \citenamefont {Mueller}, \citenamefont {Peigne},\
  and\ \citenamefont {Schiff}}]{Baier:1996kr}%
  \BibitemOpen
  \bibfield  {author} {\bibinfo {author} {\bibfnamefont {R.}~\bibnamefont
  {Baier}}, \bibinfo {author} {\bibfnamefont {Y.~L.}\ \bibnamefont
  {Dokshitzer}}, \bibinfo {author} {\bibfnamefont {A.~H.}\ \bibnamefont
  {Mueller}}, \bibinfo {author} {\bibfnamefont {S.}~\bibnamefont {Peigne}},\
  and\ \bibinfo {author} {\bibfnamefont {D.}~\bibnamefont {Schiff}},\
  }\bibfield  {title} {\bibinfo {title} {{Radiative energy loss of high-energy
  quarks and gluons in a finite volume quark - gluon plasma}},\ }\href
  {https://doi.org/10.1016/S0550-3213(96)00553-6} {\bibfield  {journal}
  {\bibinfo  {journal} {Nucl. Phys. B}\ }\textbf {\bibinfo {volume} {483}},\
  \bibinfo {pages} {291} (\bibinfo {year} {1997})},\ \Eprint
  {https://arxiv.org/abs/hep-ph/9607355} {arXiv:hep-ph/9607355} \BibitemShut
  {NoStop}%
\bibitem [{\citenamefont {Zakharov}(1996)}]{Zakharov:1996fv}%
  \BibitemOpen
  \bibfield  {author} {\bibinfo {author} {\bibfnamefont {B.}~\bibnamefont
  {Zakharov}},\ }\bibfield  {title} {\bibinfo {title} {{Fully quantum treatment
  of the Landau-Pomeranchuk-Migdal effect in QED and QCD}},\ }\href
  {https://doi.org/10.1134/1.567126} {\bibfield  {journal} {\bibinfo  {journal}
  {JETP Lett.}\ }\textbf {\bibinfo {volume} {63}},\ \bibinfo {pages} {952}
  (\bibinfo {year} {1996})},\ \Eprint {https://arxiv.org/abs/hep-ph/9607440}
  {arXiv:hep-ph/9607440} \BibitemShut {NoStop}%
\bibitem [{\citenamefont {Gyulassy}\ \emph {et~al.}(2000)\citenamefont
  {Gyulassy}, \citenamefont {Levai},\ and\ \citenamefont
  {Vitev}}]{Gyulassy:1999zd}%
  \BibitemOpen
  \bibfield  {author} {\bibinfo {author} {\bibfnamefont {M.}~\bibnamefont
  {Gyulassy}}, \bibinfo {author} {\bibfnamefont {P.}~\bibnamefont {Levai}},\
  and\ \bibinfo {author} {\bibfnamefont {I.}~\bibnamefont {Vitev}},\ }\bibfield
   {title} {\bibinfo {title} {{Jet quenching in thin quark gluon plasmas. 1.
  Formalism}},\ }\href {https://doi.org/10.1016/S0550-3213(99)00713-0}
  {\bibfield  {journal} {\bibinfo  {journal} {Nucl. Phys. B}\ }\textbf
  {\bibinfo {volume} {571}},\ \bibinfo {pages} {197} (\bibinfo {year}
  {2000})},\ \Eprint {https://arxiv.org/abs/hep-ph/9907461}
  {arXiv:hep-ph/9907461} \BibitemShut {NoStop}%
\bibitem [{\citenamefont {Wiedemann}(2000)}]{Wiedemann:2000za}%
  \BibitemOpen
  \bibfield  {author} {\bibinfo {author} {\bibfnamefont {U.~A.}\ \bibnamefont
  {Wiedemann}},\ }\bibfield  {title} {\bibinfo {title} {{Gluon radiation off
  hard quarks in a nuclear environment: Opacity expansion}},\ }\href
  {https://doi.org/10.1016/S0550-3213(00)00457-0} {\bibfield  {journal}
  {\bibinfo  {journal} {Nucl. Phys. B}\ }\textbf {\bibinfo {volume} {588}},\
  \bibinfo {pages} {303} (\bibinfo {year} {2000})},\ \Eprint
  {https://arxiv.org/abs/hep-ph/0005129} {arXiv:hep-ph/0005129} \BibitemShut
  {NoStop}%
\bibitem [{\citenamefont {Ghiglieri}\ and\ \citenamefont
  {Teaney}(2015)}]{Ghiglieri:2015zma}%
  \BibitemOpen
  \bibfield  {author} {\bibinfo {author} {\bibfnamefont {J.}~\bibnamefont
  {Ghiglieri}}\ and\ \bibinfo {author} {\bibfnamefont {D.}~\bibnamefont
  {Teaney}},\ }\bibfield  {title} {\bibinfo {title} {{Parton energy loss and
  momentum broadening at NLO in high temperature QCD plasmas}},\ }\href
  {https://doi.org/10.1142/S0218301315300131} {\bibfield  {journal} {\bibinfo
  {journal} {Int. J. Mod. Phys. E}\ }\textbf {\bibinfo {volume} {24}},\
  \bibinfo {pages} {1530013} (\bibinfo {year} {2015})},\ \Eprint
  {https://arxiv.org/abs/1502.03730} {arXiv:1502.03730 [hep-ph]} \BibitemShut
  {NoStop}%
\bibitem [{\citenamefont {Schlichting}\ and\ \citenamefont
  {Teaney}(2019)}]{Schlichting:2019abc}%
  \BibitemOpen
  \bibfield  {author} {\bibinfo {author} {\bibfnamefont {S.}~\bibnamefont
  {Schlichting}}\ and\ \bibinfo {author} {\bibfnamefont {D.}~\bibnamefont
  {Teaney}},\ }\bibfield  {title} {\bibinfo {title} {{The First fm/c of
  Heavy-Ion Collisions}},\ }\href
  {https://doi.org/10.1146/annurev-nucl-101918-023825} {\bibfield  {journal}
  {\bibinfo  {journal} {Ann. Rev. Nucl. Part. Sci.}\ }\textbf {\bibinfo
  {volume} {69}},\ \bibinfo {pages} {447} (\bibinfo {year} {2019})},\ \Eprint
  {https://arxiv.org/abs/1908.02113} {arXiv:1908.02113 [nucl-th]} \BibitemShut
  {NoStop}%
\bibitem [{\citenamefont {Berges}\ \emph {et~al.}(2021)\citenamefont {Berges},
  \citenamefont {Heller}, \citenamefont {Mazeliauskas},\ and\ \citenamefont
  {Venugopalan}}]{Berges:2020fwq}%
  \BibitemOpen
  \bibfield  {author} {\bibinfo {author} {\bibfnamefont {J.}~\bibnamefont
  {Berges}}, \bibinfo {author} {\bibfnamefont {M.~P.}\ \bibnamefont {Heller}},
  \bibinfo {author} {\bibfnamefont {A.}~\bibnamefont {Mazeliauskas}},\ and\
  \bibinfo {author} {\bibfnamefont {R.}~\bibnamefont {Venugopalan}},\
  }\bibfield  {title} {\bibinfo {title} {{QCD thermalization: Ab initio
  approaches and interdisciplinary connections}},\ }\href
  {https://doi.org/10.1103/RevModPhys.93.035003} {\bibfield  {journal}
  {\bibinfo  {journal} {Rev. Mod. Phys.}\ }\textbf {\bibinfo {volume} {93}},\
  \bibinfo {pages} {035003} (\bibinfo {year} {2021})},\ \Eprint
  {https://arxiv.org/abs/2005.12299} {arXiv:2005.12299 [hep-th]} \BibitemShut
  {NoStop}%
\bibitem [{\citenamefont {Andersson}\ \emph {et~al.}(1983)\citenamefont
  {Andersson}, \citenamefont {Gustafson}, \citenamefont {Ingelman},\ and\
  \citenamefont {Sjostrand}}]{Andersson:1983ia}%
  \BibitemOpen
  \bibfield  {author} {\bibinfo {author} {\bibfnamefont {B.}~\bibnamefont
  {Andersson}}, \bibinfo {author} {\bibfnamefont {G.}~\bibnamefont
  {Gustafson}}, \bibinfo {author} {\bibfnamefont {G.}~\bibnamefont
  {Ingelman}},\ and\ \bibinfo {author} {\bibfnamefont {T.}~\bibnamefont
  {Sjostrand}},\ }\bibfield  {title} {\bibinfo {title} {{Parton Fragmentation
  and String Dynamics}},\ }\href {https://doi.org/10.1016/0370-1573(83)90080-7}
  {\bibfield  {journal} {\bibinfo  {journal} {Phys. Rept.}\ }\textbf {\bibinfo
  {volume} {97}},\ \bibinfo {pages} {31} (\bibinfo {year} {1983})}\BibitemShut
  {NoStop}%
\bibitem [{\citenamefont {Webber}(1984)}]{Webber:1983if}%
  \BibitemOpen
  \bibfield  {author} {\bibinfo {author} {\bibfnamefont {B.~R.}\ \bibnamefont
  {Webber}},\ }\bibfield  {title} {\bibinfo {title} {{A QCD Model for Jet
  Fragmentation Including Soft Gluon Interference}},\ }\href
  {https://doi.org/10.1016/0550-3213(84)90333-X} {\bibfield  {journal}
  {\bibinfo  {journal} {Nucl. Phys. B}\ }\textbf {\bibinfo {volume} {238}},\
  \bibinfo {pages} {492} (\bibinfo {year} {1984})}\BibitemShut {NoStop}%
\bibitem [{\citenamefont {Beraudo}\ \emph {et~al.}(2012)\citenamefont
  {Beraudo}, \citenamefont {Milhano},\ and\ \citenamefont
  {Wiedemann}}]{Beraudo:2011bh}%
  \BibitemOpen
  \bibfield  {author} {\bibinfo {author} {\bibfnamefont {A.}~\bibnamefont
  {Beraudo}}, \bibinfo {author} {\bibfnamefont {J.~G.}\ \bibnamefont
  {Milhano}},\ and\ \bibinfo {author} {\bibfnamefont {U.~A.}\ \bibnamefont
  {Wiedemann}},\ }\bibfield  {title} {\bibinfo {title} {{Medium-induced color
  flow softens hadronization}},\ }\href
  {https://doi.org/10.1103/PhysRevC.85.031901} {\bibfield  {journal} {\bibinfo
  {journal} {Phys. Rev. C}\ }\textbf {\bibinfo {volume} {85}},\ \bibinfo
  {pages} {031901} (\bibinfo {year} {2012})},\ \Eprint
  {https://arxiv.org/abs/1109.5025} {arXiv:1109.5025 [hep-ph]} \BibitemShut
  {NoStop}%
\bibitem [{\citenamefont {Chesler}\ \emph {et~al.}(2009)\citenamefont
  {Chesler}, \citenamefont {Jensen}, \citenamefont {Karch},\ and\ \citenamefont
  {Yaffe}}]{Chesler:2008uy}%
  \BibitemOpen
  \bibfield  {author} {\bibinfo {author} {\bibfnamefont {P.~M.}\ \bibnamefont
  {Chesler}}, \bibinfo {author} {\bibfnamefont {K.}~\bibnamefont {Jensen}},
  \bibinfo {author} {\bibfnamefont {A.}~\bibnamefont {Karch}},\ and\ \bibinfo
  {author} {\bibfnamefont {L.~G.}\ \bibnamefont {Yaffe}},\ }\bibfield  {title}
  {\bibinfo {title} {{Light quark energy loss in strongly-coupled N = 4
  supersymmetric Yang-Mills plasma}},\ }\href
  {https://doi.org/10.1103/PhysRevD.79.125015} {\bibfield  {journal} {\bibinfo
  {journal} {Phys. Rev. D}\ }\textbf {\bibinfo {volume} {79}},\ \bibinfo
  {pages} {125015} (\bibinfo {year} {2009})},\ \Eprint
  {https://arxiv.org/abs/0810.1985} {arXiv:0810.1985 [hep-th]} \BibitemShut
  {NoStop}%
\bibitem [{\citenamefont {Gubser}\ \emph {et~al.}(2008)\citenamefont {Gubser},
  \citenamefont {Gulotta}, \citenamefont {Pufu},\ and\ \citenamefont
  {Rocha}}]{Gubser:2008as}%
  \BibitemOpen
  \bibfield  {author} {\bibinfo {author} {\bibfnamefont {S.~S.}\ \bibnamefont
  {Gubser}}, \bibinfo {author} {\bibfnamefont {D.~R.}\ \bibnamefont {Gulotta}},
  \bibinfo {author} {\bibfnamefont {S.~S.}\ \bibnamefont {Pufu}},\ and\
  \bibinfo {author} {\bibfnamefont {F.~D.}\ \bibnamefont {Rocha}},\ }\bibfield
  {title} {\bibinfo {title} {{Gluon energy loss in the gauge-string duality}},\
  }\href {https://doi.org/10.1088/1126-6708/2008/10/052} {\bibfield  {journal}
  {\bibinfo  {journal} {JHEP}\ }\textbf {\bibinfo {volume} {10}},\ \bibinfo
  {pages} {052}},\ \Eprint {https://arxiv.org/abs/0803.1470} {arXiv:0803.1470
  [hep-th]} \BibitemShut {NoStop}%
\bibitem [{\citenamefont {Hatta}\ \emph {et~al.}(2008)\citenamefont {Hatta},
  \citenamefont {Iancu},\ and\ \citenamefont {Mueller}}]{Hatta:2008tx}%
  \BibitemOpen
  \bibfield  {author} {\bibinfo {author} {\bibfnamefont {Y.}~\bibnamefont
  {Hatta}}, \bibinfo {author} {\bibfnamefont {E.}~\bibnamefont {Iancu}},\ and\
  \bibinfo {author} {\bibfnamefont {A.~H.}\ \bibnamefont {Mueller}},\
  }\bibfield  {title} {\bibinfo {title} {{Jet evolution in the N=4 SYM plasma
  at strong coupling}},\ }\href {https://doi.org/10.1088/1126-6708/2008/05/037}
  {\bibfield  {journal} {\bibinfo  {journal} {JHEP}\ }\textbf {\bibinfo
  {volume} {05}},\ \bibinfo {pages} {037}},\ \Eprint
  {https://arxiv.org/abs/0803.2481} {arXiv:0803.2481 [hep-th]} \BibitemShut
  {NoStop}%
\bibitem [{\citenamefont {Arnold}\ and\ \citenamefont
  {Vaman}(2010)}]{Arnold:2010ir}%
  \BibitemOpen
  \bibfield  {author} {\bibinfo {author} {\bibfnamefont {P.}~\bibnamefont
  {Arnold}}\ and\ \bibinfo {author} {\bibfnamefont {D.}~\bibnamefont {Vaman}},\
  }\bibfield  {title} {\bibinfo {title} {{Jet quenching in hot strongly coupled
  gauge theories revisited: 3-point correlators with gauge-gravity duality}},\
  }\href {https://doi.org/10.1007/JHEP10(2010)099} {\bibfield  {journal}
  {\bibinfo  {journal} {JHEP}\ }\textbf {\bibinfo {volume} {10}},\ \bibinfo
  {pages} {099}},\ \Eprint {https://arxiv.org/abs/1008.4023} {arXiv:1008.4023
  [hep-th]} \BibitemShut {NoStop}%
\bibitem [{\citenamefont {Chesler}\ and\ \citenamefont
  {Rajagopal}(2014)}]{Chesler:2014jva}%
  \BibitemOpen
  \bibfield  {author} {\bibinfo {author} {\bibfnamefont {P.~M.}\ \bibnamefont
  {Chesler}}\ and\ \bibinfo {author} {\bibfnamefont {K.}~\bibnamefont
  {Rajagopal}},\ }\bibfield  {title} {\bibinfo {title} {{Jet quenching in
  strongly coupled plasma}},\ }\href
  {https://doi.org/10.1103/PhysRevD.90.025033} {\bibfield  {journal} {\bibinfo
  {journal} {Phys. Rev. D}\ }\textbf {\bibinfo {volume} {90}},\ \bibinfo
  {pages} {025033} (\bibinfo {year} {2014})},\ \Eprint
  {https://arxiv.org/abs/1402.6756} {arXiv:1402.6756 [hep-th]} \BibitemShut
  {NoStop}%
\bibitem [{\citenamefont {Lokhtin}\ and\ \citenamefont
  {Snigirev}(2006)}]{Lokhtin:2005px}%
  \BibitemOpen
  \bibfield  {author} {\bibinfo {author} {\bibfnamefont {I.~P.}\ \bibnamefont
  {Lokhtin}}\ and\ \bibinfo {author} {\bibfnamefont {A.~M.}\ \bibnamefont
  {Snigirev}},\ }\bibfield  {title} {\bibinfo {title} {{A Model of jet
  quenching in ultrarelativistic heavy ion collisions and high-p(T) hadron
  spectra at RHIC}},\ }\href {https://doi.org/10.1140/epjc/s2005-02426-3}
  {\bibfield  {journal} {\bibinfo  {journal} {Eur. Phys. J. C}\ }\textbf
  {\bibinfo {volume} {45}},\ \bibinfo {pages} {211} (\bibinfo {year} {2006})},\
  \Eprint {https://arxiv.org/abs/hep-ph/0506189} {arXiv:hep-ph/0506189}
  \BibitemShut {NoStop}%
\bibitem [{\citenamefont {Zapp}\ \emph {et~al.}(2009)\citenamefont {Zapp},
  \citenamefont {Ingelman}, \citenamefont {Rathsman}, \citenamefont {Stachel},\
  and\ \citenamefont {Wiedemann}}]{Zapp:2008gi}%
  \BibitemOpen
  \bibfield  {author} {\bibinfo {author} {\bibfnamefont {K.}~\bibnamefont
  {Zapp}}, \bibinfo {author} {\bibfnamefont {G.}~\bibnamefont {Ingelman}},
  \bibinfo {author} {\bibfnamefont {J.}~\bibnamefont {Rathsman}}, \bibinfo
  {author} {\bibfnamefont {J.}~\bibnamefont {Stachel}},\ and\ \bibinfo {author}
  {\bibfnamefont {U.~A.}\ \bibnamefont {Wiedemann}},\ }\bibfield  {title}
  {\bibinfo {title} {{A Monte Carlo Model for 'Jet Quenching'}},\ }\href
  {https://doi.org/10.1140/epjc/s10052-009-0941-2} {\bibfield  {journal}
  {\bibinfo  {journal} {Eur. Phys. J. C}\ }\textbf {\bibinfo {volume} {60}},\
  \bibinfo {pages} {617} (\bibinfo {year} {2009})},\ \Eprint
  {https://arxiv.org/abs/0804.3568} {arXiv:0804.3568 [hep-ph]} \BibitemShut
  {NoStop}%
\bibitem [{\citenamefont {Armesto}\ \emph
  {et~al.}(2009{\natexlab{a}})\citenamefont {Armesto}, \citenamefont
  {Cunqueiro},\ and\ \citenamefont {Salgado}}]{Armesto:2009fj}%
  \BibitemOpen
  \bibfield  {author} {\bibinfo {author} {\bibfnamefont {N.}~\bibnamefont
  {Armesto}}, \bibinfo {author} {\bibfnamefont {L.}~\bibnamefont {Cunqueiro}},\
  and\ \bibinfo {author} {\bibfnamefont {C.~A.}\ \bibnamefont {Salgado}},\
  }\bibfield  {title} {\bibinfo {title} {{Q-PYTHIA: A Medium-modified
  implementation of final state radiation}},\ }\href
  {https://doi.org/10.1140/epjc/s10052-009-1133-9} {\bibfield  {journal}
  {\bibinfo  {journal} {Eur. Phys. J. C}\ }\textbf {\bibinfo {volume} {63}},\
  \bibinfo {pages} {679} (\bibinfo {year} {2009}{\natexlab{a}})},\ \Eprint
  {https://arxiv.org/abs/0907.1014} {arXiv:0907.1014 [hep-ph]} \BibitemShut
  {NoStop}%
\bibitem [{\citenamefont {Casalderrey-Solana}\ \emph
  {et~al.}(2012)\citenamefont {Casalderrey-Solana}, \citenamefont {Milhano},\
  and\ \citenamefont {Quiroga-Arias}}]{Casalderrey-Solana:2011fza}%
  \BibitemOpen
  \bibfield  {author} {\bibinfo {author} {\bibfnamefont {J.}~\bibnamefont
  {Casalderrey-Solana}}, \bibinfo {author} {\bibfnamefont {J.~G.}\ \bibnamefont
  {Milhano}},\ and\ \bibinfo {author} {\bibfnamefont {P.}~\bibnamefont
  {Quiroga-Arias}},\ }\bibfield  {title} {\bibinfo {title} {{Out of Medium
  Fragmentation from Long-Lived Jet Showers}},\ }\href
  {https://doi.org/10.1016/j.physletb.2012.02.066} {\bibfield  {journal}
  {\bibinfo  {journal} {Phys. Lett. B}\ }\textbf {\bibinfo {volume} {710}},\
  \bibinfo {pages} {175} (\bibinfo {year} {2012})},\ \Eprint
  {https://arxiv.org/abs/1111.0310} {arXiv:1111.0310 [hep-ph]} \BibitemShut
  {NoStop}%
\bibitem [{\citenamefont {Schenke}\ \emph {et~al.}(2009)\citenamefont
  {Schenke}, \citenamefont {Gale},\ and\ \citenamefont
  {Jeon}}]{Schenke:2009gb}%
  \BibitemOpen
  \bibfield  {author} {\bibinfo {author} {\bibfnamefont {B.}~\bibnamefont
  {Schenke}}, \bibinfo {author} {\bibfnamefont {C.}~\bibnamefont {Gale}},\ and\
  \bibinfo {author} {\bibfnamefont {S.}~\bibnamefont {Jeon}},\ }\bibfield
  {title} {\bibinfo {title} {{MARTINI: An Event generator for relativistic
  heavy-ion collisions}},\ }\href {https://doi.org/10.1103/PhysRevC.80.054913}
  {\bibfield  {journal} {\bibinfo  {journal} {Phys. Rev. C}\ }\textbf {\bibinfo
  {volume} {80}},\ \bibinfo {pages} {054913} (\bibinfo {year} {2009})},\
  \Eprint {https://arxiv.org/abs/0909.2037} {arXiv:0909.2037 [hep-ph]}
  \BibitemShut {NoStop}%
\bibitem [{\citenamefont {Majumder}(2013)}]{Majumder:2013re}%
  \BibitemOpen
  \bibfield  {author} {\bibinfo {author} {\bibfnamefont {A.}~\bibnamefont
  {Majumder}},\ }\bibfield  {title} {\bibinfo {title} {{Incorporating
  Space-Time Within Medium-Modified Jet Event Generators}},\ }\href
  {https://doi.org/10.1103/PhysRevC.88.014909} {\bibfield  {journal} {\bibinfo
  {journal} {Phys. Rev. C}\ }\textbf {\bibinfo {volume} {88}},\ \bibinfo
  {pages} {014909} (\bibinfo {year} {2013})},\ \Eprint
  {https://arxiv.org/abs/1301.5323} {arXiv:1301.5323 [nucl-th]} \BibitemShut
  {NoStop}%
\bibitem [{\citenamefont {Wang}\ and\ \citenamefont
  {Zhu}(2013)}]{Wang:2013cia}%
  \BibitemOpen
  \bibfield  {author} {\bibinfo {author} {\bibfnamefont {X.-N.}\ \bibnamefont
  {Wang}}\ and\ \bibinfo {author} {\bibfnamefont {Y.}~\bibnamefont {Zhu}},\
  }\bibfield  {title} {\bibinfo {title} {{Medium Modification of $\gamma$-jets
  in High-energy Heavy-ion Collisions}},\ }\href
  {https://doi.org/10.1103/PhysRevLett.111.062301} {\bibfield  {journal}
  {\bibinfo  {journal} {Phys. Rev. Lett.}\ }\textbf {\bibinfo {volume} {111}},\
  \bibinfo {pages} {062301} (\bibinfo {year} {2013})},\ \Eprint
  {https://arxiv.org/abs/1302.5874} {arXiv:1302.5874 [hep-ph]} \BibitemShut
  {NoStop}%
\bibitem [{\citenamefont {Casalderrey-Solana}\ \emph
  {et~al.}(2014)\citenamefont {Casalderrey-Solana}, \citenamefont {Gulhan},
  \citenamefont {Milhano}, \citenamefont {Pablos},\ and\ \citenamefont
  {Rajagopal}}]{Casalderrey-Solana:2014bpa}%
  \BibitemOpen
  \bibfield  {author} {\bibinfo {author} {\bibfnamefont {J.}~\bibnamefont
  {Casalderrey-Solana}}, \bibinfo {author} {\bibfnamefont {D.~C.}\ \bibnamefont
  {Gulhan}}, \bibinfo {author} {\bibfnamefont {J.~G.}\ \bibnamefont {Milhano}},
  \bibinfo {author} {\bibfnamefont {D.}~\bibnamefont {Pablos}},\ and\ \bibinfo
  {author} {\bibfnamefont {K.}~\bibnamefont {Rajagopal}},\ }\bibfield  {title}
  {\bibinfo {title} {{A Hybrid Strong/Weak Coupling Approach to Jet
  Quenching}},\ }\href {https://doi.org/10.1007/JHEP09(2015)175} {\bibfield
  {journal} {\bibinfo  {journal} {JHEP}\ }\textbf {\bibinfo {volume} {10}},\
  \bibinfo {pages} {019}},\ \bibinfo {note} {[Erratum: JHEP 09, 175 (2015)]},\
  \Eprint {https://arxiv.org/abs/1405.3864} {arXiv:1405.3864 [hep-ph]}
  \BibitemShut {NoStop}%
\bibitem [{\citenamefont {Bierlich}\ \emph {et~al.}(2018)\citenamefont
  {Bierlich}, \citenamefont {Gustafson}, \citenamefont {L\"onnblad},\ and\
  \citenamefont {Shah}}]{Bierlich:2018xfw}%
  \BibitemOpen
  \bibfield  {author} {\bibinfo {author} {\bibfnamefont {C.}~\bibnamefont
  {Bierlich}}, \bibinfo {author} {\bibfnamefont {G.}~\bibnamefont {Gustafson}},
  \bibinfo {author} {\bibfnamefont {L.}~\bibnamefont {L\"onnblad}},\ and\
  \bibinfo {author} {\bibfnamefont {H.}~\bibnamefont {Shah}},\ }\bibfield
  {title} {\bibinfo {title} {{The Angantyr model for Heavy-Ion Collisions in
  PYTHIA8}},\ }\href {https://doi.org/10.1007/JHEP10(2018)134} {\bibfield
  {journal} {\bibinfo  {journal} {JHEP}\ }\textbf {\bibinfo {volume} {10}},\
  \bibinfo {pages} {134}},\ \Eprint {https://arxiv.org/abs/1806.10820}
  {arXiv:1806.10820 [hep-ph]} \BibitemShut {NoStop}%
\bibitem [{\citenamefont {Ke}\ and\ \citenamefont {Wang}(2021)}]{Ke:2020clc}%
  \BibitemOpen
  \bibfield  {author} {\bibinfo {author} {\bibfnamefont {W.}~\bibnamefont
  {Ke}}\ and\ \bibinfo {author} {\bibfnamefont {X.-N.}\ \bibnamefont {Wang}},\
  }\bibfield  {title} {\bibinfo {title} {{QGP modification to single inclusive
  jets in a calibrated transport model}},\ }\href
  {https://doi.org/10.1007/JHEP05(2021)041} {\bibfield  {journal} {\bibinfo
  {journal} {JHEP}\ }\textbf {\bibinfo {volume} {05}},\ \bibinfo {pages}
  {041}},\ \Eprint {https://arxiv.org/abs/2010.13680} {arXiv:2010.13680
  [hep-ph]} \BibitemShut {NoStop}%
\bibitem [{\citenamefont {Tachibana}\ \emph {et~al.}(2023)\citenamefont
  {Tachibana} \emph {et~al.}}]{JETSCAPE:2023hqn}%
  \BibitemOpen
  \bibfield  {author} {\bibinfo {author} {\bibfnamefont {Y.}~\bibnamefont
  {Tachibana}} \emph {et~al.} (\bibinfo {collaboration} {JETSCAPE}),\
  }\bibfield  {title} {\bibinfo {title} {{Hard Jet Substructure in a
  Multi-stage Approach}},\ }\href@noop {} {\  (\bibinfo {year} {2023})},\
  \Eprint {https://arxiv.org/abs/2301.02485} {arXiv:2301.02485 [hep-ph]}
  \BibitemShut {NoStop}%
\bibitem [{\citenamefont {Armesto}\ \emph
  {et~al.}(2009{\natexlab{b}})\citenamefont {Armesto}, \citenamefont
  {Corcella}, \citenamefont {Cunqueiro},\ and\ \citenamefont
  {Salgado}}]{Armesto:2009ab}%
  \BibitemOpen
  \bibfield  {author} {\bibinfo {author} {\bibfnamefont {N.}~\bibnamefont
  {Armesto}}, \bibinfo {author} {\bibfnamefont {G.}~\bibnamefont {Corcella}},
  \bibinfo {author} {\bibfnamefont {L.}~\bibnamefont {Cunqueiro}},\ and\
  \bibinfo {author} {\bibfnamefont {C.~A.}\ \bibnamefont {Salgado}},\
  }\bibfield  {title} {\bibinfo {title} {{Angular-ordered parton showers with
  medium-modified splitting functions}},\ }\href
  {https://doi.org/10.1088/1126-6708/2009/11/122} {\bibfield  {journal}
  {\bibinfo  {journal} {JHEP}\ }\textbf {\bibinfo {volume} {11}},\ \bibinfo
  {pages} {122}},\ \Eprint {https://arxiv.org/abs/0909.5118} {arXiv:0909.5118
  [hep-ph]} \BibitemShut {NoStop}%
\bibitem [{\citenamefont {Zapp}\ \emph {et~al.}(2013)\citenamefont {Zapp},
  \citenamefont {Krauss},\ and\ \citenamefont {Wiedemann}}]{Zapp:2012ak}%
  \BibitemOpen
  \bibfield  {author} {\bibinfo {author} {\bibfnamefont {K.~C.}\ \bibnamefont
  {Zapp}}, \bibinfo {author} {\bibfnamefont {F.}~\bibnamefont {Krauss}},\ and\
  \bibinfo {author} {\bibfnamefont {U.~A.}\ \bibnamefont {Wiedemann}},\
  }\bibfield  {title} {\bibinfo {title} {{A perturbative framework for jet
  quenching}},\ }\href {https://doi.org/10.1007/JHEP03(2013)080} {\bibfield
  {journal} {\bibinfo  {journal} {JHEP}\ }\textbf {\bibinfo {volume} {03}},\
  \bibinfo {pages} {080}},\ \Eprint {https://arxiv.org/abs/1212.1599}
  {arXiv:1212.1599 [hep-ph]} \BibitemShut {NoStop}%
\bibitem [{\citenamefont {Dreyer}\ \emph {et~al.}(2018)\citenamefont {Dreyer},
  \citenamefont {Salam},\ and\ \citenamefont {Soyez}}]{Dreyer:2018nbf}%
  \BibitemOpen
  \bibfield  {author} {\bibinfo {author} {\bibfnamefont {F.~A.}\ \bibnamefont
  {Dreyer}}, \bibinfo {author} {\bibfnamefont {G.~P.}\ \bibnamefont {Salam}},\
  and\ \bibinfo {author} {\bibfnamefont {G.}~\bibnamefont {Soyez}},\ }\bibfield
   {title} {\bibinfo {title} {{The Lund Jet Plane}},\ }\href
  {https://doi.org/10.1007/JHEP12(2018)064} {\bibfield  {journal} {\bibinfo
  {journal} {JHEP}\ }\textbf {\bibinfo {volume} {12}},\ \bibinfo {pages}
  {064}},\ \Eprint {https://arxiv.org/abs/1807.04758} {arXiv:1807.04758
  [hep-ph]} \BibitemShut {NoStop}%
\bibitem [{\citenamefont {Dokshitzer}\ \emph {et~al.}(1991)\citenamefont
  {Dokshitzer}, \citenamefont {Khoze}, \citenamefont {Mueller},\ and\
  \citenamefont {Troian}}]{Dokshitzer:1991wu}%
  \BibitemOpen
  \bibfield  {author} {\bibinfo {author} {\bibfnamefont {Y.~L.}\ \bibnamefont
  {Dokshitzer}}, \bibinfo {author} {\bibfnamefont {V.~A.}\ \bibnamefont
  {Khoze}}, \bibinfo {author} {\bibfnamefont {A.~H.}\ \bibnamefont {Mueller}},\
  and\ \bibinfo {author} {\bibfnamefont {S.~I.}\ \bibnamefont {Troian}},\
  }\href@noop {} {\emph {\bibinfo {title} {{Basics of perturbative QCD}}}}\
  (\bibinfo  {publisher} {Editions Frontieres},\ \bibinfo {year}
  {1991})\BibitemShut {NoStop}%
\bibitem [{\citenamefont {Altarelli}\ and\ \citenamefont
  {Parisi}(1977)}]{Altarelli:1977zs}%
  \BibitemOpen
  \bibfield  {author} {\bibinfo {author} {\bibfnamefont {G.}~\bibnamefont
  {Altarelli}}\ and\ \bibinfo {author} {\bibfnamefont {G.}~\bibnamefont
  {Parisi}},\ }\bibfield  {title} {\bibinfo {title} {{Asymptotic Freedom in
  Parton Language}},\ }\href {https://doi.org/10.1016/0550-3213(77)90384-4}
  {\bibfield  {journal} {\bibinfo  {journal} {Nucl. Phys. B}\ }\textbf
  {\bibinfo {volume} {126}},\ \bibinfo {pages} {298} (\bibinfo {year}
  {1977})}\BibitemShut {NoStop}%
\bibitem [{\citenamefont {Wang}\ \emph {et~al.}(1995)\citenamefont {Wang},
  \citenamefont {Gyulassy},\ and\ \citenamefont {Plumer}}]{Wang:1994fx}%
  \BibitemOpen
  \bibfield  {author} {\bibinfo {author} {\bibfnamefont {X.-N.}\ \bibnamefont
  {Wang}}, \bibinfo {author} {\bibfnamefont {M.}~\bibnamefont {Gyulassy}},\
  and\ \bibinfo {author} {\bibfnamefont {M.}~\bibnamefont {Plumer}},\
  }\bibfield  {title} {\bibinfo {title} {{The LPM effect in QCD and radiative
  energy loss in a quark gluon plasma}},\ }\href
  {https://doi.org/10.1103/PhysRevD.51.3436} {\bibfield  {journal} {\bibinfo
  {journal} {Phys. Rev. D}\ }\textbf {\bibinfo {volume} {51}},\ \bibinfo
  {pages} {3436} (\bibinfo {year} {1995})},\ \Eprint
  {https://arxiv.org/abs/hep-ph/9408344} {arXiv:hep-ph/9408344} \BibitemShut
  {NoStop}%
\bibitem [{\citenamefont {Andr\'es}\ \emph {et~al.}(2016)\citenamefont
  {Andr\'es}, \citenamefont {Armesto}, \citenamefont {Luzum}, \citenamefont
  {Salgado},\ and\ \citenamefont {Zurita}}]{Andres:2016iys}%
  \BibitemOpen
  \bibfield  {author} {\bibinfo {author} {\bibfnamefont {C.}~\bibnamefont
  {Andr\'es}}, \bibinfo {author} {\bibfnamefont {N.}~\bibnamefont {Armesto}},
  \bibinfo {author} {\bibfnamefont {M.}~\bibnamefont {Luzum}}, \bibinfo
  {author} {\bibfnamefont {C.~A.}\ \bibnamefont {Salgado}},\ and\ \bibinfo
  {author} {\bibfnamefont {P.}~\bibnamefont {Zurita}},\ }\bibfield  {title}
  {\bibinfo {title} {{Energy versus centrality dependence of the jet quenching
  parameter $\hat{q}$ at RHIC and LHC: a new puzzle?}},\ }\href
  {https://doi.org/10.1140/epjc/s10052-016-4320-5} {\bibfield  {journal}
  {\bibinfo  {journal} {Eur. Phys. J. C}\ }\textbf {\bibinfo {volume} {76}},\
  \bibinfo {pages} {475} (\bibinfo {year} {2016})},\ \Eprint
  {https://arxiv.org/abs/1606.04837} {arXiv:1606.04837 [hep-ph]} \BibitemShut
  {NoStop}%
\bibitem [{\citenamefont {Feal}\ \emph {et~al.}(2021)\citenamefont {Feal},
  \citenamefont {Salgado},\ and\ \citenamefont {Vazquez}}]{Feal:2019xfl}%
  \BibitemOpen
  \bibfield  {author} {\bibinfo {author} {\bibfnamefont {X.}~\bibnamefont
  {Feal}}, \bibinfo {author} {\bibfnamefont {C.~A.}\ \bibnamefont {Salgado}},\
  and\ \bibinfo {author} {\bibfnamefont {R.~A.}\ \bibnamefont {Vazquez}},\
  }\bibfield  {title} {\bibinfo {title} {{Jet quenching test of the QCD matter
  created at RHIC and the LHC needs opacity-resummed medium induced
  radiation}},\ }\href {https://doi.org/10.1016/j.physletb.2021.136251}
  {\bibfield  {journal} {\bibinfo  {journal} {Phys. Lett. B}\ }\textbf
  {\bibinfo {volume} {816}},\ \bibinfo {pages} {136251} (\bibinfo {year}
  {2021})},\ \Eprint {https://arxiv.org/abs/1911.01309} {arXiv:1911.01309
  [hep-ph]} \BibitemShut {NoStop}%
\bibitem [{\citenamefont {Cao}\ \emph {et~al.}(2021)\citenamefont {Cao} \emph
  {et~al.}}]{JETSCAPE:2021ehl}%
  \BibitemOpen
  \bibfield  {author} {\bibinfo {author} {\bibfnamefont {S.}~\bibnamefont
  {Cao}} \emph {et~al.} (\bibinfo {collaboration} {JETSCAPE}),\ }\bibfield
  {title} {\bibinfo {title} {{Determining the jet transport coefficient $\hat
  q$ from inclusive hadron suppression measurements using Bayesian parameter
  estimation}},\ }\href {https://doi.org/10.1103/PhysRevC.104.024905}
  {\bibfield  {journal} {\bibinfo  {journal} {Phys. Rev. C}\ }\textbf {\bibinfo
  {volume} {104}},\ \bibinfo {pages} {024905} (\bibinfo {year} {2021})},\
  \Eprint {https://arxiv.org/abs/2102.11337} {arXiv:2102.11337 [nucl-th]}
  \BibitemShut {NoStop}%
\bibitem [{\citenamefont {Wang}\ and\ \citenamefont
  {Guo}(2001)}]{Wang:2001ifa}%
  \BibitemOpen
  \bibfield  {author} {\bibinfo {author} {\bibfnamefont {X.-N.}\ \bibnamefont
  {Wang}}\ and\ \bibinfo {author} {\bibfnamefont {X.-f.}\ \bibnamefont {Guo}},\
  }\bibfield  {title} {\bibinfo {title} {{Multiple parton scattering in nuclei:
  Parton energy loss}},\ }\href {https://doi.org/10.1016/S0375-9474(01)01130-7}
  {\bibfield  {journal} {\bibinfo  {journal} {Nucl. Phys. A}\ }\textbf
  {\bibinfo {volume} {696}},\ \bibinfo {pages} {788} (\bibinfo {year}
  {2001})},\ \Eprint {https://arxiv.org/abs/hep-ph/0102230}
  {arXiv:hep-ph/0102230} \BibitemShut {NoStop}%
\bibitem [{\citenamefont {D'Eramo}\ \emph {et~al.}(2019)\citenamefont
  {D'Eramo}, \citenamefont {Rajagopal},\ and\ \citenamefont
  {Yin}}]{DEramo:2018eoy}%
  \BibitemOpen
  \bibfield  {author} {\bibinfo {author} {\bibfnamefont {F.}~\bibnamefont
  {D'Eramo}}, \bibinfo {author} {\bibfnamefont {K.}~\bibnamefont {Rajagopal}},\
  and\ \bibinfo {author} {\bibfnamefont {Y.}~\bibnamefont {Yin}},\ }\bibfield
  {title} {\bibinfo {title} {{Moli\`ere scattering in quark-gluon plasma:
  finding point-like scatterers in a liquid}},\ }\href
  {https://doi.org/10.1007/JHEP01(2019)172} {\bibfield  {journal} {\bibinfo
  {journal} {JHEP}\ }\textbf {\bibinfo {volume} {01}},\ \bibinfo {pages}
  {172}},\ \Eprint {https://arxiv.org/abs/1808.03250} {arXiv:1808.03250
  [hep-ph]} \BibitemShut {NoStop}%
\bibitem [{\citenamefont {Mehtar-Tani}\ \emph {et~al.}(2011)\citenamefont
  {Mehtar-Tani}, \citenamefont {Salgado},\ and\ \citenamefont
  {Tywoniuk}}]{Mehtar-Tani:2010ebp}%
  \BibitemOpen
  \bibfield  {author} {\bibinfo {author} {\bibfnamefont {Y.}~\bibnamefont
  {Mehtar-Tani}}, \bibinfo {author} {\bibfnamefont {C.~A.}\ \bibnamefont
  {Salgado}},\ and\ \bibinfo {author} {\bibfnamefont {K.}~\bibnamefont
  {Tywoniuk}},\ }\bibfield  {title} {\bibinfo {title} {{Anti-angular ordering
  of gluon radiation in QCD media}},\ }\href
  {https://doi.org/10.1103/PhysRevLett.106.122002} {\bibfield  {journal}
  {\bibinfo  {journal} {Phys. Rev. Lett.}\ }\textbf {\bibinfo {volume} {106}},\
  \bibinfo {pages} {122002} (\bibinfo {year} {2011})},\ \Eprint
  {https://arxiv.org/abs/1009.2965} {arXiv:1009.2965 [hep-ph]} \BibitemShut
  {NoStop}%
\bibitem [{\citenamefont {Casalderrey-Solana}\ and\ \citenamefont
  {Iancu}(2011)}]{Casalderrey-Solana:2011ule}%
  \BibitemOpen
  \bibfield  {author} {\bibinfo {author} {\bibfnamefont {J.}~\bibnamefont
  {Casalderrey-Solana}}\ and\ \bibinfo {author} {\bibfnamefont
  {E.}~\bibnamefont {Iancu}},\ }\bibfield  {title} {\bibinfo {title}
  {{Interference effects in medium-induced gluon radiation}},\ }\href
  {https://doi.org/10.1007/JHEP08(2011)015} {\bibfield  {journal} {\bibinfo
  {journal} {JHEP}\ }\textbf {\bibinfo {volume} {08}},\ \bibinfo {pages}
  {015}},\ \Eprint {https://arxiv.org/abs/1105.1760} {arXiv:1105.1760 [hep-ph]}
  \BibitemShut {NoStop}%
\bibitem [{\citenamefont {Rajagopal}\ \emph {et~al.}(2016)\citenamefont
  {Rajagopal}, \citenamefont {Sadofyev},\ and\ \citenamefont {van~der
  Schee}}]{Rajagopal:2016uip}%
  \BibitemOpen
  \bibfield  {author} {\bibinfo {author} {\bibfnamefont {K.}~\bibnamefont
  {Rajagopal}}, \bibinfo {author} {\bibfnamefont {A.~V.}\ \bibnamefont
  {Sadofyev}},\ and\ \bibinfo {author} {\bibfnamefont {W.}~\bibnamefont
  {van~der Schee}},\ }\bibfield  {title} {\bibinfo {title} {{Evolution of the
  jet opening angle distribution in holographic plasma}},\ }\href
  {https://doi.org/10.1103/PhysRevLett.116.211603} {\bibfield  {journal}
  {\bibinfo  {journal} {Phys. Rev. Lett.}\ }\textbf {\bibinfo {volume} {116}},\
  \bibinfo {pages} {211603} (\bibinfo {year} {2016})},\ \Eprint
  {https://arxiv.org/abs/1602.04187} {arXiv:1602.04187 [nucl-th]} \BibitemShut
  {NoStop}%
\bibitem [{\citenamefont {Brewer}\ \emph {et~al.}(2019)\citenamefont {Brewer},
  \citenamefont {Milhano},\ and\ \citenamefont {Thaler}}]{Brewer:2018dfs}%
  \BibitemOpen
  \bibfield  {author} {\bibinfo {author} {\bibfnamefont {J.}~\bibnamefont
  {Brewer}}, \bibinfo {author} {\bibfnamefont {J.~G.}\ \bibnamefont
  {Milhano}},\ and\ \bibinfo {author} {\bibfnamefont {J.}~\bibnamefont
  {Thaler}},\ }\bibfield  {title} {\bibinfo {title} {{Sorting out quenched
  jets}},\ }\href {https://doi.org/10.1103/PhysRevLett.122.222301} {\bibfield
  {journal} {\bibinfo  {journal} {Phys. Rev. Lett.}\ }\textbf {\bibinfo
  {volume} {122}},\ \bibinfo {pages} {222301} (\bibinfo {year} {2019})},\
  \Eprint {https://arxiv.org/abs/1812.05111} {arXiv:1812.05111 [hep-ph]}
  \BibitemShut {NoStop}%
\bibitem [{\citenamefont {Brewer}\ \emph {et~al.}(2021)\citenamefont {Brewer},
  \citenamefont {Thaler},\ and\ \citenamefont {Turner}}]{Brewer:2020och}%
  \BibitemOpen
  \bibfield  {author} {\bibinfo {author} {\bibfnamefont {J.}~\bibnamefont
  {Brewer}}, \bibinfo {author} {\bibfnamefont {J.}~\bibnamefont {Thaler}},\
  and\ \bibinfo {author} {\bibfnamefont {A.~P.}\ \bibnamefont {Turner}},\
  }\bibfield  {title} {\bibinfo {title} {{Data-driven quark and gluon jet
  modification in heavy-ion collisions}},\ }\href
  {https://doi.org/10.1103/PhysRevC.103.L021901} {\bibfield  {journal}
  {\bibinfo  {journal} {Phys. Rev. C}\ }\textbf {\bibinfo {volume} {103}},\
  \bibinfo {pages} {L021901} (\bibinfo {year} {2021})},\ \Eprint
  {https://arxiv.org/abs/2008.08596} {arXiv:2008.08596 [hep-ph]} \BibitemShut
  {NoStop}%
\bibitem [{\citenamefont {Brewer}\ \emph {et~al.}(2022)\citenamefont {Brewer},
  \citenamefont {Brodsky},\ and\ \citenamefont {Rajagopal}}]{Brewer:2021hmh}%
  \BibitemOpen
  \bibfield  {author} {\bibinfo {author} {\bibfnamefont {J.}~\bibnamefont
  {Brewer}}, \bibinfo {author} {\bibfnamefont {Q.}~\bibnamefont {Brodsky}},\
  and\ \bibinfo {author} {\bibfnamefont {K.}~\bibnamefont {Rajagopal}},\
  }\bibfield  {title} {\bibinfo {title} {{Disentangling jet modification in jet
  simulations and in Z+jet data}},\ }\href
  {https://doi.org/10.1007/JHEP02(2022)175} {\bibfield  {journal} {\bibinfo
  {journal} {JHEP}\ }\textbf {\bibinfo {volume} {02}},\ \bibinfo {pages}
  {175}},\ \Eprint {https://arxiv.org/abs/2110.13159} {arXiv:2110.13159
  [hep-ph]} \BibitemShut {NoStop}%
\bibitem [{\citenamefont {Takacs}\ and\ \citenamefont
  {Tywoniuk}(2021)}]{Takacs:2021bpv}%
  \BibitemOpen
  \bibfield  {author} {\bibinfo {author} {\bibfnamefont {A.}~\bibnamefont
  {Takacs}}\ and\ \bibinfo {author} {\bibfnamefont {K.}~\bibnamefont
  {Tywoniuk}},\ }\bibfield  {title} {\bibinfo {title} {{Quenching effects in
  the cumulative jet spectrum}},\ }\href
  {https://doi.org/10.1007/JHEP10(2021)038} {\bibfield  {journal} {\bibinfo
  {journal} {JHEP}\ }\textbf {\bibinfo {volume} {10}},\ \bibinfo {pages}
  {038}},\ \Eprint {https://arxiv.org/abs/2103.14676} {arXiv:2103.14676
  [hep-ph]} \BibitemShut {NoStop}%
\bibitem [{\citenamefont {Pablos}\ and\ \citenamefont
  {Soto-Ontoso}(2023)}]{Pablos:2022mrx}%
  \BibitemOpen
  \bibfield  {author} {\bibinfo {author} {\bibfnamefont {D.}~\bibnamefont
  {Pablos}}\ and\ \bibinfo {author} {\bibfnamefont {A.}~\bibnamefont
  {Soto-Ontoso}},\ }\bibfield  {title} {\bibinfo {title} {{Pushing forward jet
  substructure measurements in heavy-ion collisions}},\ }\href
  {https://doi.org/10.1103/PhysRevD.107.094003} {\bibfield  {journal} {\bibinfo
   {journal} {Phys. Rev. D}\ }\textbf {\bibinfo {volume} {107}},\ \bibinfo
  {pages} {094003} (\bibinfo {year} {2023})},\ \Eprint
  {https://arxiv.org/abs/2210.07901} {arXiv:2210.07901 [hep-ph]} \BibitemShut
  {NoStop}%
\bibitem [{\citenamefont {Spousta}\ and\ \citenamefont
  {Cole}(2016)}]{Spousta:2015fca}%
  \BibitemOpen
  \bibfield  {author} {\bibinfo {author} {\bibfnamefont {M.}~\bibnamefont
  {Spousta}}\ and\ \bibinfo {author} {\bibfnamefont {B.}~\bibnamefont {Cole}},\
  }\bibfield  {title} {\bibinfo {title} {{Interpreting single jet measurements
  in Pb $+$ Pb collisions at the LHC}},\ }\href
  {https://doi.org/10.1140/epjc/s10052-016-3896-0} {\bibfield  {journal}
  {\bibinfo  {journal} {Eur. Phys. J. C}\ }\textbf {\bibinfo {volume} {76}},\
  \bibinfo {pages} {50} (\bibinfo {year} {2016})},\ \Eprint
  {https://arxiv.org/abs/1504.05169} {arXiv:1504.05169 [hep-ph]} \BibitemShut
  {NoStop}%
\bibitem [{\citenamefont {Qiu}\ \emph {et~al.}(2019)\citenamefont {Qiu},
  \citenamefont {Ringer}, \citenamefont {Sato},\ and\ \citenamefont
  {Zurita}}]{Qiu:2019sfj}%
  \BibitemOpen
  \bibfield  {author} {\bibinfo {author} {\bibfnamefont {J.-W.}\ \bibnamefont
  {Qiu}}, \bibinfo {author} {\bibfnamefont {F.}~\bibnamefont {Ringer}},
  \bibinfo {author} {\bibfnamefont {N.}~\bibnamefont {Sato}},\ and\ \bibinfo
  {author} {\bibfnamefont {P.}~\bibnamefont {Zurita}},\ }\bibfield  {title}
  {\bibinfo {title} {{Factorization of jet cross sections in heavy-ion
  collisions}},\ }\href {https://doi.org/10.1103/PhysRevLett.122.252301}
  {\bibfield  {journal} {\bibinfo  {journal} {Phys. Rev. Lett.}\ }\textbf
  {\bibinfo {volume} {122}},\ \bibinfo {pages} {252301} (\bibinfo {year}
  {2019})},\ \Eprint {https://arxiv.org/abs/1903.01993} {arXiv:1903.01993
  [hep-ph]} \BibitemShut {NoStop}%
\bibitem [{\citenamefont {Ringer}\ \emph {et~al.}(2020)\citenamefont {Ringer},
  \citenamefont {Xiao},\ and\ \citenamefont {Yuan}}]{Ringer:2019rfk}%
  \BibitemOpen
  \bibfield  {author} {\bibinfo {author} {\bibfnamefont {F.}~\bibnamefont
  {Ringer}}, \bibinfo {author} {\bibfnamefont {B.-W.}\ \bibnamefont {Xiao}},\
  and\ \bibinfo {author} {\bibfnamefont {F.}~\bibnamefont {Yuan}},\ }\bibfield
  {title} {\bibinfo {title} {{Can we observe jet $P_T$-broadening in heavy-ion
  collisions at the LHC?}},\ }\href
  {https://doi.org/10.1016/j.physletb.2020.135634} {\bibfield  {journal}
  {\bibinfo  {journal} {Phys. Lett. B}\ }\textbf {\bibinfo {volume} {808}},\
  \bibinfo {pages} {135634} (\bibinfo {year} {2020})},\ \Eprint
  {https://arxiv.org/abs/1907.12541} {arXiv:1907.12541 [hep-ph]} \BibitemShut
  {NoStop}%
\bibitem [{\citenamefont {Caucal}\ \emph {et~al.}(2022)\citenamefont {Caucal},
  \citenamefont {Soto-Ontoso},\ and\ \citenamefont {Takacs}}]{Caucal:2021cfb}%
  \BibitemOpen
  \bibfield  {author} {\bibinfo {author} {\bibfnamefont {P.}~\bibnamefont
  {Caucal}}, \bibinfo {author} {\bibfnamefont {A.}~\bibnamefont
  {Soto-Ontoso}},\ and\ \bibinfo {author} {\bibfnamefont {A.}~\bibnamefont
  {Takacs}},\ }\bibfield  {title} {\bibinfo {title} {{Dynamically groomed jet
  radius in heavy-ion collisions}},\ }\href
  {https://doi.org/10.1103/PhysRevD.105.114046} {\bibfield  {journal} {\bibinfo
   {journal} {Phys. Rev. D}\ }\textbf {\bibinfo {volume} {105}},\ \bibinfo
  {pages} {114046} (\bibinfo {year} {2022})},\ \Eprint
  {https://arxiv.org/abs/2111.14768} {arXiv:2111.14768 [hep-ph]} \BibitemShut
  {NoStop}%
\bibitem [{\citenamefont {Aad}\ \emph {et~al.}(2023)\citenamefont {Aad} \emph
  {et~al.}}]{ATLAS:2022vii}%
  \BibitemOpen
  \bibfield  {author} {\bibinfo {author} {\bibfnamefont {G.}~\bibnamefont
  {Aad}} \emph {et~al.} (\bibinfo {collaboration} {ATLAS}),\ }\bibfield
  {title} {\bibinfo {title} {{Measurement of substructure-dependent jet
  suppression in Pb+Pb collisions at 5.02 TeV with the ATLAS detector}},\
  }\href {https://doi.org/10.1103/PhysRevC.107.054909} {\bibfield  {journal}
  {\bibinfo  {journal} {Phys. Rev. C}\ }\textbf {\bibinfo {volume} {107}},\
  \bibinfo {pages} {054909} (\bibinfo {year} {2023})},\ \Eprint
  {https://arxiv.org/abs/2211.11470} {arXiv:2211.11470 [nucl-ex]} \BibitemShut
  {NoStop}%
\bibitem [{\citenamefont {Liou}\ \emph {et~al.}(2013)\citenamefont {Liou},
  \citenamefont {Mueller},\ and\ \citenamefont {Wu}}]{Liou:2013qya}%
  \BibitemOpen
  \bibfield  {author} {\bibinfo {author} {\bibfnamefont {T.}~\bibnamefont
  {Liou}}, \bibinfo {author} {\bibfnamefont {A.~H.}\ \bibnamefont {Mueller}},\
  and\ \bibinfo {author} {\bibfnamefont {B.}~\bibnamefont {Wu}},\ }\bibfield
  {title} {\bibinfo {title} {{Radiative $p_\bot$-broadening of high-energy
  quarks and gluons in QCD matter}},\ }\href
  {https://doi.org/10.1016/j.nuclphysa.2013.08.005} {\bibfield  {journal}
  {\bibinfo  {journal} {Nucl. Phys. A}\ }\textbf {\bibinfo {volume} {916}},\
  \bibinfo {pages} {102} (\bibinfo {year} {2013})},\ \Eprint
  {https://arxiv.org/abs/1304.7677} {arXiv:1304.7677 [hep-ph]} \BibitemShut
  {NoStop}%
\bibitem [{\citenamefont {Blaizot}\ and\ \citenamefont
  {Mehtar-Tani}(2014)}]{Blaizot:2014bha}%
  \BibitemOpen
  \bibfield  {author} {\bibinfo {author} {\bibfnamefont {J.-P.}\ \bibnamefont
  {Blaizot}}\ and\ \bibinfo {author} {\bibfnamefont {Y.}~\bibnamefont
  {Mehtar-Tani}},\ }\bibfield  {title} {\bibinfo {title} {{Renormalization of
  the jet-quenching parameter}},\ }\href
  {https://doi.org/10.1016/j.nuclphysa.2014.05.018} {\bibfield  {journal}
  {\bibinfo  {journal} {Nucl. Phys. A}\ }\textbf {\bibinfo {volume} {929}},\
  \bibinfo {pages} {202} (\bibinfo {year} {2014})},\ \Eprint
  {https://arxiv.org/abs/1403.2323} {arXiv:1403.2323 [hep-ph]} \BibitemShut
  {NoStop}%
\bibitem [{\citenamefont {Cacciari}\ \emph {et~al.}(2008)\citenamefont
  {Cacciari}, \citenamefont {Salam},\ and\ \citenamefont
  {Soyez}}]{Cacciari:2008gp}%
  \BibitemOpen
  \bibfield  {author} {\bibinfo {author} {\bibfnamefont {M.}~\bibnamefont
  {Cacciari}}, \bibinfo {author} {\bibfnamefont {G.~P.}\ \bibnamefont
  {Salam}},\ and\ \bibinfo {author} {\bibfnamefont {G.}~\bibnamefont {Soyez}},\
  }\bibfield  {title} {\bibinfo {title} {{The anti-$k_t$ jet clustering
  algorithm}},\ }\href {https://doi.org/10.1088/1126-6708/2008/04/063}
  {\bibfield  {journal} {\bibinfo  {journal} {JHEP}\ }\textbf {\bibinfo
  {volume} {04}},\ \bibinfo {pages} {063}},\ \Eprint
  {https://arxiv.org/abs/0802.1189} {arXiv:0802.1189 [hep-ph]} \BibitemShut
  {NoStop}%
\bibitem [{\citenamefont {Cacciari}\ \emph {et~al.}(2012)\citenamefont
  {Cacciari}, \citenamefont {Salam},\ and\ \citenamefont
  {Soyez}}]{Cacciari:2011ma}%
  \BibitemOpen
  \bibfield  {author} {\bibinfo {author} {\bibfnamefont {M.}~\bibnamefont
  {Cacciari}}, \bibinfo {author} {\bibfnamefont {G.~P.}\ \bibnamefont
  {Salam}},\ and\ \bibinfo {author} {\bibfnamefont {G.}~\bibnamefont {Soyez}},\
  }\bibfield  {title} {\bibinfo {title} {{FastJet User Manual}},\ }\href
  {https://doi.org/10.1140/epjc/s10052-012-1896-2} {\bibfield  {journal}
  {\bibinfo  {journal} {Eur. Phys. J. C}\ }\textbf {\bibinfo {volume} {72}},\
  \bibinfo {pages} {1896} (\bibinfo {year} {2012})},\ \Eprint
  {https://arxiv.org/abs/1111.6097} {arXiv:1111.6097 [hep-ph]} \BibitemShut
  {NoStop}%
\bibitem [{\citenamefont {Dokshitzer}\ \emph {et~al.}(1997)\citenamefont
  {Dokshitzer}, \citenamefont {Leder}, \citenamefont {Moretti},\ and\
  \citenamefont {Webber}}]{Dokshitzer:1997in}%
  \BibitemOpen
  \bibfield  {author} {\bibinfo {author} {\bibfnamefont {Y.~L.}\ \bibnamefont
  {Dokshitzer}}, \bibinfo {author} {\bibfnamefont {G.~D.}\ \bibnamefont
  {Leder}}, \bibinfo {author} {\bibfnamefont {S.}~\bibnamefont {Moretti}},\
  and\ \bibinfo {author} {\bibfnamefont {B.~R.}\ \bibnamefont {Webber}},\
  }\bibfield  {title} {\bibinfo {title} {{Better jet clustering algorithms}},\
  }\href {https://doi.org/10.1088/1126-6708/1997/08/001} {\bibfield  {journal}
  {\bibinfo  {journal} {JHEP}\ }\textbf {\bibinfo {volume} {08}},\ \bibinfo
  {pages} {001}},\ \Eprint {https://arxiv.org/abs/hep-ph/9707323}
  {arXiv:hep-ph/9707323} \BibitemShut {NoStop}%
\bibitem [{\citenamefont {Wobisch}\ and\ \citenamefont
  {Wengler}(1998)}]{Wobisch:1998wt}%
  \BibitemOpen
  \bibfield  {author} {\bibinfo {author} {\bibfnamefont {M.}~\bibnamefont
  {Wobisch}}\ and\ \bibinfo {author} {\bibfnamefont {T.}~\bibnamefont
  {Wengler}},\ }\bibfield  {title} {\bibinfo {title} {{Hadronization
  corrections to jet cross-sections in deep inelastic scattering}},\ }in\
  \href@noop {} {\emph {\bibinfo {booktitle} {{Workshop on Monte Carlo
  Generators for HERA Physics}}}}\ (\bibinfo {year} {1998})\ pp.\ \bibinfo
  {pages} {270--279},\ \Eprint {https://arxiv.org/abs/hep-ph/9907280}
  {arXiv:hep-ph/9907280} \BibitemShut {NoStop}%
\bibitem [{\citenamefont {Casalderrey-Solana}\ \emph
  {et~al.}(2013)\citenamefont {Casalderrey-Solana}, \citenamefont
  {Mehtar-Tani}, \citenamefont {Salgado},\ and\ \citenamefont
  {Tywoniuk}}]{Casalderrey-Solana:2012evi}%
  \BibitemOpen
  \bibfield  {author} {\bibinfo {author} {\bibfnamefont {J.}~\bibnamefont
  {Casalderrey-Solana}}, \bibinfo {author} {\bibfnamefont {Y.}~\bibnamefont
  {Mehtar-Tani}}, \bibinfo {author} {\bibfnamefont {C.~A.}\ \bibnamefont
  {Salgado}},\ and\ \bibinfo {author} {\bibfnamefont {K.}~\bibnamefont
  {Tywoniuk}},\ }\bibfield  {title} {\bibinfo {title} {{New picture of jet
  quenching dictated by color coherence}},\ }\href
  {https://doi.org/10.1016/j.physletb.2013.07.046} {\bibfield  {journal}
  {\bibinfo  {journal} {Phys. Lett. B}\ }\textbf {\bibinfo {volume} {725}},\
  \bibinfo {pages} {357} (\bibinfo {year} {2013})},\ \Eprint
  {https://arxiv.org/abs/1210.7765} {arXiv:1210.7765 [hep-ph]} \BibitemShut
  {NoStop}%
\bibitem [{\citenamefont {Mehtar-Tani}\ \emph {et~al.}(2020)\citenamefont
  {Mehtar-Tani}, \citenamefont {Soto-Ontoso},\ and\ \citenamefont
  {Tywoniuk}}]{Mehtar-Tani:2019rrk}%
  \BibitemOpen
  \bibfield  {author} {\bibinfo {author} {\bibfnamefont {Y.}~\bibnamefont
  {Mehtar-Tani}}, \bibinfo {author} {\bibfnamefont {A.}~\bibnamefont
  {Soto-Ontoso}},\ and\ \bibinfo {author} {\bibfnamefont {K.}~\bibnamefont
  {Tywoniuk}},\ }\bibfield  {title} {\bibinfo {title} {{Dynamical grooming of
  QCD jets}},\ }\href {https://doi.org/10.1103/PhysRevD.101.034004} {\bibfield
  {journal} {\bibinfo  {journal} {Phys. Rev. D}\ }\textbf {\bibinfo {volume}
  {101}},\ \bibinfo {pages} {034004} (\bibinfo {year} {2020})},\ \Eprint
  {https://arxiv.org/abs/1911.00375} {arXiv:1911.00375 [hep-ph]} \BibitemShut
  {NoStop}%
\bibitem [{\citenamefont {Wang}\ \emph {et~al.}(2023)\citenamefont {Wang},
  \citenamefont {Kang}, \citenamefont {Zhang}, \citenamefont {Shen},
  \citenamefont {Dai}, \citenamefont {Zhang},\ and\ \citenamefont
  {Wang}}]{Wang:2022yrp}%
  \BibitemOpen
  \bibfield  {author} {\bibinfo {author} {\bibfnamefont {L.}~\bibnamefont
  {Wang}}, \bibinfo {author} {\bibfnamefont {J.-W.}\ \bibnamefont {Kang}},
  \bibinfo {author} {\bibfnamefont {Q.}~\bibnamefont {Zhang}}, \bibinfo
  {author} {\bibfnamefont {S.}~\bibnamefont {Shen}}, \bibinfo {author}
  {\bibfnamefont {W.}~\bibnamefont {Dai}}, \bibinfo {author} {\bibfnamefont
  {B.-W.}\ \bibnamefont {Zhang}},\ and\ \bibinfo {author} {\bibfnamefont
  {E.}~\bibnamefont {Wang}},\ }\bibfield  {title} {\bibinfo {title} {{Jet
  Radius and Momentum Splitting Fraction with Dynamical Grooming in Heavy-Ion
  Collisions}},\ }\href {https://doi.org/10.1088/0256-307X/40/3/032101}
  {\bibfield  {journal} {\bibinfo  {journal} {Chin. Phys. Lett.}\ }\textbf
  {\bibinfo {volume} {40}},\ \bibinfo {pages} {032101} (\bibinfo {year}
  {2023})},\ \Eprint {https://arxiv.org/abs/2211.13674} {arXiv:2211.13674
  [nucl-th]} \BibitemShut {NoStop}%
\bibitem [{\citenamefont {Acharya}\ \emph {et~al.}(2023)\citenamefont {Acharya}
  \emph {et~al.}}]{ALICE:2022hyz}%
  \BibitemOpen
  \bibfield  {author} {\bibinfo {author} {\bibfnamefont {S.}~\bibnamefont
  {Acharya}} \emph {et~al.} (\bibinfo {collaboration} {ALICE}),\ }\bibfield
  {title} {\bibinfo {title} {{Measurements of the groomed jet radius and
  momentum splitting fraction with the soft drop and dynamical grooming
  algorithms in pp collisions at $ \sqrt{s} $ = 5.02 TeV}},\ }\href
  {https://doi.org/10.1007/JHEP05(2023)244} {\bibfield  {journal} {\bibinfo
  {journal} {JHEP}\ }\textbf {\bibinfo {volume} {05}},\ \bibinfo {pages}
  {244}},\ \Eprint {https://arxiv.org/abs/2204.10246} {arXiv:2204.10246
  [nucl-ex]} \BibitemShut {NoStop}%
\bibitem [{\citenamefont {Apolin\'ario}\ \emph {et~al.}(2021)\citenamefont
  {Apolin\'ario}, \citenamefont {Cordeiro},\ and\ \citenamefont
  {Zapp}}]{Apolinario:2020uvt}%
  \BibitemOpen
  \bibfield  {author} {\bibinfo {author} {\bibfnamefont {L.}~\bibnamefont
  {Apolin\'ario}}, \bibinfo {author} {\bibfnamefont {A.}~\bibnamefont
  {Cordeiro}},\ and\ \bibinfo {author} {\bibfnamefont {K.}~\bibnamefont
  {Zapp}},\ }\bibfield  {title} {\bibinfo {title} {{Time reclustering for jet
  quenching studies}},\ }\href
  {https://doi.org/10.1140/epjc/s10052-021-09346-8} {\bibfield  {journal}
  {\bibinfo  {journal} {Eur. Phys. J. C}\ }\textbf {\bibinfo {volume} {81}},\
  \bibinfo {pages} {561} (\bibinfo {year} {2021})},\ \Eprint
  {https://arxiv.org/abs/2012.02199} {arXiv:2012.02199 [hep-ph]} \BibitemShut
  {NoStop}%
\bibitem [{\citenamefont {Apolin\'ario}\ \emph
  {et~al.}(2022{\natexlab{b}})\citenamefont {Apolin\'ario}, \citenamefont
  {Kunnawalkam~Elayavalli},\ and\ \citenamefont {Olavo}}]{Apolinario:2022guz}%
  \BibitemOpen
  \bibfield  {author} {\bibinfo {author} {\bibfnamefont {L.}~\bibnamefont
  {Apolin\'ario}}, \bibinfo {author} {\bibfnamefont {R.}~\bibnamefont
  {Kunnawalkam~Elayavalli}},\ and\ \bibinfo {author} {\bibfnamefont
  {N.}~\bibnamefont {Olavo}},\ }\bibfield  {title} {\bibinfo {title}
  {{Transitioning from perturbative to non-perturbative splittings within QCD
  jets}},\ }\href@noop {} {\  (\bibinfo {year} {2022}{\natexlab{b}})},\ \Eprint
  {https://arxiv.org/abs/2212.11846} {arXiv:2212.11846 [hep-ph]} \BibitemShut
  {NoStop}%
\bibitem [{\citenamefont {Cunqueiro}\ \emph {et~al.}(2023)\citenamefont
  {Cunqueiro}, \citenamefont {Napoletano},\ and\ \citenamefont
  {Soto-Ontoso}}]{Cunqueiro:2022svx}%
  \BibitemOpen
  \bibfield  {author} {\bibinfo {author} {\bibfnamefont {L.}~\bibnamefont
  {Cunqueiro}}, \bibinfo {author} {\bibfnamefont {D.}~\bibnamefont
  {Napoletano}},\ and\ \bibinfo {author} {\bibfnamefont {A.}~\bibnamefont
  {Soto-Ontoso}},\ }\bibfield  {title} {\bibinfo {title} {{Dead-cone searches
  in heavy-ion collisions using the jet tree}},\ }\href
  {https://doi.org/10.1103/PhysRevD.107.094008} {\bibfield  {journal} {\bibinfo
   {journal} {Phys. Rev. D}\ }\textbf {\bibinfo {volume} {107}},\ \bibinfo
  {pages} {094008} (\bibinfo {year} {2023})},\ \Eprint
  {https://arxiv.org/abs/2211.11789} {arXiv:2211.11789 [hep-ph]} \BibitemShut
  {NoStop}%
\bibitem [{\citenamefont {Larkoski}\ \emph {et~al.}(2014)\citenamefont
  {Larkoski}, \citenamefont {Marzani}, \citenamefont {Soyez},\ and\
  \citenamefont {Thaler}}]{Larkoski:2014wba}%
  \BibitemOpen
  \bibfield  {author} {\bibinfo {author} {\bibfnamefont {A.~J.}\ \bibnamefont
  {Larkoski}}, \bibinfo {author} {\bibfnamefont {S.}~\bibnamefont {Marzani}},
  \bibinfo {author} {\bibfnamefont {G.}~\bibnamefont {Soyez}},\ and\ \bibinfo
  {author} {\bibfnamefont {J.}~\bibnamefont {Thaler}},\ }\bibfield  {title}
  {\bibinfo {title} {{Soft Drop}},\ }\href
  {https://doi.org/10.1007/JHEP05(2014)146} {\bibfield  {journal} {\bibinfo
  {journal} {JHEP}\ }\textbf {\bibinfo {volume} {05}},\ \bibinfo {pages}
  {146}},\ \Eprint {https://arxiv.org/abs/1402.2657} {arXiv:1402.2657 [hep-ph]}
  \BibitemShut {NoStop}%
\bibitem [{\citenamefont {Mehtar-Tani}\ and\ \citenamefont
  {Tywoniuk}(2017)}]{Mehtar-Tani:2016aco}%
  \BibitemOpen
  \bibfield  {author} {\bibinfo {author} {\bibfnamefont {Y.}~\bibnamefont
  {Mehtar-Tani}}\ and\ \bibinfo {author} {\bibfnamefont {K.}~\bibnamefont
  {Tywoniuk}},\ }\bibfield  {title} {\bibinfo {title} {{Groomed jets in
  heavy-ion collisions: sensitivity to medium-induced bremsstrahlung}},\ }\href
  {https://doi.org/10.1007/JHEP04(2017)125} {\bibfield  {journal} {\bibinfo
  {journal} {JHEP}\ }\textbf {\bibinfo {volume} {04}},\ \bibinfo {pages}
  {125}},\ \Eprint {https://arxiv.org/abs/1610.08930} {arXiv:1610.08930
  [hep-ph]} \BibitemShut {NoStop}%
\bibitem [{\citenamefont {Chien}\ and\ \citenamefont
  {Vitev}(2017)}]{Chien:2016led}%
  \BibitemOpen
  \bibfield  {author} {\bibinfo {author} {\bibfnamefont {Y.-T.}\ \bibnamefont
  {Chien}}\ and\ \bibinfo {author} {\bibfnamefont {I.}~\bibnamefont {Vitev}},\
  }\bibfield  {title} {\bibinfo {title} {{Probing the Hardest Branching within
  Jets in Heavy-Ion Collisions}},\ }\href
  {https://doi.org/10.1103/PhysRevLett.119.112301} {\bibfield  {journal}
  {\bibinfo  {journal} {Phys. Rev. Lett.}\ }\textbf {\bibinfo {volume} {119}},\
  \bibinfo {pages} {112301} (\bibinfo {year} {2017})},\ \Eprint
  {https://arxiv.org/abs/1608.07283} {arXiv:1608.07283 [hep-ph]} \BibitemShut
  {NoStop}%
\bibitem [{\citenamefont {Milhano}\ \emph {et~al.}(2018)\citenamefont
  {Milhano}, \citenamefont {Wiedemann},\ and\ \citenamefont
  {Zapp}}]{Milhano:2017nzm}%
  \BibitemOpen
  \bibfield  {author} {\bibinfo {author} {\bibfnamefont {G.}~\bibnamefont
  {Milhano}}, \bibinfo {author} {\bibfnamefont {U.~A.}\ \bibnamefont
  {Wiedemann}},\ and\ \bibinfo {author} {\bibfnamefont {K.~C.}\ \bibnamefont
  {Zapp}},\ }\bibfield  {title} {\bibinfo {title} {{Sensitivity of jet
  substructure to jet-induced medium response}},\ }\href
  {https://doi.org/10.1016/j.physletb.2018.01.029} {\bibfield  {journal}
  {\bibinfo  {journal} {Phys. Lett. B}\ }\textbf {\bibinfo {volume} {779}},\
  \bibinfo {pages} {409} (\bibinfo {year} {2018})},\ \Eprint
  {https://arxiv.org/abs/1707.04142} {arXiv:1707.04142 [hep-ph]} \BibitemShut
  {NoStop}%
\bibitem [{\citenamefont {Casalderrey-Solana}\ \emph
  {et~al.}(2020)\citenamefont {Casalderrey-Solana}, \citenamefont {Milhano},
  \citenamefont {Pablos},\ and\ \citenamefont
  {Rajagopal}}]{Casalderrey-Solana:2019ubu}%
  \BibitemOpen
  \bibfield  {author} {\bibinfo {author} {\bibfnamefont {J.}~\bibnamefont
  {Casalderrey-Solana}}, \bibinfo {author} {\bibfnamefont {G.}~\bibnamefont
  {Milhano}}, \bibinfo {author} {\bibfnamefont {D.}~\bibnamefont {Pablos}},\
  and\ \bibinfo {author} {\bibfnamefont {K.}~\bibnamefont {Rajagopal}},\
  }\bibfield  {title} {\bibinfo {title} {{Modification of Jet Substructure in
  Heavy Ion Collisions as a Probe of the Resolution Length of Quark-Gluon
  Plasma}},\ }\href {https://doi.org/10.1007/JHEP01(2020)044} {\bibfield
  {journal} {\bibinfo  {journal} {JHEP}\ }\textbf {\bibinfo {volume} {01}},\
  \bibinfo {pages} {044}},\ \Eprint {https://arxiv.org/abs/1907.11248}
  {arXiv:1907.11248 [hep-ph]} \BibitemShut {NoStop}%
\bibitem [{\citenamefont {Sirunyan}\ \emph
  {et~al.}(2018{\natexlab{a}})\citenamefont {Sirunyan} \emph
  {et~al.}}]{CMS:2017qlm}%
  \BibitemOpen
  \bibfield  {author} {\bibinfo {author} {\bibfnamefont {A.~M.}\ \bibnamefont
  {Sirunyan}} \emph {et~al.} (\bibinfo {collaboration} {CMS}),\ }\bibfield
  {title} {\bibinfo {title} {{Measurement of the Splitting Function in $pp$ and
  Pb-Pb Collisions at $\sqrt{s_{_{\mathrm{NN}}}} =$ 5.02 TeV}},\ }\href
  {https://doi.org/10.1103/PhysRevLett.120.142302} {\bibfield  {journal}
  {\bibinfo  {journal} {Phys. Rev. Lett.}\ }\textbf {\bibinfo {volume} {120}},\
  \bibinfo {pages} {142302} (\bibinfo {year} {2018}{\natexlab{a}})},\ \Eprint
  {https://arxiv.org/abs/1708.09429} {arXiv:1708.09429 [nucl-ex]} \BibitemShut
  {NoStop}%
\bibitem [{\citenamefont {Adam}\ \emph {et~al.}(2020)\citenamefont {Adam} \emph
  {et~al.}}]{STAR:2020ejj}%
  \BibitemOpen
  \bibfield  {author} {\bibinfo {author} {\bibfnamefont {J.}~\bibnamefont
  {Adam}} \emph {et~al.} (\bibinfo {collaboration} {STAR}),\ }\bibfield
  {title} {\bibinfo {title} {{Measurement of groomed jet substructure
  observables in p+p collisions at $\sqrt {s}$ =200 GeV with STAR}},\ }\href
  {https://doi.org/10.1016/j.physletb.2020.135846} {\bibfield  {journal}
  {\bibinfo  {journal} {Phys. Lett. B}\ }\textbf {\bibinfo {volume} {811}},\
  \bibinfo {pages} {135846} (\bibinfo {year} {2020})},\ \Eprint
  {https://arxiv.org/abs/2003.02114} {arXiv:2003.02114 [hep-ex]} \BibitemShut
  {NoStop}%
\bibitem [{\citenamefont {Sirunyan}\ \emph
  {et~al.}(2018{\natexlab{b}})\citenamefont {Sirunyan} \emph
  {et~al.}}]{CMS:2018fof}%
  \BibitemOpen
  \bibfield  {author} {\bibinfo {author} {\bibfnamefont {A.~M.}\ \bibnamefont
  {Sirunyan}} \emph {et~al.} (\bibinfo {collaboration} {CMS}),\ }\bibfield
  {title} {\bibinfo {title} {{Measurement of the groomed jet mass in PbPb and
  pp collisions at $ \sqrt{s_{\mathrm{NN}}}=5.02 $ TeV}},\ }\href
  {https://doi.org/10.1007/JHEP10(2018)161} {\bibfield  {journal} {\bibinfo
  {journal} {JHEP}\ }\textbf {\bibinfo {volume} {10}},\ \bibinfo {pages}
  {161}},\ \Eprint {https://arxiv.org/abs/1805.05145} {arXiv:1805.05145
  [hep-ex]} \BibitemShut {NoStop}%
\bibitem [{\citenamefont {Andrews}\ \emph {et~al.}(2020)\citenamefont {Andrews}
  \emph {et~al.}}]{Andrews:2018jcm}%
  \BibitemOpen
  \bibfield  {author} {\bibinfo {author} {\bibfnamefont {H.~A.}\ \bibnamefont
  {Andrews}} \emph {et~al.},\ }\bibfield  {title} {\bibinfo {title} {{Novel
  tools and observables for jet physics in heavy-ion collisions}},\ }\href
  {https://doi.org/10.1088/1361-6471/ab7cbc} {\bibfield  {journal} {\bibinfo
  {journal} {J. Phys. G}\ }\textbf {\bibinfo {volume} {47}},\ \bibinfo {pages}
  {065102} (\bibinfo {year} {2020})},\ \Eprint
  {https://arxiv.org/abs/1808.03689} {arXiv:1808.03689 [hep-ph]} \BibitemShut
  {NoStop}%
\bibitem [{\citenamefont {Andres}\ \emph
  {et~al.}(2023{\natexlab{a}})\citenamefont {Andres}, \citenamefont
  {Dominguez}, \citenamefont {Kunnawalkam~Elayavalli}, \citenamefont {Holguin},
  \citenamefont {Marquet},\ and\ \citenamefont {Moult}}]{Andres:2022ovj}%
  \BibitemOpen
  \bibfield  {author} {\bibinfo {author} {\bibfnamefont {C.}~\bibnamefont
  {Andres}}, \bibinfo {author} {\bibfnamefont {F.}~\bibnamefont {Dominguez}},
  \bibinfo {author} {\bibfnamefont {R.}~\bibnamefont {Kunnawalkam~Elayavalli}},
  \bibinfo {author} {\bibfnamefont {J.}~\bibnamefont {Holguin}}, \bibinfo
  {author} {\bibfnamefont {C.}~\bibnamefont {Marquet}},\ and\ \bibinfo {author}
  {\bibfnamefont {I.}~\bibnamefont {Moult}},\ }\bibfield  {title} {\bibinfo
  {title} {{Resolving the Scales of the Quark-Gluon Plasma with Energy
  Correlators}},\ }\href {https://doi.org/10.1103/PhysRevLett.130.262301}
  {\bibfield  {journal} {\bibinfo  {journal} {Phys. Rev. Lett.}\ }\textbf
  {\bibinfo {volume} {130}},\ \bibinfo {pages} {262301} (\bibinfo {year}
  {2023}{\natexlab{a}})},\ \Eprint {https://arxiv.org/abs/2209.11236}
  {arXiv:2209.11236 [hep-ph]} \BibitemShut {NoStop}%
\bibitem [{\citenamefont {Andres}\ \emph
  {et~al.}(2023{\natexlab{b}})\citenamefont {Andres}, \citenamefont
  {Dominguez}, \citenamefont {Holguin}, \citenamefont {Marquet},\ and\
  \citenamefont {Moult}}]{Andres:2023xwr}%
  \BibitemOpen
  \bibfield  {author} {\bibinfo {author} {\bibfnamefont {C.}~\bibnamefont
  {Andres}}, \bibinfo {author} {\bibfnamefont {F.}~\bibnamefont {Dominguez}},
  \bibinfo {author} {\bibfnamefont {J.}~\bibnamefont {Holguin}}, \bibinfo
  {author} {\bibfnamefont {C.}~\bibnamefont {Marquet}},\ and\ \bibinfo {author}
  {\bibfnamefont {I.}~\bibnamefont {Moult}},\ }\bibfield  {title} {\bibinfo
  {title} {{A coherent view of the quark-gluon plasma from energy
  correlators}},\ }\href {https://doi.org/10.1007/JHEP09(2023)088} {\bibfield
  {journal} {\bibinfo  {journal} {JHEP}\ }\textbf {\bibinfo {volume} {09}},\
  \bibinfo {pages} {088}},\ \Eprint {https://arxiv.org/abs/2303.03413}
  {arXiv:2303.03413 [hep-ph]} \BibitemShut {NoStop}%
\bibitem [{\citenamefont {Alioli}\ \emph {et~al.}(2011)\citenamefont {Alioli},
  \citenamefont {Hamilton}, \citenamefont {Nason}, \citenamefont {Oleari},\
  and\ \citenamefont {Re}}]{Alioli:2010xa}%
  \BibitemOpen
  \bibfield  {author} {\bibinfo {author} {\bibfnamefont {S.}~\bibnamefont
  {Alioli}}, \bibinfo {author} {\bibfnamefont {K.}~\bibnamefont {Hamilton}},
  \bibinfo {author} {\bibfnamefont {P.}~\bibnamefont {Nason}}, \bibinfo
  {author} {\bibfnamefont {C.}~\bibnamefont {Oleari}},\ and\ \bibinfo {author}
  {\bibfnamefont {E.}~\bibnamefont {Re}},\ }\bibfield  {title} {\bibinfo
  {title} {{Jet pair production in POWHEG}},\ }\href
  {https://doi.org/10.1007/JHEP04(2011)081} {\bibfield  {journal} {\bibinfo
  {journal} {JHEP}\ }\textbf {\bibinfo {volume} {04}},\ \bibinfo {pages}
  {081}},\ \Eprint {https://arxiv.org/abs/1012.3380} {arXiv:1012.3380 [hep-ph]}
  \BibitemShut {NoStop}%
\bibitem [{\citenamefont {Bierlich}\ \emph {et~al.}(2022)\citenamefont
  {Bierlich} \emph {et~al.}}]{Bierlich:2022pfr}%
  \BibitemOpen
  \bibfield  {author} {\bibinfo {author} {\bibfnamefont {C.}~\bibnamefont
  {Bierlich}} \emph {et~al.},\ }\bibfield  {title} {\bibinfo {title} {{A
  comprehensive guide to the physics and usage of PYTHIA 8.3}}\ }\href
  {https://doi.org/10.21468/SciPostPhysCodeb.8} {10.21468/SciPostPhysCodeb.8}
  (\bibinfo {year} {2022}),\ \Eprint {https://arxiv.org/abs/2203.11601}
  {arXiv:2203.11601 [hep-ph]} \BibitemShut {NoStop}%
\bibitem [{\citenamefont {Alwall}\ \emph {et~al.}(2014)\citenamefont {Alwall},
  \citenamefont {Frederix}, \citenamefont {Frixione}, \citenamefont {Hirschi},
  \citenamefont {Maltoni}, \citenamefont {Mattelaer}, \citenamefont {Shao},
  \citenamefont {Stelzer}, \citenamefont {Torrielli},\ and\ \citenamefont
  {Zaro}}]{Alwall:2014hca}%
  \BibitemOpen
  \bibfield  {author} {\bibinfo {author} {\bibfnamefont {J.}~\bibnamefont
  {Alwall}}, \bibinfo {author} {\bibfnamefont {R.}~\bibnamefont {Frederix}},
  \bibinfo {author} {\bibfnamefont {S.}~\bibnamefont {Frixione}}, \bibinfo
  {author} {\bibfnamefont {V.}~\bibnamefont {Hirschi}}, \bibinfo {author}
  {\bibfnamefont {F.}~\bibnamefont {Maltoni}}, \bibinfo {author} {\bibfnamefont
  {O.}~\bibnamefont {Mattelaer}}, \bibinfo {author} {\bibfnamefont {H.~S.}\
  \bibnamefont {Shao}}, \bibinfo {author} {\bibfnamefont {T.}~\bibnamefont
  {Stelzer}}, \bibinfo {author} {\bibfnamefont {P.}~\bibnamefont {Torrielli}},\
  and\ \bibinfo {author} {\bibfnamefont {M.}~\bibnamefont {Zaro}},\ }\bibfield
  {title} {\bibinfo {title} {{The automated computation of tree-level and
  next-to-leading order differential cross sections, and their matching to
  parton shower simulations}},\ }\href
  {https://doi.org/10.1007/JHEP07(2014)079} {\bibfield  {journal} {\bibinfo
  {journal} {JHEP}\ }\textbf {\bibinfo {volume} {07}},\ \bibinfo {pages}
  {079}},\ \Eprint {https://arxiv.org/abs/1405.0301} {arXiv:1405.0301 [hep-ph]}
  \BibitemShut {NoStop}%
\bibitem [{\citenamefont {Bellm}\ \emph {et~al.}(2016)\citenamefont {Bellm}
  \emph {et~al.}}]{Bellm:2015jjp}%
  \BibitemOpen
  \bibfield  {author} {\bibinfo {author} {\bibfnamefont {J.}~\bibnamefont
  {Bellm}} \emph {et~al.},\ }\bibfield  {title} {\bibinfo {title} {{Herwig
  7.0/Herwig++ 3.0 release note}},\ }\href
  {https://doi.org/10.1140/epjc/s10052-016-4018-8} {\bibfield  {journal}
  {\bibinfo  {journal} {Eur. Phys. J. C}\ }\textbf {\bibinfo {volume} {76}},\
  \bibinfo {pages} {196} (\bibinfo {year} {2016})},\ \Eprint
  {https://arxiv.org/abs/1512.01178} {arXiv:1512.01178 [hep-ph]} \BibitemShut
  {NoStop}%
\bibitem [{\citenamefont {Caucal}\ \emph {et~al.}(2019)\citenamefont {Caucal},
  \citenamefont {Iancu},\ and\ \citenamefont {Soyez}}]{Caucal:2019uvr}%
  \BibitemOpen
  \bibfield  {author} {\bibinfo {author} {\bibfnamefont {P.}~\bibnamefont
  {Caucal}}, \bibinfo {author} {\bibfnamefont {E.}~\bibnamefont {Iancu}},\ and\
  \bibinfo {author} {\bibfnamefont {G.}~\bibnamefont {Soyez}},\ }\bibfield
  {title} {\bibinfo {title} {{Deciphering the $z_g$ distribution in
  ultrarelativistic heavy ion collisions}},\ }\href
  {https://doi.org/10.1007/JHEP10(2019)273} {\bibfield  {journal} {\bibinfo
  {journal} {JHEP}\ }\textbf {\bibinfo {volume} {10}},\ \bibinfo {pages}
  {273}},\ \Eprint {https://arxiv.org/abs/1907.04866} {arXiv:1907.04866
  [hep-ph]} \BibitemShut {NoStop}%
\bibitem [{\citenamefont {Sjostrand}\ \emph {et~al.}(2006)\citenamefont
  {Sjostrand}, \citenamefont {Mrenna},\ and\ \citenamefont
  {Skands}}]{Sjostrand:2006za}%
  \BibitemOpen
  \bibfield  {author} {\bibinfo {author} {\bibfnamefont {T.}~\bibnamefont
  {Sjostrand}}, \bibinfo {author} {\bibfnamefont {S.}~\bibnamefont {Mrenna}},\
  and\ \bibinfo {author} {\bibfnamefont {P.~Z.}\ \bibnamefont {Skands}},\
  }\bibfield  {title} {\bibinfo {title} {{PYTHIA 6.4 Physics and Manual}},\
  }\href {https://doi.org/10.1088/1126-6708/2006/05/026} {\bibfield  {journal}
  {\bibinfo  {journal} {JHEP}\ }\textbf {\bibinfo {volume} {05}},\ \bibinfo
  {pages} {026}},\ \Eprint {https://arxiv.org/abs/hep-ph/0603175}
  {arXiv:hep-ph/0603175} \BibitemShut {NoStop}%
\bibitem [{\citenamefont {Caucal}\ \emph
  {et~al.}(2021{\natexlab{a}})\citenamefont {Caucal}, \citenamefont {Iancu},\
  and\ \citenamefont {Soyez}}]{Caucal:2020uic}%
  \BibitemOpen
  \bibfield  {author} {\bibinfo {author} {\bibfnamefont {P.}~\bibnamefont
  {Caucal}}, \bibinfo {author} {\bibfnamefont {E.}~\bibnamefont {Iancu}},\ and\
  \bibinfo {author} {\bibfnamefont {G.}~\bibnamefont {Soyez}},\ }\bibfield
  {title} {\bibinfo {title} {{Jet radiation in a longitudinally expanding
  medium}},\ }\href {https://doi.org/10.1007/JHEP04(2021)209} {\bibfield
  {journal} {\bibinfo  {journal} {JHEP}\ }\textbf {\bibinfo {volume} {04}},\
  \bibinfo {pages} {209}},\ \Eprint {https://arxiv.org/abs/2012.01457}
  {arXiv:2012.01457 [hep-ph]} \BibitemShut {NoStop}%
\bibitem [{\citenamefont {Casalderrey-Solana}\ \emph
  {et~al.}(2016)\citenamefont {Casalderrey-Solana}, \citenamefont {Gulhan},
  \citenamefont {Milhano}, \citenamefont {Pablos},\ and\ \citenamefont
  {Rajagopal}}]{Casalderrey-Solana:2015vaa}%
  \BibitemOpen
  \bibfield  {author} {\bibinfo {author} {\bibfnamefont {J.}~\bibnamefont
  {Casalderrey-Solana}}, \bibinfo {author} {\bibfnamefont {D.~C.}\ \bibnamefont
  {Gulhan}}, \bibinfo {author} {\bibfnamefont {J.~G.}\ \bibnamefont {Milhano}},
  \bibinfo {author} {\bibfnamefont {D.}~\bibnamefont {Pablos}},\ and\ \bibinfo
  {author} {\bibfnamefont {K.}~\bibnamefont {Rajagopal}},\ }\bibfield  {title}
  {\bibinfo {title} {{Predictions for Boson-Jet Observables and Fragmentation
  Function Ratios from a Hybrid Strong/Weak Coupling Model for Jet
  Quenching}},\ }\href {https://doi.org/10.1007/JHEP03(2016)053} {\bibfield
  {journal} {\bibinfo  {journal} {JHEP}\ }\textbf {\bibinfo {volume} {03}},\
  \bibinfo {pages} {053}},\ \Eprint {https://arxiv.org/abs/1508.00815}
  {arXiv:1508.00815 [hep-ph]} \BibitemShut {NoStop}%
\bibitem [{\citenamefont {Chesler}\ and\ \citenamefont
  {Rajagopal}(2016)}]{Chesler:2015nqz}%
  \BibitemOpen
  \bibfield  {author} {\bibinfo {author} {\bibfnamefont {P.~M.}\ \bibnamefont
  {Chesler}}\ and\ \bibinfo {author} {\bibfnamefont {K.}~\bibnamefont
  {Rajagopal}},\ }\bibfield  {title} {\bibinfo {title} {{On the Evolution of
  Jet Energy and Opening Angle in Strongly Coupled Plasma}},\ }\href
  {https://doi.org/10.1007/JHEP05(2016)098} {\bibfield  {journal} {\bibinfo
  {journal} {JHEP}\ }\textbf {\bibinfo {volume} {05}},\ \bibinfo {pages}
  {098}},\ \Eprint {https://arxiv.org/abs/1511.07567} {arXiv:1511.07567
  [hep-th]} \BibitemShut {NoStop}%
\bibitem [{\citenamefont {Shen}\ \emph {et~al.}(2016)\citenamefont {Shen},
  \citenamefont {Qiu}, \citenamefont {Song}, \citenamefont {Bernhard},
  \citenamefont {Bass},\ and\ \citenamefont {Heinz}}]{Shen:2014vra}%
  \BibitemOpen
  \bibfield  {author} {\bibinfo {author} {\bibfnamefont {C.}~\bibnamefont
  {Shen}}, \bibinfo {author} {\bibfnamefont {Z.}~\bibnamefont {Qiu}}, \bibinfo
  {author} {\bibfnamefont {H.}~\bibnamefont {Song}}, \bibinfo {author}
  {\bibfnamefont {J.}~\bibnamefont {Bernhard}}, \bibinfo {author}
  {\bibfnamefont {S.}~\bibnamefont {Bass}},\ and\ \bibinfo {author}
  {\bibfnamefont {U.}~\bibnamefont {Heinz}},\ }\bibfield  {title} {\bibinfo
  {title} {{The iEBE-VISHNU code package for relativistic heavy-ion
  collisions}},\ }\href {https://doi.org/10.1016/j.cpc.2015.08.039} {\bibfield
  {journal} {\bibinfo  {journal} {Comput. Phys. Commun.}\ }\textbf {\bibinfo
  {volume} {199}},\ \bibinfo {pages} {61} (\bibinfo {year} {2016})},\ \Eprint
  {https://arxiv.org/abs/1409.8164} {arXiv:1409.8164 [nucl-th]} \BibitemShut
  {NoStop}%
\bibitem [{\citenamefont {Miller}\ \emph {et~al.}(2007)\citenamefont {Miller},
  \citenamefont {Reygers}, \citenamefont {Sanders},\ and\ \citenamefont
  {Steinberg}}]{Miller:2007ri}%
  \BibitemOpen
  \bibfield  {author} {\bibinfo {author} {\bibfnamefont {M.~L.}\ \bibnamefont
  {Miller}}, \bibinfo {author} {\bibfnamefont {K.}~\bibnamefont {Reygers}},
  \bibinfo {author} {\bibfnamefont {S.~J.}\ \bibnamefont {Sanders}},\ and\
  \bibinfo {author} {\bibfnamefont {P.}~\bibnamefont {Steinberg}},\ }\bibfield
  {title} {\bibinfo {title} {{Glauber modeling in high energy nuclear
  collisions}},\ }\href {https://doi.org/10.1146/annurev.nucl.57.090506.123020}
  {\bibfield  {journal} {\bibinfo  {journal} {Ann. Rev. Nucl. Part. Sci.}\
  }\textbf {\bibinfo {volume} {57}},\ \bibinfo {pages} {205} (\bibinfo {year}
  {2007})},\ \Eprint {https://arxiv.org/abs/nucl-ex/0701025}
  {arXiv:nucl-ex/0701025} \BibitemShut {NoStop}%
\bibitem [{\citenamefont {Hulcher}\ \emph {et~al.}(2018)\citenamefont
  {Hulcher}, \citenamefont {Pablos},\ and\ \citenamefont
  {Rajagopal}}]{Hulcher:2017cpt}%
  \BibitemOpen
  \bibfield  {author} {\bibinfo {author} {\bibfnamefont {Z.}~\bibnamefont
  {Hulcher}}, \bibinfo {author} {\bibfnamefont {D.}~\bibnamefont {Pablos}},\
  and\ \bibinfo {author} {\bibfnamefont {K.}~\bibnamefont {Rajagopal}},\
  }\bibfield  {title} {\bibinfo {title} {{Resolution Effects in the Hybrid
  Strong/Weak Coupling Model}},\ }\href
  {https://doi.org/10.1007/JHEP03(2018)010} {\bibfield  {journal} {\bibinfo
  {journal} {JHEP}\ }\textbf {\bibinfo {volume} {03}},\ \bibinfo {pages}
  {010}},\ \Eprint {https://arxiv.org/abs/1707.05245} {arXiv:1707.05245
  [hep-ph]} \BibitemShut {NoStop}%
\bibitem [{\citenamefont {Casalderrey-Solana}\ \emph
  {et~al.}(2017)\citenamefont {Casalderrey-Solana}, \citenamefont {Gulhan},
  \citenamefont {Milhano}, \citenamefont {Pablos},\ and\ \citenamefont
  {Rajagopal}}]{Casalderrey-Solana:2016jvj}%
  \BibitemOpen
  \bibfield  {author} {\bibinfo {author} {\bibfnamefont {J.}~\bibnamefont
  {Casalderrey-Solana}}, \bibinfo {author} {\bibfnamefont {D.}~\bibnamefont
  {Gulhan}}, \bibinfo {author} {\bibfnamefont {G.}~\bibnamefont {Milhano}},
  \bibinfo {author} {\bibfnamefont {D.}~\bibnamefont {Pablos}},\ and\ \bibinfo
  {author} {\bibfnamefont {K.}~\bibnamefont {Rajagopal}},\ }\bibfield  {title}
  {\bibinfo {title} {{Angular Structure of Jet Quenching Within a Hybrid
  Strong/Weak Coupling Model}},\ }\href
  {https://doi.org/10.1007/JHEP03(2017)135} {\bibfield  {journal} {\bibinfo
  {journal} {JHEP}\ }\textbf {\bibinfo {volume} {03}},\ \bibinfo {pages}
  {135}},\ \Eprint {https://arxiv.org/abs/1609.05842} {arXiv:1609.05842
  [hep-ph]} \BibitemShut {NoStop}%
\bibitem [{\citenamefont {Ovanesyan}\ and\ \citenamefont
  {Vitev}(2011)}]{Ovanesyan:2011xy}%
  \BibitemOpen
  \bibfield  {author} {\bibinfo {author} {\bibfnamefont {G.}~\bibnamefont
  {Ovanesyan}}\ and\ \bibinfo {author} {\bibfnamefont {I.}~\bibnamefont
  {Vitev}},\ }\bibfield  {title} {\bibinfo {title} {{An effective theory for
  jet propagation in dense QCD matter: jet broadening and medium-induced
  bremsstrahlung}},\ }\href {https://doi.org/10.1007/JHEP06(2011)080}
  {\bibfield  {journal} {\bibinfo  {journal} {JHEP}\ }\textbf {\bibinfo
  {volume} {06}},\ \bibinfo {pages} {080}},\ \Eprint
  {https://arxiv.org/abs/1103.1074} {arXiv:1103.1074 [hep-ph]} \BibitemShut
  {NoStop}%
\bibitem [{\citenamefont {Kumar}\ \emph {et~al.}(2023)\citenamefont {Kumar}
  \emph {et~al.}}]{JETSCAPE:2022jer}%
  \BibitemOpen
  \bibfield  {author} {\bibinfo {author} {\bibfnamefont {A.}~\bibnamefont
  {Kumar}} \emph {et~al.} (\bibinfo {collaboration} {JETSCAPE}),\ }\bibfield
  {title} {\bibinfo {title} {{Inclusive jet and hadron suppression in a
  multistage approach}},\ }\href {https://doi.org/10.1103/PhysRevC.107.034911}
  {\bibfield  {journal} {\bibinfo  {journal} {Phys. Rev. C}\ }\textbf {\bibinfo
  {volume} {107}},\ \bibinfo {pages} {034911} (\bibinfo {year} {2023})},\
  \Eprint {https://arxiv.org/abs/2204.01163} {arXiv:2204.01163 [hep-ph]}
  \BibitemShut {NoStop}%
\bibitem [{\citenamefont {Du}\ \emph {et~al.}(2020)\citenamefont {Du},
  \citenamefont {Pablos},\ and\ \citenamefont {Tywoniuk}}]{Du:2020pmp}%
  \BibitemOpen
  \bibfield  {author} {\bibinfo {author} {\bibfnamefont {Y.-L.}\ \bibnamefont
  {Du}}, \bibinfo {author} {\bibfnamefont {D.}~\bibnamefont {Pablos}},\ and\
  \bibinfo {author} {\bibfnamefont {K.}~\bibnamefont {Tywoniuk}},\ }\bibfield
  {title} {\bibinfo {title} {{Deep learning jet modifications in heavy-ion
  collisions}},\ }\href {https://doi.org/10.1007/JHEP03(2021)206} {\bibfield
  {journal} {\bibinfo  {journal} {JHEP}\ }\textbf {\bibinfo {volume} {21}},\
  \bibinfo {pages} {206}},\ \Eprint {https://arxiv.org/abs/2012.07797}
  {arXiv:2012.07797 [hep-ph]} \BibitemShut {NoStop}%
\bibitem [{\citenamefont {Mehtar-Tani}\ \emph {et~al.}(2021)\citenamefont
  {Mehtar-Tani}, \citenamefont {Pablos},\ and\ \citenamefont
  {Tywoniuk}}]{Mehtar-Tani:2021fud}%
  \BibitemOpen
  \bibfield  {author} {\bibinfo {author} {\bibfnamefont {Y.}~\bibnamefont
  {Mehtar-Tani}}, \bibinfo {author} {\bibfnamefont {D.}~\bibnamefont
  {Pablos}},\ and\ \bibinfo {author} {\bibfnamefont {K.}~\bibnamefont
  {Tywoniuk}},\ }\bibfield  {title} {\bibinfo {title} {{Cone-Size Dependence of
  Jet Suppression in Heavy-Ion Collisions}},\ }\href
  {https://doi.org/10.1103/PhysRevLett.127.252301} {\bibfield  {journal}
  {\bibinfo  {journal} {Phys. Rev. Lett.}\ }\textbf {\bibinfo {volume} {127}},\
  \bibinfo {pages} {252301} (\bibinfo {year} {2021})},\ \Eprint
  {https://arxiv.org/abs/2101.01742} {arXiv:2101.01742 [hep-ph]} \BibitemShut
  {NoStop}%
\bibitem [{CMS(2023)}]{CMS:2023cka}%
  \BibitemOpen
  \bibfield  {title} {\bibinfo {title} {{Groomed jet radius and girth of jets
  recoiling against isolated photons in PbPb and pp collisions at
  $5.02~\mathrm{TeV}$}},\ }\href@noop {} {\  (\bibinfo {year}
  {2023})}\BibitemShut {NoStop}%
\bibitem [{\citenamefont {Armesto}\ \emph {et~al.}(2004)\citenamefont
  {Armesto}, \citenamefont {Salgado},\ and\ \citenamefont
  {Wiedemann}}]{Armesto:2003jh}%
  \BibitemOpen
  \bibfield  {author} {\bibinfo {author} {\bibfnamefont {N.}~\bibnamefont
  {Armesto}}, \bibinfo {author} {\bibfnamefont {C.~A.}\ \bibnamefont
  {Salgado}},\ and\ \bibinfo {author} {\bibfnamefont {U.~A.}\ \bibnamefont
  {Wiedemann}},\ }\bibfield  {title} {\bibinfo {title} {{Medium induced gluon
  radiation off massive quarks fills the dead cone}},\ }\href
  {https://doi.org/10.1103/PhysRevD.69.114003} {\bibfield  {journal} {\bibinfo
  {journal} {Phys. Rev. D}\ }\textbf {\bibinfo {volume} {69}},\ \bibinfo
  {pages} {114003} (\bibinfo {year} {2004})},\ \Eprint
  {https://arxiv.org/abs/hep-ph/0312106} {arXiv:hep-ph/0312106} \BibitemShut
  {NoStop}%
\bibitem [{\citenamefont {Andres}\ \emph
  {et~al.}(2023{\natexlab{c}})\citenamefont {Andres}, \citenamefont
  {Dominguez}, \citenamefont {Holguin}, \citenamefont {Marquet},\ and\
  \citenamefont {Moult}}]{Andres:2023ymw}%
  \BibitemOpen
  \bibfield  {author} {\bibinfo {author} {\bibfnamefont {C.}~\bibnamefont
  {Andres}}, \bibinfo {author} {\bibfnamefont {F.}~\bibnamefont {Dominguez}},
  \bibinfo {author} {\bibfnamefont {J.}~\bibnamefont {Holguin}}, \bibinfo
  {author} {\bibfnamefont {C.}~\bibnamefont {Marquet}},\ and\ \bibinfo {author}
  {\bibfnamefont {I.}~\bibnamefont {Moult}},\ }\bibfield  {title} {\bibinfo
  {title} {{Seeing Beauty in the Quark-Gluon Plasma with Energy Correlators}},\
  }\href@noop {} {\  (\bibinfo {year} {2023}{\natexlab{c}})},\ \Eprint
  {https://arxiv.org/abs/2307.15110} {arXiv:2307.15110 [hep-ph]} \BibitemShut
  {NoStop}%
\bibitem [{\citenamefont {Caucal}\ \emph
  {et~al.}(2021{\natexlab{b}})\citenamefont {Caucal}, \citenamefont
  {Soto-Ontoso},\ and\ \citenamefont {Takacs}}]{Caucal:2021bae}%
  \BibitemOpen
  \bibfield  {author} {\bibinfo {author} {\bibfnamefont {P.}~\bibnamefont
  {Caucal}}, \bibinfo {author} {\bibfnamefont {A.}~\bibnamefont
  {Soto-Ontoso}},\ and\ \bibinfo {author} {\bibfnamefont {A.}~\bibnamefont
  {Takacs}},\ }\bibfield  {title} {\bibinfo {title} {{Dynamical Grooming meets
  LHC data}},\ }\href {https://doi.org/10.1007/JHEP07(2021)020} {\bibfield
  {journal} {\bibinfo  {journal} {JHEP}\ }\textbf {\bibinfo {volume} {07}},\
  \bibinfo {pages} {020}},\ \Eprint {https://arxiv.org/abs/2103.06566}
  {arXiv:2103.06566 [hep-ph]} \BibitemShut {NoStop}%
\bibitem [{\citenamefont {Mulligan}\ and\ \citenamefont
  {Ploskon}(2020)}]{Mulligan:2020tim}%
  \BibitemOpen
  \bibfield  {author} {\bibinfo {author} {\bibfnamefont {J.}~\bibnamefont
  {Mulligan}}\ and\ \bibinfo {author} {\bibfnamefont {M.}~\bibnamefont
  {Ploskon}},\ }\bibfield  {title} {\bibinfo {title} {{Identifying groomed jet
  splittings in heavy-ion collisions}},\ }\href
  {https://doi.org/10.1103/PhysRevC.102.044913} {\bibfield  {journal} {\bibinfo
   {journal} {Phys. Rev. C}\ }\textbf {\bibinfo {volume} {102}},\ \bibinfo
  {pages} {044913} (\bibinfo {year} {2020})},\ \Eprint
  {https://arxiv.org/abs/2006.01812} {arXiv:2006.01812 [hep-ph]} \BibitemShut
  {NoStop}%
\bibitem [{\citenamefont {Berta}\ \emph {et~al.}(2014)\citenamefont {Berta},
  \citenamefont {Spousta}, \citenamefont {Miller},\ and\ \citenamefont
  {Leitner}}]{Berta:2014eza}%
  \BibitemOpen
  \bibfield  {author} {\bibinfo {author} {\bibfnamefont {P.}~\bibnamefont
  {Berta}}, \bibinfo {author} {\bibfnamefont {M.}~\bibnamefont {Spousta}},
  \bibinfo {author} {\bibfnamefont {D.~W.}\ \bibnamefont {Miller}},\ and\
  \bibinfo {author} {\bibfnamefont {R.}~\bibnamefont {Leitner}},\ }\bibfield
  {title} {\bibinfo {title} {{Particle-level pileup subtraction for jets and
  jet shapes}},\ }\href {https://doi.org/10.1007/JHEP06(2014)092} {\bibfield
  {journal} {\bibinfo  {journal} {JHEP}\ }\textbf {\bibinfo {volume} {06}},\
  \bibinfo {pages} {092}},\ \Eprint {https://arxiv.org/abs/1403.3108}
  {arXiv:1403.3108 [hep-ex]} \BibitemShut {NoStop}%
\bibitem [{\citenamefont {Ehlers}(2023)}]{Ehlers:2023jbf}%
  \BibitemOpen
  \bibfield  {author} {\bibinfo {author} {\bibfnamefont {R.}~\bibnamefont
  {Ehlers}} (\bibinfo {collaboration} {ALICE}),\ }\bibfield  {title} {\bibinfo
  {title} {{Exploring medium properties with hard transverse momentum
  splittings using groomed jet substructure measurements in Pb--Pb collisions
  with ALICE}}\ }(\bibinfo {year} {2023})\ \Eprint
  {https://arxiv.org/abs/2310.07065} {arXiv:2310.07065 [nucl-ex]} \BibitemShut
  {NoStop}%
\bibitem [{\citenamefont {{ATLAS Collaboration}}(2020)}]{ATLAS:2020bbn}%
  \BibitemOpen
  \bibfield  {author} {\bibinfo {author} {\bibnamefont {{ATLAS Collaboration}}}
  (\bibinfo {collaboration} {ATLAS}),\ }\bibfield  {title} {\bibinfo {title}
  {{Measurement of the Lund Jet Plane Using Charged Particles in 13 TeV
  Proton-Proton Collisions with the ATLAS Detector}},\ }\href
  {https://doi.org/10.1103/PhysRevLett.124.222002} {\bibfield  {journal}
  {\bibinfo  {journal} {Phys. Rev. Lett.}\ }\textbf {\bibinfo {volume} {124}},\
  \bibinfo {pages} {222002} (\bibinfo {year} {2020})},\ \Eprint
  {https://arxiv.org/abs/2004.03540} {arXiv:2004.03540 [hep-ex]} \BibitemShut
  {NoStop}%
\bibitem [{\citenamefont {{CMS Collaboration}}(2023)}]{CMS:2023ovl}%
  \BibitemOpen
  \bibfield  {author} {\bibinfo {author} {\bibnamefont {{CMS Collaboration}}}
  (\bibinfo {collaboration} {CMS}),\ }\bibfield  {title} {\bibinfo {title}
  {{Measurement of the primary Lund jet plane density in proton-proton
  collisions at $\sqrt{s} = 13~\mathrm{TeV}$}},\ }\href@noop {} {\  (\bibinfo
  {year} {2023})}\BibitemShut {NoStop}%
\bibitem [{ALI(2021)}]{ALICE:2021yet}%
  \BibitemOpen
  \bibfield  {title} {\bibinfo {title} {{Physics Preliminary Summary:
  Measurement of the primary Lund plane density in pp collisions at $\sqrt{s} =
  \rm{13}$ TeV with ALICE}},\ }\href@noop {} {\  (\bibinfo {year}
  {2021})}\BibitemShut {NoStop}%
\bibitem [{\citenamefont {CERN}(2023)}]{Chamonix2023}%
  \BibitemOpen
  \bibfield  {author} {\bibinfo {author} {\bibnamefont {CERN}},\ }\href
  {https://en.web.cern.ch/event/lhc-chamonix-workshop-2023} {\bibinfo {title}
  {Lhc chamonix workshop 2023}} (\bibinfo {year} {2023})\BibitemShut {NoStop}%
\bibitem [{\citenamefont {{ATLAS Collaboration}}(2023)}]{ATLAS:2023hso}%
  \BibitemOpen
  \bibfield  {author} {\bibinfo {author} {\bibnamefont {{ATLAS Collaboration}}}
  (\bibinfo {collaboration} {ATLAS}),\ }\bibfield  {title} {\bibinfo {title}
  {{Measurement of suppression of large-radius jets and its dependence on
  substructure in Pb+Pb collisions at $\sqrt{s_\mathrm{NN}} = 5.02$ TeV with
  the ATLAS detector}},\ }\href@noop {} {\  (\bibinfo {year} {2023})},\ \Eprint
  {https://arxiv.org/abs/2301.05606} {arXiv:2301.05606 [nucl-ex]} \BibitemShut
  {NoStop}%
\bibitem [{\citenamefont {Mehtar-Tani}\ and\ \citenamefont
  {Tywoniuk}(2018)}]{Mehtar-Tani:2017web}%
  \BibitemOpen
  \bibfield  {author} {\bibinfo {author} {\bibfnamefont {Y.}~\bibnamefont
  {Mehtar-Tani}}\ and\ \bibinfo {author} {\bibfnamefont {K.}~\bibnamefont
  {Tywoniuk}},\ }\bibfield  {title} {\bibinfo {title} {{Sudakov suppression of
  jets in QCD media}},\ }\href {https://doi.org/10.1103/PhysRevD.98.051501}
  {\bibfield  {journal} {\bibinfo  {journal} {Phys. Rev. D}\ }\textbf {\bibinfo
  {volume} {98}},\ \bibinfo {pages} {051501} (\bibinfo {year} {2018})},\
  \Eprint {https://arxiv.org/abs/1707.07361} {arXiv:1707.07361 [hep-ph]}
  \BibitemShut {NoStop}%
\bibitem [{\citenamefont {Blaizot}\ and\ \citenamefont
  {Iancu}(2002)}]{Blaizot:2001nr}%
  \BibitemOpen
  \bibfield  {author} {\bibinfo {author} {\bibfnamefont {J.-P.}\ \bibnamefont
  {Blaizot}}\ and\ \bibinfo {author} {\bibfnamefont {E.}~\bibnamefont
  {Iancu}},\ }\bibfield  {title} {\bibinfo {title} {{The Quark gluon plasma:
  Collective dynamics and hard thermal loops}},\ }\href
  {https://doi.org/10.1016/S0370-1573(01)00061-8} {\bibfield  {journal}
  {\bibinfo  {journal} {Phys. Rept.}\ }\textbf {\bibinfo {volume} {359}},\
  \bibinfo {pages} {355} (\bibinfo {year} {2002})},\ \Eprint
  {https://arxiv.org/abs/hep-ph/0101103} {arXiv:hep-ph/0101103} \BibitemShut
  {NoStop}%
\bibitem [{\citenamefont {Kapusta}\ and\ \citenamefont
  {Gale}(2006)}]{kapusta_gale_2006}%
  \BibitemOpen
  \bibfield  {author} {\bibinfo {author} {\bibfnamefont {J.~I.}\ \bibnamefont
  {Kapusta}}\ and\ \bibinfo {author} {\bibfnamefont {C.}~\bibnamefont {Gale}},\
  }\href {https://doi.org/10.1017/CBO9780511535130} {\emph {\bibinfo {title}
  {Finite-Temperature Field Theory}}},\ \bibinfo {edition} {2nd}\ ed.,\
  Cambridge Monographs on Mathematical Physics\ (\bibinfo  {publisher}
  {Cambridge University Press},\ \bibinfo {year} {2006})\BibitemShut {NoStop}%
\bibitem [{\citenamefont {Milhano}\ and\ \citenamefont
  {Zapp}(2022)}]{Milhano:2022kzx}%
  \BibitemOpen
  \bibfield  {author} {\bibinfo {author} {\bibfnamefont {J.~G.}\ \bibnamefont
  {Milhano}}\ and\ \bibinfo {author} {\bibfnamefont {K.}~\bibnamefont {Zapp}},\
  }\bibfield  {title} {\bibinfo {title} {{Improved background subtraction and a
  fresh look at jet sub-structure in JEWEL}},\ }\href
  {https://doi.org/10.1140/epjc/s10052-022-10954-1} {\bibfield  {journal}
  {\bibinfo  {journal} {Eur. Phys. J. C}\ }\textbf {\bibinfo {volume} {82}},\
  \bibinfo {pages} {1010} (\bibinfo {year} {2022})},\ \Eprint
  {https://arxiv.org/abs/2207.14814} {arXiv:2207.14814 [hep-ph]} \BibitemShut
  {NoStop}%
\bibitem [{\citenamefont {Mehtar-Tani}\ \emph {et~al.}(2023)\citenamefont
  {Mehtar-Tani}, \citenamefont {Schlichting},\ and\ \citenamefont
  {Soudi}}]{Mehtar-Tani:2022zwf}%
  \BibitemOpen
  \bibfield  {author} {\bibinfo {author} {\bibfnamefont {Y.}~\bibnamefont
  {Mehtar-Tani}}, \bibinfo {author} {\bibfnamefont {S.}~\bibnamefont
  {Schlichting}},\ and\ \bibinfo {author} {\bibfnamefont {I.}~\bibnamefont
  {Soudi}},\ }\bibfield  {title} {\bibinfo {title} {{Jet thermalization in QCD
  kinetic theory}},\ }\href {https://doi.org/10.1007/JHEP05(2023)091}
  {\bibfield  {journal} {\bibinfo  {journal} {JHEP}\ }\textbf {\bibinfo
  {volume} {05}},\ \bibinfo {pages} {091}},\ \Eprint
  {https://arxiv.org/abs/2209.10569} {arXiv:2209.10569 [hep-ph]} \BibitemShut
  {NoStop}%
\end{thebibliography}%

%%%%%%%%%%%%%%%%%%%%%%%%%%%%%%%%%%%%%%%%%%
%%%%%%%%%%%%%%%%%%%%%%%%%%%%%%%%%%%%%%%%%%

\end{document}